\documentclass[twocolumn,showpacs,aps,prb,amsmath,amssymb,floatfix,eqsecnum]{revtex4-1}
\usepackage{amsmath,amssymb}
\usepackage{bm}
\usepackage{graphicx}
\usepackage{psfrag}
\usepackage{color}

\newcommand{\bd}{\bm}

\begin{document}

\title{Non-equilibrium time evolution of bosons from the functional renormalization
group}

\author{Thomas Kloss$^{1,2}$ and Peter Kopietz$^1$}

\affiliation{
  $^1$Institut f\"{u}r Theoretische Physik, Universit\"{a}t
  Frankfurt,  Max-von-Laue Strasse 1, 60438 Frankfurt, Germany \\
  $^2$Univ. Grenoble 1/CNRS, LPMMC UMR 5493, Maison des Magist\`{e}res, 38042 Grenoble, France
}

\date{Mai 20, 2011}

 \begin{abstract}
We develop a functional renormalization group
approach to obtain the time evolution
of the momentum distribution function of interacting bosons out of equilibrium.
Using an external out-scattering rate as
flow parameter,
we derive formally exact renormalization group flow equations for
the non-equilibrium self-energies in the Keldysh basis.
A simple perturbative truncation of these  flow equations
leads to an approximate solution of the
quantum Boltzmann equation which
does not suffer from secular terms and gives accurate
results even for long times.
We demonstrate this explicitly within a simple
exactly solvable toy model describing  a quartic oscillator
with off-diagonal pairing terms.

\end{abstract}

\pacs{05.70.Ln, 05.10.Cc, 05.30.Jp, 76.20.+q}

\maketitle


\section{Introduction}
\label{sec:intro}

Quantum mechanical many-body systems
out of equilibrium pose extraordinary  challenges to theory.
Although
powerful  field-theoretical methods to
formulate non-equilibrium
problems
in terms of  Green functions and Feynman diagrams are available,\cite{Schwinger61,Bakshi63,*Bakshi63a,Kadanoff62,Keldysh64,Rammer07,Haug08,Kamenev04,Bonitz10}
in practice new concepts and approximation strategies are needed.
Because systems under the influence of external time-dependent fields
are not time-translationally invariant,
one has to formulate theoretical descriptions in the time domain.
Moreover, even if after a sufficiently long time
the system has approached a stationary non-equilibrium state,
such a state can exhibit
properties which are rather different from a thermal equilibrium state.
For example, the Fourier transform
$n_{\bd{k}}$ of the distribution function at such a non-thermal fixed point
can exhibit a scaling behavior as
function of momentum $\bd{k}$
which is characterized by a completely different exponent than
under equilibrium conditions.~\cite{Berges08,Berges09}
In this case a simple perturbative approach based on the quantum Boltzmann equation
with collision integrals calculated in lowest order Born approximation
is not sufficient.  The scaling behavior close
to non-thermal fixed points in simple models
has been studied within a next-to-leading order
$1/N$-approximation.\cite{Berges09}
This method has also been used to study the real-time dynamics of
quantum many-body systems far from equilibrium. \cite{Berges02a,*Berges02b,*Berges04,*Berges07}

Another useful method to investigate strongly correlated many-body systems
is the renormalization group (RG), which has been extensively used to study
the scaling properties of systems in the vicinity of continuous phase transitions.\cite{Goldenfeld}
Although there are many successful applications
of RG methods to systems in thermal equilibrium, there are only a few examples
where RG methods have been used to study
quantum mechanical many-body systems out of 
equilibrium.
Some authors\cite{Berges09,Mitra06,*Takei10} have focused 
on stationary non-equilibrium states, where the system is time-translationally
invariant so that the quantum dynamic equations can be formulated 
in the frequency domain.
On the other hand, the more difficult problem of obtaining the
time-evolution of quantum many-body systems out of equilibrium 
has been studied using various implementations of the
RG idea, such as the  numerical renormalization group approach,\cite{Anders05}
the density matrix renormalization group,\cite{Schmittecker04} 
real-time RG formulations in Liouville space,\cite{Schoeller00,*Keil01}
and a flow equation approach employing
continuous unitary transformations.\cite{Hackl09,Moeckel09}
In recent years a number of authors have also applied
 functional renormalization
group (FRG) methods to study many-body systems out of
equilibrium.\cite{canet04,*canet07,*canet10,Gezzi07,Jakobs07,Jakobs10,Karrasch10a,JakobsDr09,*KarraschDr10,Schoeller09,*Pletyukhov10,*Karrasch10,Gasenzer08,*Gasenzer10,Berges09}
While the non-equilibrium FRG approach to quantum dots \cite{Gezzi07,Jakobs07,Jakobs10} has so far
only been applied to study stationary non-equilibrium states,
Gasenzer and Pawlowski \cite{Gasenzer08,*Gasenzer10} have written down
a formally exact hierarchy of FRG flow equations describing the time evolution of the
one-particle irreducible vertices of interacting bosons out of equilibrium.
Using a sharp time cutoff  as RG flow parameter,
they showed that
a simple truncation of the FRG vertex expansion at the level of the four-point
vertex reproduces the results of the next-to-leading order $1/N$-expansion.\cite{Berges08}

The specific choice of the cutoff procedure is a crucial point when performing RG calculations and different schemes had been proposed in earlier attempts to 
describe non-equilibrium problems. \cite{Jakobs07,Schoeller09,*Pletyukhov10,*Karrasch10,Schoeller00,*Keil01}
In this work we shall propose an alternative
version of the non-equilibrium FRG which uses
an external out-scattering rate as flow parameter.
Such a cutoff scheme is closely related to the
``hybridization cutoff scheme'' recently proposed by Jakobs, Pletyukhov and Schoeller. \cite{Jakobs10a,Jakobs10,Karrasch10}
Technically,
we implement this cutoff scheme by replacing
the infinitesimal $\pm i \eta$ defining the boundary
condition of the retarded and advanced Green functions by finite quantities
$\pm i \Lambda$.
Given this cutoff procedure,
a simple  substitution in the usual hierarchy of FRG flow equations
for the irreducible vertices\cite{Kopietz10} yields the
FRG flow equations describing the
evolution of the irreducible vertices as the flow parameter $\Lambda$ is reduced.
An important property of our cutoff scheme is that
it preserves the triangular structure
in the Keldysh basis and hence
does not violate causality.

To be specific, we shall develop our formalism for the following
time-dependent interacting boson Hamiltonian,
\begin{align}
 {\cal{H}} ( t )  = &
\sum_{\bd{k}} \left[ \epsilon_{\bd{k}} a_{\bd{k}}^{\dagger} a_{\bd{k}}
+
\frac{\gamma_{\bd{k}}}{2} e^{ -i \omega_0 t }  a^{\dagger}_{\bd{k}} a^{\dagger}_{ - \bd{k}}
+ \frac{\gamma_{\bd{k}}^{\ast}}{2} e^{  i \omega_0 t }  a_{- \bd{k}} a_{\bd{k}}
 \right]
 \nonumber
 \\
 + & \frac{1}{2V} \sum_{ \bd{k}_1  \bd{k}_2  \bd{k}_3  \bd{k}_4}
\delta_{\bd{k}_1 + \bd{k}_2 + \bd{k}_3 + \bd{k}_4,0 }
 \nonumber
 \\
 &  \times U (  \bd{k}_1 ,   \bd{k}_2  ; \bd{k}_3 , \bd{k}_4 )
 a^{\dagger}_{-\bd{k}_1}
a^{\dagger}_{ - \bd{k}_2  }    a_{\bd{k}_3} a_{ \bd{k}_4} ,
 \label{eq:Hpr}
 \end{align}
where $a^{\dagger}_{\bd{k}}$ creates a boson with (crystal) momentum $\bd{k}$ and energy
dispersion
$\epsilon_{\bd{k}}$, the volume of the system is denoted by $V$, and
$ U (  \bd{k}_1 ,   \bd{k}_2  ; \bd{k}_3 , \bd{k}_4 ) $ is some momentum-dependent interaction
function; the minus signs in front of the momentum labels
of the creation operators in the last line of Eq.~(\ref{eq:Hpr}) are introduced for later convenience.
As explained in Refs.~[\onlinecite{Zakharov70,*Zakharov74},\,\onlinecite{Lvov94}]
the  Hamiltonian (\ref{eq:Hpr}) describes the non-equilibrium dynamics of
magnons in ordered dipolar ferromagnets such as
yttrium-iron garnet \cite{Cherepanov93}
subject to an external harmonically
oscillating microwave field with frequency $\omega_0$.
The energy scale $\gamma_{\bd{k}}$ is then
proportional to the amplitude of the microwave field.
The non-equilibrium dynamics generated by the Hamiltonian
(\ref{eq:Hpr}) is very rich and exhibits the phenomenon of
parametric resonance for sufficiently strong pumping.\cite{Zakharov70,*Zakharov74,Lvov94}

In practice
 it is often useful to remove the explicit time dependence
from the Hamiltonian $ {\cal{H}} ( t )$ in Eq.~(\ref{eq:Hpr}) by going to the
rotating reference frame.
The effective time-independent Hamiltonian
$\tilde{\cal{H}}$ in the rotating reference frame (denoted by a tilde) is obtained as follows:
Given the unitary time evolution operator
${\cal{U}} ( t )$ defined by
 \begin{equation}
 i \partial_t {\cal{U}} ( t ) = {\cal{H}} ( t ) {\cal{U}} ( t ),
 \end{equation}
we make the factorization ansatz
 \begin{equation}
{\cal{U}} ( t ) = {\cal{U}}_0 ( t ) \tilde{\cal{U}} ( t ),
 \label{eq:Ufactor}
 \end{equation}
 with
 \begin{equation}
 {\cal{U}}_0 ( t )
 =
 e^{ - \frac{i}{2}  \sum_{\bd{k}} ( \omega_0 t - \varphi_{\bd{k}} ) a_{\bd{k}}^{\dagger}
 a_{\bd{k}} } ,
 \end{equation}
where $\varphi_{\bd{k}}$ is the phase of
$\gamma_{\bd{k}} = | \gamma_{\bd{k}} | e^{i \varphi_{\bd{k}} }$.
The time evolution operator $\tilde{\cal{U}} ( t )$
in the rotating reference frame then satisfies
 \begin{equation}
 i \partial_t \tilde{\cal{U}} ( t ) = \tilde{\cal{H}} \,  \tilde{\cal{U}} ( t ),
 \label{eq:tildeU}
\end{equation}
where the transformed Hamiltonian $\tilde{\cal{H}}$ does not explicitly depend on time,
 \begin{eqnarray}
  \tilde{\cal{H}} & = &  {\cal{U}}_0^{\dagger} ( t ) [ {\cal{H}} ( t ) - i \partial_t ] {\cal{U}}_0 ( t )
 \nonumber
 \\
  &  = & \sum_{\bd{k}} \left[ \tilde{\epsilon}_{\bd{k}} a_{\bd{k}}^{\dagger} a_{\bd{k}}
 +  \frac{| \gamma_{\bd{k}} |}{2} \left(
  a^{\dagger}_{\bd{k}} a^{\dagger}_{ - \bd{k}} +    a_{- \bd{k}} a_{\bd{k}} \right)
 \right]
 \nonumber
 \\
& + & \frac{1}{2V} \sum_{ \bd{k}_1  \bd{k}_2  \bd{k}_3  \bd{k}_4}
\delta_{\bd{k}_1 + \bd{k}_2 + \bd{k}_3 + \bd{k}_4,0 }
 \nonumber
 \\
 & & \times U (  \bd{k}_1 ,   \bd{k}_2  ; \bd{k}_3 , \bd{k}_4 )
 a^{\dagger}_{- \bd{k}_1}
a^{\dagger}_{ - \bd{k}_2  }    a_{\bd{k}_3} a_{ \bd{k}_4} ,
 \label{eq:tildeHdef}
 \end{eqnarray}
with the shifted energy
 \begin{equation}
 \tilde{\epsilon}_{\bd{k}} = \epsilon_{\bd{k}} - \frac{\omega_0}{2}.
 \label{eq:tildeepsilon}
\end{equation}
The solution of Eq.~(\ref{eq:tildeU}) is simply
$ \tilde{\cal{U}} ( t ) = e^{ - i \tilde{\cal{H}} t }$,
so that the total time evolution operator of our system can be explicitly written down,
 \begin{equation}
{\cal{U}} ( t ) =    e^{ - \frac{i}{2}  \sum_{\bd{k}} ( \omega_0 t - \varphi_{\bd{k}} ) a_{\bd{k}}^{\dagger}
 a_{\bd{k}} }
 e^{ - i \tilde{\cal{H}} t }.
 \end{equation}
Throughout this article we shall work in the rotating reference frame
were the effective Hamiltonian (\ref{eq:tildeHdef}) is time
independent. To simplify the notation we shall  rename
$\tilde{\epsilon}_{\bd{k}} \rightarrow \epsilon_{\bd{k}}$ and give all Green functions and distribution functions in the rotating reference frame. Explicit prescriptions to relate
these functions in the original and the rotating reference frame are given in
appendix A. We emphasize that the general FRG formalism developed
in this work remains also valid if the Hamiltonian depends explicitly on time.

The rest of this paper is organized as follows: In
Sec.~\ref{sec:Green} we shall define various types of
non-equilibrium Green functions and represent them in terms of
functional integrals involving a properly symmetrized Keldysh
action. Due to the off-diagonal terms in our Hamiltonian
(\ref{eq:Hpr}) the quantum dynamics is also characterized by
anomalous Green functions involving the simultaneous creation and
annihilation of two bosons. To keep track of these off-diagonal
correlations together with the usual single-particle
correlations, we introduce in Sec.~\ref{sec:Green} a compact matrix
notation. In Sec.~\ref{sec:quantumkinetic} we then
derive several equivalent quantum kinetic equations for the diagonal
and off-diagonal distribution functions. In Sec.~\ref{sec:FRG} we
write down formally exact FRG flow equations for the self-energies
which appear in the collision integrals of the quantum kinetic
equations discussed in Sec.~\ref{sec:quantumkinetic}. We also
discuss several cutoff schemes. In Sec.~\ref{sec:toy} we then use
our non-equilibrium FRG flow equations to study the time evolution
of a simple exactly solvable toy model which is obtained by
retaining in our Hamiltonian (\ref{eq:tildeHdef}) only a single
momentum mode. We show that a rather simple truncation of the FRG
flow equations yields already quite accurate results for the time
evolution. Finally, in Sec.~\ref{sec:summary} we summarize our
results and discuss some open problems. There are four appendices
with additional technical details.

\section{Non-equilibrium Green functions}
\label{sec:Green}

Our goal is to develop methods to calculate the time evolution of the diagonal and
off-diagonal distribution functions
 \begin{subequations}
 \begin{eqnarray}
 n_{\bd{k}} ( t ) & = & \langle a^{\dagger}_{\bd{k}} ( t ) a_{\bd{k}} (t )  \rangle ,
 \label{eq:nkdef}
 \\
p_{\bd{k}} ( t ) & = & \langle a_{- \bd{k}} ( t ) a_{\bd{k}} (t )  \rangle ,
 \label{eq:pkdef}
 \end{eqnarray}
\end{subequations}
where all  operators are in the Heisenberg picture and the
expectation values are with respect to some initial density matrix
$\hat{\rho} ( t_0 ) $ specified at some time $t_0$,
 \begin{equation}
 \langle \ldots \rangle = {\rm Tr} [ \hat{\rho} ( t_0 ) \ldots ].
 \end{equation}
Note that in our Hamiltonian (\ref{eq:tildeHdef})
the combinations $a^{\dagger}_{\bd{k}} a^{\dagger}_{- \bd{k}}$
and  $a_{-\bd{k}} a_{ \bd{k}}$
do not conserve particle number, so that we should also consider
the anomalous distribution function (\ref{eq:pkdef}) and its complex conjugate.
Our final goal is to derive renormalization group equations for the self-energy
appearing in the collision integrals of the
quantum kinetic equations for these distribution functions.
In order to do this, it is useful to collect the various types of
non-equilibrium Green functions into a symmetric matrix, as described in the following section.

\subsection{Keldysh (RAK) basis}
\label{subsec:RAK}

In the Keldysh technique \cite{Keldysh64} one doubles the degrees of
freedom to distinguish between forward and backward propagation in
time. As a consequence, all quantities carry extra indices $p \in
\{+,-\} $ which label the branches of the Keldysh contour associated
with the forward and backward propagation. The single-particle Green
function is then a $2 \times 2$ matrix in Keldysh space.
Alternative formulations of the Keldysh technique in combination with diagonal and off-diagonal terms are known from the theory of non-equilibrium superconductivity, see e.g.\ [\onlinecite{SupercondBook}].
In order to
formulate the Keldysh technique in terms of functional
integrals,\cite{Kamenev04} it is convenient to work in a basis where
causality is manifestly implemented via the vanishing of the lower
diagonal element of the Green function matrices in Keldysh
space. The other matrix elements can then be identified with the
usual retarded ($R$), advanced ($A$) and Keldysh ($K$) Green functions.
Keeping in  mind that our model has also anomalous
correlations, we define the following non-equilibrium Green
functions,
\begin{subequations}
\begin{eqnarray}
g^{R}_{\bd{k}} ( t , t^{\prime} ) & = &
 - i \Theta ( t - t^{\prime} ) \langle [ a_{\bd{k}} ( t ) , a_{\bd{k}}^{\dagger} ( t^{\prime} ) ]
 \rangle ,
 \label{eq:GR}
 \\
 g^{A}_{\bd{k}} ( t , t^{\prime} ) & = &
 i \Theta (  t^{\prime} - t ) \langle [ a_{\bd{k}} ( t ) , a_{\bd{k}}^{\dagger} ( t^{\prime} ) ]
 \rangle ,
 \label{eq:GA}
 \\
g^{K}_{\bd{k}} ( t , t^{\prime} )   & = &
 - i  \langle \bigl\{ a_{\bd{k}} ( t ) , a_{\bd{k}}^{\dagger} ( t^{\prime} ) \bigr\}
 \rangle ,
 \label{eq:GK}
 \end{eqnarray}
 \end{subequations}
where $ [ \; , \;  ]$ denotes the commutator and
$\{ \;  , \;  \}$ is the anticommutator. The  corresponding off-diagonal
Green functions are
 \begin{subequations}
 \begin{eqnarray}
 p^{R}_{\bd{k}} ( t , t^{\prime} ) & = &
 - i \Theta ( t - t^{\prime} ) \langle [ a_{\bd{k}} ( t ) , a_{- \bd{k}} ( t^{\prime} ) ]
 \rangle ,
 \label{eq:DR}
 \\
p^{A}_{\bd{k}} ( t , t^{\prime} ) & = &
 i \Theta (  t^{\prime} - t ) \langle [ a_{\bd{k}} ( t ) , a_{- \bd{k}} ( t^{\prime} ) ]
 \rangle ,
 \label{eq:DA}
 \\
 p^{K}_{\bd{k}} ( t , t^{\prime} ) & = &
 - i  \langle \left\{ a_{\bd{k}} ( t ) , a_{- \bd{k}} ( t^{\prime} ) \right\}
 \rangle .
 \label{eq:DK}
 \end{eqnarray}
 \end{subequations}
In Sec.~\ref{sec:func} we represent these Green functions as functional integrals
involving a suitably defined
Keldysh action.
To write the Gaussian part of this action in a compact form,
it is convenient
to introduce infinite matrices
$\hat{g}^{X}$ and $\hat{p}^X$ in the momentum and time labels
 (where $X=R,A, K$ labels the three types
of Green functions)
whose matrix elements
are related to the Green functions (\ref{eq:GR}--\ref{eq:DK}) as follows,
\footnote{
Our notation is as follows:
large bold-face letters such as $\mathbf{G}$ and $\mathbf{I}$
denote matrices in all labels:  Keldysh labels $\lambda = C, Q$, field-type labels $\sigma = a , \bar{a}$
(which we call flavor labels), as well as momentum and time labels.
If we take matrix elements of $\mathbf{G}$
in Keldysh space the corresponding sub-blocks are denoted by capital letters with a hat, such as
 $\hat{G}^{R}$, $\hat{G}^A$, $\hat{G}^K$  and $\hat{I}$.
If we further specify the flavor labels
the resulting infinite matrices in the momentum- and time labels are
denoted by small letters with a hat, such as
$\hat{g}^R$ and $\hat{p}^R$ and $\hat{1}$, as defined in Eq.~(\ref{eq:GRmat}). The
matrix-elements of these sub-blocks in the momentum- and time labels are the
Green functions defined in Eqs.~(\ref{eq:GR}--\ref{eq:DK}), i.e.
$[\hat{g}^R]_{ \bd{k} t , \bd{k}^{\prime}t^{\prime} } =  \delta_{ \bd{k}, - \bd{k}^{\prime}}
g^{R}_{\bd{k}} (  t, t^{\prime})$  and
$[\hat{p}^R]_{ \bd{k}  t , \bd{k}^{\prime} t^{\prime} } =
 \delta_{ \bd{k}, - \bd{k}^{\prime}}
p^{R}_{ \bd{k}} (  t, t^{\prime})$.
On the other hand, if we take matrix-elements of
$\hat{G}^R$   in the momentum- and time labels we obtain $2 \times 2$-matrices in flavor space,
$ [ \hat{G}^R ]_{ \bd{k} t ,\bd{k}^{\prime} t^{\prime}}
= \delta_{ \bd{k}, - \bd{k}^{\prime}} G^{R}  (\bd{k},  t , t^{\prime} )
$
whose matrix elements consist of  the functions $g^{R}_{\bd{k}}(  t, t^{\prime})$,
$p^{R}_{\bd{k}} (  t, t^{\prime})$ and their complex conjugates, as
defined in Eqs.~(\ref{eq:GRmat}--\ref{eq:GKmat}).
In general, $2 \times 2$ matrices in flavor space are denoted by capital letters such as
$G, Z, M$.
}
 \begin{subequations}
 \begin{eqnarray}
 [ \hat{g}^{X} ]_{ \bd{k} t ,  \bd{k}^{\prime} t^{\prime}} & = &
 \delta_{\bd{k} , -\bd{k}^{\prime} }
g^{X}_{\bd{k}} ( t , t^{\prime} ),
\label{eq:gXmat}
 \\
 {[} \hat{p}^{X} ]_{  \bd{k} t  , \bd{k}^{\prime} t^{\prime}} & = &
 \delta_{ \bd{k} ,- \bd{k}^{\prime}  }
p^{X}_{\bd{k}} ( t , t^{\prime} ).
 \label{eq:pXmat}
 \end{eqnarray}
 \end{subequations}
We have assumed spatial homogeneity so that the Green functions are diagonal matrices in momentum space.
From the above definitions it is easy to show that
the normal blocks satisfy the usual relations \cite{Kamenev04}
\begin{eqnarray}
 ( \hat{g}^R )^{\dagger} =  \hat{g}^A \; \; & , &   \; \;
( \hat{g}^K )^{\dagger} =  - \hat{g}^K,
 \label{eq:symtimenormal}
\end{eqnarray}
while the pairing blocks have the symmetries
 \begin{eqnarray}
( \hat{p}^R )^{T} =  \hat{p}^A  \; \; & , &   \; \;
( \hat{p}^K )^{T} =   \hat{p}^K .
 \label{eq:symtimeano}
\end{eqnarray}
For each type of Green function, we collect
the normal and anomalous components into
larger matrices,
\begin{subequations}
 \begin{eqnarray}
 \hat{G}^{R}  & = &
\left(
 \begin{array}{cc}
    \hat{G}^R_{ a a}   &   \hat{G}^R_{ a \bar{a}}    \\
  \hat{G}^R_{ \bar{a} a}   &   \hat{G}^R_{ \bar{a} \bar{a}}
 \end{array}
 \right)
=
\left(
 \begin{array}{cc}
 \hat{p}^{R} & \hat{g}^R \\
 ( \hat{g}^R)^{\ast} &  ( \hat{p}^R)^{\ast}
 \end{array}
 \right),
 \label{eq:GRmat}
 \\
    \hat{G}^{A}  & = &
\left(
 \begin{array}{cc}
    \hat{G}^A_{ a a}   &   \hat{G}^A_{ a \bar{a}}    \\
  \hat{G}^A_{ \bar{a} a}   &   \hat{G}^A_{ \bar{a} \bar{a}}
 \end{array}
 \right) =
\left(
 \begin{array}{cc}
 \hat{p}^{A} & \hat{g}^A \\
 ( \hat{g}^A)^{\ast} &  ( \hat{p}^A)^{\ast}
 \end{array}
 \right) ,
 \label{eq:GAmat}
 \\
 \hat{G}^{K} & = &
\left(
 \begin{array}{cc}
    \hat{G}^K_{ a a}   &   \hat{G}^K_{ a \bar{a}}    \\
  \hat{G}^K_{ \bar{a} a}   &   \hat{G}^K_{ \bar{a} \bar{a}}
 \end{array}
 \right) =
\left(
 \begin{array}{cc}
 \hat{p}^{K} & \hat{g}^K \\
 - ( \hat{g}^K)^{\ast} &  - ( \hat{p}^K)^{\ast}
 \end{array}
 \right) , \hspace{7mm}
 \label{eq:GKmat}
 \end{eqnarray}
 \end{subequations}
whose blocks $\hat{G}^X_{\sigma \sigma^{\prime}}$ are
infinite matrices in the momentum and time labels.
The subscripts $\sigma , \sigma^{\prime} \in \{ a , \bar{a} \}$
indicate whether the associated Green functions involve
annihilation operators  $a$ or creation operators $a^{\dagger}$.
We shall refer to these subscripts as flavor labels.
The symmetries (\ref{eq:symtimenormal}) and (\ref{eq:symtimeano})
imply
 \begin{equation}
 ( \hat{G}^R )^T  =  \hat{G}^A \; \; , \; \;
 ( \hat{G}^K )^T  =  \hat{G}^K .
 \label{eq:GKT}
\end{equation}
Finally, we collect the blocks (\ref{eq:GRmat}--\ref{eq:GKmat})
into an even larger matrix Green function,
\begin{equation}
 \mathbf{{{G}}}  =
\left(
 \begin{array}{cc}
 {[ \mathbf{{G}}]}^{CC} & {[ \mathbf{{G}}]}^{CQ} \\
 {[ \mathbf{{G}}]}^{QC} & 0
 \end{array}
 \right)  =
\left(
 \begin{array}{cc}
 \hat{G}^{K} & \hat{G}^R \\
 \hat{G}^A & 0
 \end{array}
 \right),
 \label{eq:Gmatdef}
 \end{equation}
where the superscripts
$C$ and $Q$ anticipate that in the functional integral approach
we shall identify the corresponding block with
correlation functions involving the  classical ($C$) and quantum ($Q$)
component of the field, see Eqs.\ (\ref{eq:acq},\ref{eq:abarcq}) below. The symmetries (\ref{eq:GKT}) imply that
the infinite matrix $\mathbf{G}$ is symmetric,
 \begin{equation}
 \mathbf{G}  = \mathbf{G}^T.
 \end{equation}
As emphasized by Vasiliev \cite{Vasilievbook} (see also
Refs.~[\onlinecite{Kopietz10},\,\onlinecite{Schuetz05}]) the symmetrization of the Green function
and all vertices greatly facilitates the derivation of the proper combinatorial factors
in perturbation theory and in the functional renormalization group equations.
The definitions (\ref{eq:gXmat}, \ref{eq:pXmat}, \ref{eq:GKmat}) imply
that at equal times and vanishing total momentum the matrix elements of the Keldysh block $\hat{G}^K$
contain
the diagonal and off-diagonal
distribution functions defined in Eqs.~(\ref{eq:nkdef},\ref{eq:pkdef}),
 \begin{eqnarray}
 [ \hat{G}^K ]_{ \bd{k} t , - \bd{k} t } & = &
\left(
 \begin{array}{cc}
 {p}^{K}_{\bd{k}} ( t,t) & {g}^K_{\bd{k}} (t,t) \\
 g^K_{\bd{k}} (t,t) &   {p}^K_{\bd{k}} (t,t)^{\ast}
 \end{array}
 \right)
 \nonumber
 \\
 & = & - 2 i
\left(
 \begin{array}{cc} p_{\bd{k}} ( t ) & n_{\bd{k}} (t) + \frac{1}{2} \\
  n_{\bd{k}} (t) + \frac{1}{2} & p_{\bd{k}}^{\ast} ( t )
 \end{array}
 \right). \hspace{7mm}
 \end{eqnarray}
For later reference we note that the inverse of the matrix $\mathbf{G} $
in Eq.~(\ref{eq:Gmatdef}) has the block structure
\begin{equation}
 \mathbf{G}^{-1}  = \left(
 \begin{array}{cc}
 0  & (\hat{G}^A)^{-1} \\
 (\hat{G}^R)^{-1} & -   (\hat{G}^R)^{-1}  \hat{G}^{K} (\hat{G}^A)^{-1}
 \end{array}
 \right),
 \label{eq:Gmatinv}
 \end{equation}
where the products in the lower diagonal block
denote the usual matrix multiplication, i.e.,
 \begin{equation}
 [ \hat{A} \hat{B} ]_{ \sigma \bd{k} t , \sigma^{\prime} \bd{k}^{\prime} t^{\prime} }
 = \sum_{\sigma_1}\sum_{ \bd{k}_1} \int d t_1
  [\hat{A}]_{ \sigma \bd{k} t , \sigma_1 \bd{k}_1 t_1 }
 [\hat{B}]_{ \sigma_1 \bd{k}_1 t_1  , \sigma^{\prime} \bd{k}^{\prime} t^{\prime} }.
 \end{equation}
The symmetry relations (\ref{eq:GKT}) guarantee that
the lower diagonal block in Eq.~(\ref{eq:Gmatinv}) is again symmetric.
To introduce a matrix $\hat{F}$ in flavor space
which contains both the normal and the anomalous
components of the distribution function, we
parametrize the Keldysh block in the form
 \begin{equation}
 \hat{G}^K = \hat{G}^R \hat{F} \hat{Z} -  \hat{Z}
 \hat{F}^T \hat{G}^A,
 \label{eq:kelF}
 \end{equation}
where the antisymmetric matrix $\hat{Z} =  - \hat{Z}^T  $ is defined by
 \begin{equation}
\hat{Z} = Z \otimes  \hat{1} =
\left( \begin{array}{cc} 0 & \hat{1} \\
 - \hat{1} & 0 \end{array} \right) .
 \label{eq:Zdef}
 \end{equation}
Here $Z$ is the following antisymmetric $2 \times 2$-matrix in flavor space,
 \begin{equation}
Z = i \sigma_2  = \left( \begin{array}{cc}
  0 & 1 \\  -1 & 0 \end{array} \right),
\end{equation}
and
$\hat{1}$ is the unit matrix in the momentum- and time-labels,
i.e. $[ \hat{1} ]_{ \bd{k} t ,  \bd{k}^{\prime} t^{\prime}} = \delta_{ \bd{k}, \bd{k}^{\prime}}
\delta ( t - t^{\prime} )$.
Substituting Eq.~(\ref{eq:kelF}) into (\ref{eq:Gmatinv}), the lower diagonal block of the matrix (\ref{eq:Gmatinv}) can be written as
 \begin{eqnarray}
 [\mathbf{G}^{-1}]^{QQ} & = &
 -   (\hat{G}^R)^{-1}  \hat{G}^{K} (\hat{G}^A)^{-1}
 \nonumber
 \\
 & = &  (\hat{G}^R)^{-1} \hat{Z} \hat{F}^T -  \hat{F}  \hat{Z}
( \hat{G}^A)^{-1},
 \label{eq:GinvQQ}
 \end{eqnarray}
so that Eq.~(\ref{eq:Gmatinv}) takes the form
\begin{equation}
\mathbf{G}^{-1}  = \left(
 \begin{array}{cc}
  0  & (\hat{G}^A)^{-1} \\
 (\hat{G}^R)^{-1} & (\hat{G}^R)^{-1} \hat{Z} \hat{F}^T -
 \hat{F} \hat{Z} ( \hat{G}^A)^{-1}
 \end{array}
 \right).
 \label{eq:Gmatinv2}
 \end{equation}
In Sec.~\ref{sec:nonint} we shall explicitly calculate
the matrix elements of $\hat{F}$ in the non-interacting limit,
see Eqs.~(\ref{eq:Fdiag}, \ref{eq:Ft}) below.

Introducing the self-energy matrix $\mathbf{\Sigma}$ in all indices via the matrix Dyson
equation
 \begin{equation}
 \label{eq:dyson}
 \mathbf{G}^{-1} = \mathbf{G}_0^{-1} - \mathbf{\Sigma},
 \end{equation}
the non-equilibrium self-energy in the Keldysh basis acquires the
form
\begin{equation}
 \mathbf{\Sigma}  =
\left(
 \begin{array}{cc}
 0                    & {[\mathbf{\Sigma}]}^{CQ} \\
 {[\mathbf{\Sigma}]}^{QC} & {[\mathbf{\Sigma}]}^{QQ}
 \end{array}
 \right)  =
\left(
 \begin{array}{cc}
 0                 & \hat{\Sigma}^A \\
 \hat{\Sigma}^R & \hat{\Sigma}^K
 \end{array}
 \right) .
 \label{eq:Sigmamatdef}
 \end{equation}
The sub-blocks contain the normal and anomalous self-energy matrices
 \begin{subequations}
\begin{align}
    \hat{\Sigma}^{R} &=
 \left(
   \begin{array}{cc}
     {\hat{\Sigma}^R}_{ a a}         &   {\hat{\Sigma}^R}_{ a \bar{a}}       \\
     {\hat{\Sigma}^R}_{ \bar{a} a}   &   {\hat{\Sigma}^R}_{ \bar{a} \bar{a}}
   \end{array}
 \right) =
 \left(
   \begin{array}{cc}
     \hat{\pi}^R         &   \hat{\sigma}^R       \\
    (\hat{\sigma}^R)^*   &  (\hat{\pi}^R)^*
   \end{array}
 \right)
 ,
 \\
  \hat{\Sigma}^{A}  &=
 \left(
   \begin{array}{cc}
     {\hat{\Sigma}^A}_{ a a}         &   {\hat{\Sigma}^A}_{ a \bar{a}}       \\
     {\hat{\Sigma}^A}_{ \bar{a} a}   &   {\hat{\Sigma}^A}_{ \bar{a} \bar{a}}
   \end{array}
 \right) =
 \left(
   \begin{array}{cc}
     \hat{\pi}^A         &   \hat{\sigma}^A       \\
    (\hat{\sigma}^A)^*   &  (\hat{\pi}^A)^*
   \end{array}
 \right)
,
 \\
  \hat{\Sigma}^{K} &=
 \left(
   \begin{array}{cc}
     {\hat{\Sigma}^K}_{ a a}         &   {\hat{\Sigma}^K}_{ a \bar{a}}       \\
     {\hat{\Sigma}^K}_{ \bar{a} a}   &   {\hat{\Sigma}^K}_{ \bar{a} \bar{a}}
   \end{array}
 \right)
=
 \left(
   \begin{array}{cc}
     \hat{\pi}^K         &   \hat{\sigma}^K       \\
   - (\hat{\sigma}^K)^*   &  - (\hat{\pi}^K)^*
   \end{array}
 \right).
\end{align}
\end{subequations}
The Dyson equation (\ref{eq:dyson}) and the symmetries (\ref{eq:symtimenormal},\ref{eq:symtimeano},\ref{eq:GKT}) imply that the sub-blocks satisfy the
symmetries
\begin{eqnarray}
 ( \hat{\sigma}^R )^{\dagger} =  \hat{\sigma}^A \; \; & , &   \; \;
( \hat{\sigma}^K )^{\dagger} =  - \hat{\sigma}^K,
 \label{eq:symtimenormalsig}
\end{eqnarray}
and
 \begin{eqnarray}
( \hat{\pi}^R )^{T} =  \hat{\pi}^A  \; \; & , &   \; \;
( \hat{\pi}^K )^{T} =   \hat{\pi}^K .
 \label{eq:symtimeanosig}
\end{eqnarray}
Since the self-energy blocks satisfy the same symmetry relations as
the Green functions (\ref{eq:symtimenormal},\ref{eq:symtimeano}),
the symmetries (\ref{eq:GKT}) hold also for the self-energy blocks
in Keldysh space,
\begin{equation}
  (\hat{\Sigma}^{R})^T =  \hat{\Sigma}^{A} \; \; , \; \;
  (\hat{\Sigma}^{K})^T =  \hat{\Sigma}^{K}.
\end{equation}
The full self-energy matrix is therefore again symmetric,
\begin{equation}
 \mathbf{\Sigma} =  \mathbf{\Sigma}^T .
\end{equation}
In the presence of interactions the lower-diagonal block of the inverse
propagator is given by the negative of the Keldysh component of the self-energy,
 \begin{equation}
  \hat{\Sigma}^K =   (\hat{G}^R)^{-1}
\hat{G}^{K} (\hat{G}^A )^{-1}
 = - [ \mathbf{G}^{-1} ]^{QQ}.
 \label{eq:sigmaK}
\end{equation}

\subsection{Contour basis}

In the Keldysh technique
all operators are considered as functions of the time-argument on the
Keldysh contour. The time contour runs in real-time direction from some initial time $t_0$ to some
upper limit $t_{>}$, which is larger than all other times of interest and which is slightly shifted in the upper positive imaginary branch of the contour, and then back  to $t_0$ in the lower, negative imaginary branch.
Alternatively, all time integrations can be restricted to
the interval $[t_0, t_{>}]$ and one can keep track of the two
branches of the Keldysh contour using an extra label $p = \pm $,
where $p = +$ corresponds to the forward part of the
contour and $p =-$ denotes its backward part.
In the functional integral formulation of the Keldysh
technique, \cite{Kamenev04} the bosonic annihilation and creation operators
are then represented by pairs of complex fields
$a_{\bd{k}, p } ( t )$
and $\bar{a}_{\bd{k}, p } ( t )$
carrying the contour label $p$.
The contour ordered expectation values of these fields define
four different propagators $\hat{G}^{p p^{\prime}}$,
which are related to the usual time-ordered ($\hat{G}^{\cal{T}}$),
anti-time-ordered ($\hat{G}^{\bar{\cal{T}}}$), lesser ($\hat{G}^{<}$) and
greater  ($\hat{G}^{>}$) Green functions and their RAK-counterparts as follows, \cite{Rammer07,Haug08}
 \begin{equation}
\left( \begin{array}{cc} \hat{G}^{\cal{T}} & \hat{G}^{<} \\ \hat{G}^{>} & \hat{G}^{\bar{\cal{T}}}
 \end{array} \right) =
 \left( \begin{array}{cc} \hat{G}^{++} & \hat{G}^{+-} \\ \hat{G}^{-+} & \hat{G}^{--}
 \end{array} \right) =
 \mathbf{R}  \left( \begin{array}{cc} \hat{G}^{K} & \hat{G}^{R} \\ \hat{G}^{A} & 0
 \end{array} \right) \mathbf{R} ,
 \label{eq:GconGK}
 \end{equation}
where the transformation matrix $\mathbf{R}$ has the block structure
 \begin{equation}
\mathbf{R} = \frac{1}{\sqrt{2}} \left( \begin{array}{cc} \hat{I} & \hat{I} \\ \hat{I} & - \hat{I}
 \end{array} \right) = \mathbf{R}^{-1} =   \mathbf{R}^{T}.
\end{equation}
Here $\hat{I}$ is the unit matrix in the flavor-, momentum-, and time labels,
i.e.,
 $[ \hat{I} ]_{ \sigma \bd{k} t, \sigma^{\prime} \bd{k}^{\prime} t^{\prime}} =
\delta_{\sigma , \sigma^{\prime}} \delta_{ \bd{k}, \bd{k}^{\prime}}
\delta ( t - t^{\prime} )$.
The matrix equation~(\ref{eq:GconGK}) implies for the blocks in the contour basis
 \begin{equation}
  \label{eq:Gcontour}
  \hat{G}^{ p p^{\prime}} = \frac{1}{2} \left[ p^{\prime} \hat{G}^R + p \hat{G}^A + \hat{G}^K \right],
 \end{equation}
where $ p, p^{\prime} \in\{ +,-\} $. From Eq.~(\ref{eq:Gcontour}) one easily verifies the
inverse relations,
 \begin{subequations}
 \begin{eqnarray}
 \hat{G}^R & = &  [\mathbf{G}]^{CQ} =
\frac{1}{2} \sum_{ p p^{\prime}} p^{\prime} \hat{G}^{p p^{\prime}}  ,
 \\
 \hat{G}^A & = &   [\mathbf{G}]^{QC}  =   \frac{1}{2} \sum_{ p p^{\prime}} p \hat{G}^{p p^{\prime}}
 ,
 \\
\hat{G}^K & = &   [\mathbf{G}]^{CC} =  \frac{1}{2} \sum_{ p p^{\prime}}  \hat{G}^{p p^{\prime}}  ,
 \\
0  & = &   \sum_{ p p^{\prime}} p p^{\prime} \hat{G}^{p p^{\prime}}    .
\end{eqnarray}
\end{subequations}

The corresponding relations for the self-energy are
\begin{equation}
\left( \begin{array}{cc} \hat{\Sigma}^{\cal{T}} & \hat{\Sigma}^{<} \\ \hat{\Sigma}^{>} &
\hat{\Sigma}^{\bar{\cal{T}}}
 \end{array} \right) =
 \left( \begin{array}{cc} \hat{\Sigma}^{++} & \hat{\Sigma}^{+-} \\ \hat{\Sigma}^{-+} &
 \hat{\Sigma}^{--}
 \end{array} \right) =
 \mathbf{R}  \left( \begin{array}{cc} 0 & \hat{\Sigma}^{A} \\ \hat{\Sigma}^{R} & \hat{\Sigma}^K
 \end{array} \right) \mathbf{R} ,
 \label{eq:SigmaconGK}
 \end{equation}
which gives
 \begin{equation}
  \hat{\Sigma}^{ p p^{\prime}} = \frac{1}{2} \left[ p \hat{\Sigma}^R + p^{\prime} \hat{\Sigma}^A
 + p p^{\prime} \hat{\Sigma}^K \right],
 \end{equation}
and the inverse relations
 \begin{subequations}
 \begin{eqnarray}
 \hat{\Sigma}^R & = & [\bm{\Sigma}]^{QC} =
\frac{1}{2} \sum_{p p^{\prime}} p \hat{\Sigma}^{p p^{\prime}},
 \label{eq:Rpm}
 \\
  \hat{\Sigma}^A & = & [\bm{\Sigma}]^{CQ} =
\frac{1}{2} \sum_{p p^{\prime}} p^{\prime} \hat{\Sigma}^{p p^{\prime}} ,
 \label{eq:Apm}
 \\
\hat{\Sigma}^K & = & [\bm{\Sigma}]^{QQ}
=
\frac{1}{2} \sum_{p p^{\prime}} p p^{\prime} \hat{\Sigma}^{p p^{\prime}},
 \label{eq:Kpm}
 \\
0 & = & \sum_{p p^{\prime}}  \hat{\Sigma}^{p p^{\prime}}.
 \label{eq:causality}
\end{eqnarray}
\end{subequations}

\subsection{Functional integral representation of Green functions
in the RAK-basis}
\label{sec:func}

To define the proper boundary conditions
in the functional integral formulation of the Keldysh technique, \cite{Kamenev04}
it is convenient to work  in the RAK basis.
The transition from the contour basis to the RAK basis is achieved by
introducing the classical ($C$) and quantum ($Q$) components
of the fields,  which are related to the corresponding fields
$a_{\bd{k} , \pm}$ and $\bar{a}_{\bd{k} , \pm}$ in the contour basis
via
 \begin{subequations}
 \begin{eqnarray}
 \label{eq:acq}
 a^C_{\bd{k}} ( t ) & = & \frac{1}{\sqrt{2}}  \left[  a_{\bd{k}, +} ( t ) +  a_{\bd{k}, -} ( t ) \right],
 \\
 \bar{a}^C_{\bd{k}} ( t ) & = & \frac{1}{\sqrt{2}}  \left[  \bar{a}_{\bd{k}, +} ( t )
+  \bar{a}_{\bd{k}, -} ( t ) \right],
 \\
 a^Q_{\bd{k}} ( t ) & = & \frac{1}{\sqrt{2}}  \left[  a_{\bd{k}, +} ( t ) -  a_{\bd{k}, -} ( t ) \right],
 \\
 \bar{a}^Q_{\bd{k}} ( t ) & = & \frac{1}{\sqrt{2}}  \left[  \bar{a}_{\bd{k}, +} ( t )
-  \bar{a}_{\bd{k}, -} ( t ) \right].
 \label{eq:abarcq}
 \end{eqnarray}
\end{subequations}
Introducing a four-component ``super-field'',
\begin{equation}
\left(
 \begin{array}{c}
 \Phi^{C}_a ( {\bd{k} } , t ) \\
 \Phi^C_{\bar{a} } (  \bd{k}  ,  t ) \\
 \Phi^Q_{a} ( \bd{k} , t ) \\
\Phi^Q_{ \bar{a} } (  \bd{k}  ,  t )
 \end{array}
 \right)
=
\left(
 \begin{array}{c}
 a^{C}_{\bd{k} } ( t ) \\
\bar{a}^C_{ - \bd{k} } (  t ) \\
 a^Q_{\bd{k}} ( t ) \\
\bar{a}^Q_{ - \bd{k} } (  t )
 \end{array}
 \right),
 \end{equation}
the matrix elements of the symmetrized matrix Green function
$\mathbf{G}$ defined in Sec.~\ref{subsec:RAK} can be represented as
the following functional average,
 \begin{eqnarray}
i [\mathbf{G}]^{\lambda \lambda^{\prime} }_{ \sigma \bd{k} t,
\sigma^{\prime} \bd{k}^{\prime} t^{\prime}} & \equiv &
 i  {G}_{\sigma \sigma^{\prime}}^{\lambda \lambda^{\prime}} (\bd{k} t , \bd{k}^{\prime}
t^{\prime})
 =  \langle \Phi_{\sigma}^{\lambda} ( \bd{k} t )
 \Phi_{\sigma^{\prime}}^{\lambda^{\prime}} ( \bd{k}^{\prime} t^{\prime} )  \rangle
 \nonumber
 \\
 & = &
\int {\cal{D}} [ \Phi ] e^{ i S [  \Phi ]} \Phi_{\sigma}^{\lambda} ( \bd{k} t )
 \Phi_{\sigma^{\prime}}^{\lambda^{\prime}} ( \bd{k}^{\prime} t^{\prime} ).
 \hspace{5mm}
 \label{eq:Greendef}
 \end{eqnarray}
For simplicity, we shall assume throughout this
work that the initial state at $t= t_0$ is not correlated.
In this case the corresponding Gaussian initial density matrix
does not explicitly appear in the above Keldysh action 
$ S [  \Phi ]$, but enters the dynamics via the initial conditions
for the independent one- and two-point functions.\cite{Berges04,Danielewicz84a,Semkat00,Garny09}
In principle,
non-Gaussian initial correlations can be considered either explicitly 
by assuming a non-Gaussian initial density matrix,
or implicitly by introducing multiple sources in the generating functional for the $n$-point functions. \cite{Garny09}
The latter approach will also change the matrix structure of the theory;
for instance if the system is initially in thermal equilibrium the 
Keldysh matrices expand from a $2\times2$ to a $3\times3$ form. As shown in
Ref.\ [\onlinecite{Kwong00}], initial correlations can lead to additional damping effects which modify
the amplitude and phase of the oscillatory evolution at intermediate times.
In the absence of initial correlations
the  form of the Keldysh action $S [ \Phi ]$  in Eq.~(\ref{eq:Greendef})
can be directly obtained from the corresponding Hamiltonian, so that it has
has the following two contributions,
 \begin{equation}
 S [ \Phi ] = S_0 [\Phi ] + S_1 [\Phi ],
 \label{eq:Kelaction}
 \end{equation}
where after proper symmetrization
the Gaussian part  $S_0 [\Phi ]$ can be written as
 \begin{eqnarray}
 S_0 [ \Phi ]  & = & \frac{1}{2}  \sum_{ \sigma \sigma^{\prime} }
\sum_{ \lambda \lambda^{\prime}}
\sum_{ \bd{k}  \bd{k}^{\prime}}  \int dt \, d t^{\prime}
 \nonumber
 \\
 &  \times &
\Phi_{\sigma}^{\lambda} (\bd{k}  t ) [ \mathbf{G}^{-1}_0 ]^{\lambda
\lambda^{\prime}}_{ \sigma   \bd{k} t  , \sigma^{\prime}
\bd{k}^{\prime} t^{\prime} }
\Phi_{\sigma^{\prime}}^{\lambda^{\prime}} ( \bd{k}^{\prime}
t^{\prime} ) .
 \label{eq:SGauss}
 \end{eqnarray}
Here   $\mathbf{G}_0^{-1}$
is the non-interacting inverse Green function matrix in Keldysh space, which is associated with
the non-interacting  part of the Hamiltonian (\ref{eq:tildeHdef}).
The matrix   $\mathbf{G}_0^{-1}$
has the same block structure as $\mathbf{G}^{-1}$ in Eq.~(\ref{eq:Gmatinv}),
with retarded and advanced blocks given by
\begin{equation}
  ( {\hat{G}_0^{R}} )^{-1}  =  \hat{D} - i \eta \hat{Z}
 \quad , \quad
  ( {\hat{G}_0^{A}} )^{-1}  =   \hat{D} + i \eta \hat{Z} ,
 \label{eq:GRAM}
 \end{equation}
where the antisymmetric
matrix $\hat{Z}$ is given in Eq.~(\ref{eq:Zdef}),
and the symmetric matrix  $\hat{D}$ is defined by
 \begin{equation}
  [ \hat{D} ]_{ \bd{k} t , \bd{k}^{\prime} t^{\prime}}
= \delta_{ \bd{k} , - \bd{k}^{\prime}  } Z
 \left[  - i  \delta^{\prime} ( t - t^{\prime})  + \delta ( t - t^{\prime})
 M_{\bd{k}^\prime} \right].
 \label{eq:Ddef}
 \end{equation}
Here
 $\delta^{\prime} ( t ) = \frac{ d }{dt} \delta ( t ) $
is the derivative of the Dirac $\delta$-function and
$M_{\bd{k}}$ is the following matrix in flavor space,
\footnote{
Recall that we work in the rotating reference frame and have redefined
$\tilde{\epsilon}_{\bd{k}} \equiv \epsilon_{\bd{k}} - \frac{\omega_0}{2} \rightarrow \epsilon_{\bd{k}}$ and removed the phase of $\gamma_{\bd{k}}$.
In the original frame the matrix $M_{\bd{k}}$ in Eq.~(\ref{eq:tildeMkdef}) should be replaced by the matrix
\[
 M_{\bd{k}}(t) = \left( \begin{array}{cc}  \epsilon_{\bd{k}} &
\gamma_{\bd{k}} e^{- i \omega_0 t}  \\
 - \gamma^{\ast}_{\bd{k}} e^{i \omega_0 t}  & - \epsilon_{\bd{k}}
 \end{array}
 \right),
 \label{eq:Mkdef}
 \]
 which depends explicitly on time.
}
 \begin{equation}
 M_{\bd{k}}  = \left( \begin{array}{cc}  \epsilon_{\bd{k}} &
| \gamma_{\bd{k}} | \\
 - | \gamma_{\bd{k}} | & - \epsilon_{\bd{k}}
 \end{array}
 \right).
 \label{eq:tildeMkdef}
 \end{equation}
Recall that we are working in the rotating reference frame where we have redefined $\tilde{\epsilon}_{\bd{k}} \equiv \epsilon_{\bd{k}}- \frac{\omega_0}{2} \rightarrow \epsilon_{\bd{k}} $,see Eq.~(\ref{eq:tildeepsilon}).
Keeping in mind that $i \delta^{\prime} ( t - t^{\prime}  ) = - i
\delta^{\prime} (  t^{\prime} -t  )$,
it is obvious that
$[ (\hat{G}_0^R)^{-1} ]^T = (\hat{G}_0^A)^{-1}$,
in agreement with Eq.~(\ref{eq:GKT}).

Although the Keldysh block $ [ \mathbf{G}_0^{-1}]^{QQ}$
of the inverse Gaussian propagator
in Eq.~(\ref{eq:SGauss}) vanishes in continuum notation,
it is actually finite if the path integral is properly discretized. \cite{Kamenev04}
It is, however, more convenient to stick with the continuum notation and
take the discretization effectively into account by adding an infinitesimal
regularization $\eta$. To derive this regularization, we note that
the relations $\mathbf{G}_0^{-1} \mathbf{G}_0 =
\mathbf{G}_0 \mathbf{G}_0^{-1} = \mathbf{I}$ imply
that  in the non-interacting limit the Keldysh block satisfies
 \begin{equation}
  \hat{D} \hat{G}_0^K  =  \hat{G}_0^K \hat{D} =0.
 \label{eq:MK}
 \end{equation}
Introducing the non-interacting distribution matrix $\hat{F}_0$ as in
Eq.~(\ref{eq:kelF}) this implies for $\eta \rightarrow 0$,
 \begin{equation}
    \hat{D} \hat{G}_0^K \hat{D} =
  \hat{F}_0 \hat{Z} \hat{D}
- \hat{D} \hat{Z}\hat{F}_0^T =0.
 \label{eq:nullrelation}
\end{equation}
Using Eq.~(\ref{eq:GinvQQ}) we thus obtain for the
lower diagonal
block of $\mathbf{G}_0^{-1} $  in the non-interacting limit,
 \begin{eqnarray}
 [\mathbf{G}_0^{-1}]^{QQ} & = &
 -   (\hat{G}_0^R)^{-1}  \hat{G}_0^{K} (\hat{G}_0^A)^{-1}
 \nonumber
 \\
 & =  &  (\hat{G}_0^R)^{-1} \hat{Z} \hat{F}_0^T -  \hat{F}_0  \hat{Z}
( \hat{G}_0^A)^{-1}
 \nonumber
 \\
& = & i \eta ( \hat{F}_0 + \hat{F}_0^T) = 2 i \eta \hat{F}_0,
 \label{eq:pureregu}
 \end{eqnarray}
where we have used the fact that $\hat{F}_0$ is symmetric, which is
easily verified by explicit calculation, see Sec.~\ref{sec:nonint}.
The lower diagonal block of $\mathbf{G}_0^{-1}$ is thus a pure
regularization, which guarantees that in the non-interacting limit
the functional integral (\ref{eq:Greendef}) is well defined. In the
presence of interactions the Keldysh component of the self-energy is
finite due to Eq.\ (\ref{eq:sigmaK}), so that in this case the
infinitesimal regularization (\ref{eq:pureregu}) can be omitted.

To write down the interaction part of the Keldysh action associated
with the interaction part of the Hamiltonian (\ref{eq:Hpr}), one
should first symmetrize the Hamiltonian \cite{Wetterich01,Kreisel08}
before formally replacing the operators by complex fields, since $a$
and ${a}^{\dagger}$ are treated symmetrically. Noting that the
symmetrized product of $n$ bosonic operators $A_1, \ldots, A_n$ is
defined by
 \begin{equation}
 \left\{ A_1 \ldots A_n \right\} = \frac{1}{n!} \sum_{ P}
 A_{ P_1} \ldots A_{P_n},
 \end{equation}
where the sum is over all $n!$ permutations,
we have
 \begin{eqnarray}
a_{\bd{k}_1}^{\dagger} a_{\bd{k}_2}^{\dagger} a_{\bd{k}_3} a_{\bd{k}_4} & = &
\{  a_{\bd{k}_1}^{\dagger} a_{\bd{k}_2}^{\dagger} a_{\bd{k}_3} a_{\bd{k}_4}   \}
 \nonumber
 \\
 &-  & \frac{1}{2} \Bigl[
 \delta_{\bd{k}_1, \bd{k}_4}  \{
a_{\bd{k}_2}^{\dagger} a_{\bd{k}_3} \}
+ \delta_{\bd{k}_1, \bd{k}_3}  \{ a_{\bd{k}_2}^{\dagger} a_{\bd{k}_4} \}
 \nonumber
 \\
 & & \hspace{2mm}
+ \delta_{\bd{k}_2, \bd{k}_3}  \{
a_{\bd{k}_1}^{\dagger} a_{\bd{k}_4} \}
+ \delta_{\bd{k}_2, \bd{k}_4}  \{ a_{\bd{k}_1}^{\dagger} a_{\bd{k}_3} \}
 \Bigr]
\nonumber
 \\
 &+ &   \frac{1}{4} \Bigl[   \delta_{\bd{k}_1, \bd{k}_3}   \delta_{\bd{k}_2, \bd{k}_4}
+ \delta_{\bd{k}_2, \bd{k}_3}   \delta_{\bd{k}_1, \bd{k}_4}
  \Bigr]. \hspace{7mm}
 \label{eq:symres}
 \end{eqnarray}
The quadratic terms on the right-hand side of Eq.~(\ref{eq:symres})
lead to a time-independent first-order shift in the bare energy dispersion,
\begin{equation}
 \epsilon_{\bd{k}} \rightarrow
\epsilon_{\bd{k}} -
 \frac{1}{V} \sum_{\bd{k}^{\prime}} U ( - \bd{k} , - \bd{k}^{\prime} ; \bd{k} ,
 \bd{k}^{\prime} )  .
 \label{eq:epsilonshift}
\end{equation}
This shift can be absorbed by re-defining the energy
$\epsilon_{\bd{k}}$ in the matrix  $M_{\bd{k}}$ introduced in
Eq.~(\ref{eq:tildeMkdef}).
The first term  on the right-hand side of Eq.~(\ref{eq:symres}) leads to the following
interaction part of the Keldysh action in the RAK-basis,
 \begin{widetext}
\begin{eqnarray}
 S_1 [ \Phi ]
 & = &
 - \frac{1}{2V}
  \sum_{ \bd{k}_1  \bd{k}_2 \bd{k}_3  \bd{k}_4}
\int dt \; \delta_{ \bd{k}_1 + \bd{k}_2 + \bd{k}_3 + \bd{k}_4 , 0 }
 U ( \bd{k}_1 ,  \bd{k}_2 ; \bd{k}_3 , \bd{k}_4 )
 \nonumber
 \\
 & \times &
\biggl\{
 \Phi_{\bar{a}}^{C}  (  \bd{k}_1  t )  \Phi_{\bar{a}}^{Q} (  \bd{k}_2 t )
 \Big[   \Phi_{ a}^C (\bd{k}_3  t)   \Phi_{ a}^C (\bd{k}_4  t)   +
  \Phi_{ a}^Q (  \bd{k}_3   t)    \Phi_{ a}^Q (  \bd{k}_4   t) \Big]
 \nonumber
 \\
 & & \hspace{-2mm}
+\Big[   \Phi_{ \bar{a}}^C (\bd{k}_1  t)   \Phi_{ \bar{a}}^C (\bd{k}_2  t)   +
  \Phi_{ \bar{a}}^Q (  \bd{k}_1   t)    \Phi_{ \bar{a}}^Q (  \bd{k}_2   t) \Big]
 \Phi_{{a}}^{C}  (  \bd{k}_3  t )  \Phi_{{a}}^{Q} (  \bd{k}_4 t ) \biggr\} .
\label{eq:S1}
 \end{eqnarray}
To eliminate complicated combinatorial factors
in the FRG flow equations derived in Sec.~\ref{sec:FRG}
it is convenient to symmetrize
the interaction vertices in Eq.~(\ref{eq:S1}) with respect to the interchange
of any two labels\cite{Vasilievbook,Schuetz05,Kopietz10} and write
\begin{eqnarray}
 S_1 [ \Phi ]
 & = & - \frac{1}{4 ! V}  \sum_{ \sigma_1 \ldots \sigma_4}
 \sum_{ \lambda_1 \ldots \lambda_4}   \sum_{\bd{k}_1 \ldots \bd{k}_4}  \int dt
\delta_{\bd{k}_1 + \bd{k}_2 + \bd{k}_3 + \bd{k}_4,0 } \,
 U^{\lambda_1 \lambda_2 \lambda_3 \lambda_4}_{
 \sigma_1 \sigma_2 \sigma_3 \sigma_4}  ( \bd{k}_1  , \bd{k}_2, \bd{k}_3 , \bd{k}_4 )
 \Phi_{ \sigma_1}^{\lambda_1} ( \bd{k}_1 t )
  \Phi_{ \sigma_2}^{\lambda_2} (\bd{k}_2  t )
\Phi_{ \sigma_3}^{\lambda_3} ( \bd{k}_3  t ) \Phi_{ \sigma_4}^{\lambda_4} ( \bd{k}_4 t ),
 \nonumber
 \\
 & &
 \label{eq:S1kel}
\end{eqnarray}
\end{widetext}
where the interaction vertex
 $U^{\lambda_1 \lambda_2 \lambda_3 \lambda_4}_{
 \sigma_1 \sigma_2 \sigma_3 \sigma_4}  ( \bd{k}_1  , \bd{k}_2, \bd{k}_3 , \bd{k}_4 )$
is symmetric with respect to any pair of indices.
Up to permutations of the  indices, the non-zero vertices are
 \begin{eqnarray}
& & U_{\;  \bar{a} \, \bar{a} \,  a \, a}^{CQCC} ( \bd{k}_1 , \bd{k}_2 , \bd{k}_3 , \bd{k}_4 )
\nonumber
\\ & = &
 U_{\;  \bar{a} \, \bar{a} \,  a \, a}^{CQQQ} ( \bd{k}_1 , \bd{k}_2 , \bd{k}_3 , \bd{k}_4 )
\nonumber
  \\ &= &
 U_{\;  \bar{a} \, \bar{a} \,  a \, a}^{CCCQ} ( \bd{k}_1 , \bd{k}_2 , \bd{k}_3 , \bd{k}_4 )
 \nonumber
 \\ & = &U_{\;  \bar{a} \, \bar{a} \,  a \, a}^{QQCQ} ( \bd{k}_1 , \bd{k}_2 , \bd{k}_3 , \bd{k}_4 )
 = U ( \bd{k}_1 , \bd{k}_2 ; \bd{k}_3 , \bd{k}_4 )     .
 \hspace{7mm}
 \label{eq:ubare}
\end{eqnarray}

\subsection{Non-interacting Green functions}
\label{sec:nonint}

To conclude this section, let us explicitly construct the $2 \times 2$ matrix Green functions in flavor space
in the non-interacting limit. In general, we define the Green functions in
flavor space in terms of the matrix elements
 \begin{equation}
 [ \hat{G}^X ]_{ \bd{k} t, - \bd{k} t^{\prime} }
= G^X ( \bd{k} , t , t^{\prime} ) \; \; , \; \; X = R,A, K.
 \label{eq:GXk}
 \end{equation}
In the absence of interactions, we can obtain explicit expressions
for these Green functions. Then the non-interacting part of the
Hamiltonian (\ref{eq:tildeHdef}) reduces to$^{40}$
\begin{equation}
  \tilde{\cal{H}}_0
    =  \sum_{\bd{k}}
 \left[
\epsilon_{\bd{k}} a_{\bd{k}}^{\dagger} a_{\bd{k}}
 + \frac{ | \gamma_{\bd{k}} |}{2}
\left(
  a^{\dagger}_{\bd{k}} a^{\dagger}_{ - \bd{k}} +    a_{- \bd{k}} a_{\bd{k}}
 \right) \right].
 \label{eq:H0rot}
 \end{equation}
The retarded and advanced matrix Green functions in the non-interacting limit
are now easily obtained. Consider first the retarded Green function,
which satisfies
 \begin{equation}
 i \partial_t G_0^{R} (\bd{k},  t , t^{\prime} )   =  \delta ( t - t^{\prime} ) Z  +
 M_{\bd{k}}
 G^{R}_0 (   \bd{k},  t, t^{\prime} ) .
 \label{eq:GRdif}
 \end{equation}
The solution of Eq.~(\ref{eq:GRdif})  with proper boundary condition is
 \begin{equation}
  G^{R}_0 ( \bd{k},  t , t^{\prime} ) = - i \Theta ( t - t^{\prime} )
e^{ - i M_{\bd{k}} ( t - t^{\prime} ) } Z.
 \label{eq:GRres}
\end{equation}
The matrix exponential is
 \begin{equation}
 e^{ - i M_{\bd{k}} t } =  I \cos ( \mu_{\bd{k}} t )
- i M_{\bd{k}}  \frac{ \sin ( \mu_{\bd{k}} t )}{\mu_{\bd{k}}  } ,
 \end{equation}
where $I$ is the $2 \times 2$ unit matrix, and
 \begin{equation}
 \mu_{\bd{k}} = \left\{
 \begin{array}{cc}
  \sqrt{ \epsilon_{\bd{k}}^2 - | \gamma_{\bd{k}} |^2  }
 & \mbox{if} \;  | \epsilon_{\bd{k}} | > | \gamma_{\bd{k}} |, \\
 i \sqrt{  | \gamma_{\bd{k}} |^2  - \epsilon_{\bd{k}}^2 }
 & \mbox{if}  \;  | \gamma_{\bd{k}} | > | \epsilon_{\bd{k}} | .
 \end{array}
 \right.
 \end{equation}
Using the symmetry
relation $(\hat{G}^R )^T = \hat{G}^A$ given in Eq.~(\ref{eq:GKT}),
we obtain for the corresponding advanced Green function,
\begin{equation}
  G^{A}_0 (\bd{k},  t , t^{\prime} )  =   G^{R}_0 ( - \bd{k} ,  t^{\prime} , t )^T =
 - i \Theta ( t^{\prime} - t )  Z^T e^{ - i M_{\bd{k}}^T ( t^{\prime} - t )}.
 \end{equation}
Using the identities
 \begin{subequations}
 \begin{eqnarray}
 Z^2 & = & - I, \\
 Z^{-1} & = & Z^T = - Z, \\
M_{\bd{k}}^T & = & Z M_{\bd{k}} Z = - Z^T M_{\bd{k}} Z,
 \\
 Z^T e^{ - i M_{\bd{k}}^T (  t^{\prime} - t )}  & = &  -  e^{ - i M_{\bd{k}}
( t - t^{\prime} )} Z,
 \end{eqnarray}
 \end{subequations}
we may also write
 \begin{equation}
  G^{A}_0 (\bd{k},  t , t^{\prime} ) =  i \Theta (  t^{\prime} - t )  e^{ - i
 M_{\bd{k}} ( t - t^{\prime} ) } Z.
 \label{eq:GAres}
\end{equation}
Because the retarded and advanced Green functions depend only on the time difference,
it is useful to perform a Fourier transformation
to frequency space,
 \begin{equation}
 G^{X}_0  ( \bd{k} , \omega ) =
\int_{ - \infty}^{\infty} d t e^{ i \omega t } G^X_0 ( \bd{k}, t, 0 ).
 \end{equation}
Substituting Eqs.~(\ref{eq:GAres}) and (\ref{eq:GRres}) into this
expression and representing the step-functions as
 \begin{equation}
 \Theta ( t ) = \int_{ - \infty}^{\infty} \frac{ d \omega^{\prime}}{ 2 \pi i }
 \frac{ e^{ i \omega^{\prime} t }}{ \omega^{\prime} - i \eta },
 \end{equation}
it is easy to show that
 \begin{subequations}
 \begin{eqnarray}
 G^R_0 ( \bd{k} , \omega ) & = & [ \omega - M_{\bd{k}} + i \eta]^{-1} Z
 \label{eq:GRzerorot},
 \\
G^A_0 (\bd{k},  \omega ) & = & [ \omega - M_{\bd{k}} - i \eta]^{-1} Z
 \nonumber
\\
 & = & Z  [ \omega + M_{\bd{k}}^T - i \eta]^{-1},
 \label{eq:GAzerorot}
\end{eqnarray}
\end{subequations}
or explicitly,
 \begin{eqnarray}
  G^{R/A}_0 (\bd{k},  \omega ) & = & \frac{ 1}{ ( \omega \pm i \eta )^2 -
\epsilon_{\bd{k}}^2 +
 | \gamma_{\bd{k}} |^2 }
 \nonumber
 \\
 &  \times &
 \left( \begin{array}{cc} - | \gamma_{\bd{k}} | & \omega \pm i \eta + \epsilon_{\bd{k}} \\
  - \omega \mp i \eta + \epsilon_{\bd{k}} & -   | \gamma_{\bd{k}} | \end{array} \right).
 \hspace{7mm}
 \end{eqnarray}

Next, consider the Keldysh component $G^K_0 ( \bd{k} , t , t^{\prime} ) $
of our $2 \times 2$ matrix Green function in flavor space.
It satisfies the matrix equations
 \begin{subequations}
 \begin{eqnarray}
 i \partial_t G^K_0 ( \bd{k},  t , t^{\prime} ) & = &
 M_{\bd{k}} G^K_0 ( \bd{k}, t , t^{\prime} ),
 \label{eq:GKmat1}
 \\
i \partial_{t^{\prime}}G^K_0 ( \bd{k}, t , t^{\prime} ) & = &
G^K_0 (\bd{k},  t , t^{\prime} ) M_{\bd{k}}^T,
 \label{eq:GKmat2}
 \end{eqnarray}
\end{subequations}
and hence
 \begin{equation}
 ( i \partial_t + i \partial_{t^{\prime}} )
G^K_0 (\bd{k},  t , t^{\prime} )  =  M_{\bd{k}}
 G^K_0 (\bd{k},  t , t^{\prime} )
+ G^K_0 (\bd{k},  t , t^{\prime} ) M_{\bd{k}}^T.
 \label{eq:GKmat3}
\end{equation}
These equations are solved by
 \begin{equation}
G^K_0 (\bd{k},  t , t^{\prime}) =
e^{ - i M_{\bd{k}} t } G_0^K (\bd{k}, 0,0 ) e^{ - i M_{\bd{k}}^T t^{\prime} },
 \end{equation}
with an arbitrary initial matrix
$G_0^K (\bd{k}, 0,0 )$ which defines the distribution functions at
$t=0$.
To explicitly construct the distribution matrix $\hat{F}_0$
defined via Eq.~(\ref{eq:kelF}), we note that
in the non-interacting limit the distribution matrix is diagonal in time,
 \begin{equation}
 [ \hat{F}_0 ]_{\bd{k} t, \bd{k'} t^{\prime} } = \delta_{\bd{k},-\bd{k'}} F_0 ( \bd{k} , t , t^{\prime} ) = \delta_{\bd{k},-\bd{k'}} \delta ( t - t^{\prime} )
 F_0 ( \bd{k} , t ),
 \label{eq:Fdiag}
 \end{equation}
so that Eq.~(\ref{eq:kelF}) reduces to the $2 \times 2$ matrix relation
\begin{eqnarray}
 G^K_0 (\bd{k},  t , t^{\prime} ) & = &
G^R_0 (\bd{k} ,  t , t^{\prime} ) F_0 (\bd{k},  t^{\prime} )
 Z - Z F_0 (\bd{k},  t ) G^A_0 (\bd{k},  t , t^{\prime} )
 \nonumber
 \\
 & = & - i \Theta ( t - t^{\prime} ) e^{ - i M_{\bd{k}}
( t - t^{\prime} ) } Z F_0 ( \bd{k}, t^{\prime} ) Z
 \nonumber
 \\
 & &
  - i \Theta (  t^{\prime} - t ) Z F_0 (\bd{k},  t ) Z
 e^{ - i M_{\bd{k}}^T (  t^{\prime} - t) }.
 \label{eq:GKzero}
\end{eqnarray}
It follows that in the non-interacting limit the
time-diagonal element $F_0 (\bd{k},  t )$
of the distribution matrix
contains the normal and anomalous
distribution functions defined in Eqs.~(\ref{eq:nkdef}, \ref{eq:pkdef})
in the following way,
 \begin{equation}
 F_0 (\bd{k},  t ) =
 i Z G^K_0 (\bd{k},  t , t ) Z
=   \left( \begin{array}{cc} - 2 p_{\bd{k}}^{\ast} ( t ) & 2 n_{\bd{k}} ( t ) +1 \\
 2 n_{\bd{k}} ( t ) +1 &  - 2 p_{\bd{k}} ( t ) \end{array} \right).
 \label{eq:Ft}
 \end{equation}
Combining this relation with Eq.~(\ref{eq:GKmat3}), we see that
our diagonal distribution function matrix satisfies the kinetic equation
 \begin{equation}
 i \partial_t  F_0 (\bd{k},  t ) =  - M_{\bd{k}}^T  F_0 ( \bd{k}, t )
 - F_0 (\bd{k},  t )  M_{\bd{k}}  .
 \label{eq:kinfree}
 \end{equation}
Note that the non-interacting Keldysh Green function
(\ref{eq:GKzero}) can also be written as
 \begin{eqnarray}
 G^K_0 (\bd{k},  t , t^{\prime} ) & = &  - i [
 G_0^R (\bd{k},  t , t^{\prime} ) Z G_0^K (\bd{k},  t^{\prime} , t^{\prime} )
 \nonumber
 \\
 & &  \hspace{2mm} -
G_0^K (\bd{k},  t , t ) Z G_0^A (\bd{k},  t , t^{\prime} )] ,
 \label{eq:KadanoffBaym0}
 \end{eqnarray}
which relates the matrix elements of the Keldysh Green function
at different times to the corresponding equal-time matrix elements.

\section{Quantum kinetic equations}
\label{sec:quantumkinetic}

From the Keldysh component of  the Dyson equation
we obtain quantum kinetic equations for the distribution function.
In this section we shall derive several equivalent
versions of these equations.
Although matrix generalizations of quantum kinetic equations are standard, \cite{Haug08,Rammer07,Kwong98} 
we present here a special matrix structure of the
kinetic equations which takes into account off-diagonal bosonic correlations.

\subsection{Non-equilibrium evolution equations for two-time Keldysh Green functions}
\label{sec:kinexact}

In order to derive quantum kinetic equations we start with the matrix Dyson equation (\ref{eq:dyson}), which can be written as
\begin{equation}
   \left( \mathbf{G}_0^{-1} -  \mathbf{\Sigma} \right)  \mathbf{G}  =  \mathbf{I} .
\end{equation}
This ``left Dyson equation'' is equivalent with the following three
equations for the sub-blocks,
 \begin{subequations}
 \begin{eqnarray}
   [ ( \hat{G}_{0}^{R} )^{-1} - \hat{\Sigma}^R ]
\hat{G}^R & = & {\hat{I}},
 \label{eq:DysonR}
 \end{eqnarray}
 \begin{eqnarray}
   [ ( \hat{G}_{0}^{A} )^{-1} - \hat{\Sigma}^A ]
\hat{G}^A & = & {\hat{I}},
 \label{eq:DysonA}
 \end{eqnarray}
 \begin{eqnarray}
 {[} ( \hat{G}_{0}^{R} )^{-1} - \hat{\Sigma}^R {]} \hat{G}^K
 & = &    \hat{\Sigma}^K \hat{G}^A,
 \label{eq:GKdyson}
 \end{eqnarray}
 \end{subequations}
where $\hat{I}$ is again the unit matrix in
the flavor-, momentum- and  time labels.
Alternatively we can also consider the corresponding
``right Dyson equation'',
\begin{equation}
   \mathbf{G} \left( \mathbf{G}_0^{-1} -  \mathbf{\Sigma} \right)  =  \mathbf{I} ,
\end{equation}
which implies the following relations,
 \begin{subequations}
 \begin{eqnarray}
  \hat{G}^R    [ ( \hat{G}_{0}^{R} )^{-1} - \hat{\Sigma}^R ]  & = & {\hat{I}},
 \label{eq:DysonRright}
 \end{eqnarray}
 \begin{eqnarray}
  \hat{G}^A   [ ( \hat{G}_{0}^{A} )^{-1} - \hat{\Sigma}^A ]  & = & {\hat{I}},
 \label{eq:DysonAright}
 \end{eqnarray}
 \begin{eqnarray}
   \hat{G}^K [ ( \hat{G}_{0}^{A} )^{-1} - \hat{\Sigma}^A ]
 & = &
    \hat{G}^R \hat{\Sigma}^K.
 \label{eq:GKdysonright}
 \end{eqnarray}
 \end{subequations}
In order to solve the coupled set of equations
(\ref{eq:DysonR}--\ref{eq:GKdyson}) and
(\ref{eq:DysonRright}--\ref{eq:GKdysonright}), it is sometimes
useful to rewrite them as integral equations. Therefore we should
take into account that in the non-interacting limit the Keldysh
self-energy is actually an infinitesimal regularization,
$\hat{\Sigma}^K_0 = - 2 i \eta \hat{F}_0$, see
Eq.~(\ref{eq:pureregu}). In the non-interacting limit the Keldysh
component therefore satisfies
 \begin{equation}
 ( \hat{G}_{0}^{R} )^{-1} \hat{G}_{0}^K
  =    - 2 i \eta \hat{F}_0 \hat{G}_{0}^A,
 \end{equation}
which is equivalent with
 \begin{equation}
 \hat{G}_{0}^K
  =    - 2 i \eta \hat{G}_{0}^R   \hat{F}_0 \hat{G}_{0}^A.
 \end{equation}
Using the integral forms of the Dyson equations (\ref{eq:DysonR},\ref{eq:DysonA})
for the retarded and advanced Green functions,
 \begin{subequations}
 \label{eq:graInt}
 \begin{eqnarray}
 \hat{G}^R & = & \hat{G}_{0}^R +
\hat{G}_{0}^R \hat{\Sigma}^R \hat{G}^R,
 \\
\hat{G}^A & = & \hat{G}_{0}^A + \hat{G}_{0}^A
\hat{\Sigma}^A \hat{G}^{A},
 \end{eqnarray}
 \end{subequations}
the equation (\ref{eq:GKdyson}) for the Keldysh-block
can alternatively be written in integral form as
 \begin{eqnarray}
 \label{eq:gkInt}
 \hat{G}^K & = & \hat{G}_{0}^R \hat{\Sigma}^R
 \hat{G}^K + \hat{G}_{0}^R
 \hat{\Sigma}^K  \hat{G}^A
 \nonumber
\\
 &= &\hat{G}^K_{0}
 + \hat{G}_{0}^K \hat{\Sigma}^A \hat{G}^A +
\hat{G}_{0}^R \hat{\Sigma}^R \hat{G}^K
  +
\hat{G}_{0}^R \hat{\Sigma}^K
\hat{G}^A. \hspace{7mm}
\end{eqnarray}
Given an approximate expression for the self-energies, the integral equations (\ref{eq:graInt},\ref{eq:gkInt}) can be solved by an appropriate iteration to obtain the time evolution of the Keldysh Green function.
Here we follow a different strategy which is similar to the one proposed in Ref.~[\onlinecite{Berges04}]. We represent the inverse propagators (\ref{eq:GRAM}) as differential operators to derive evolution equations in differential form. We will see that the resulting initial value problem allows for approximate solutions which manifestly preserves causality.
Using the advanced and retarded components of the Dyson equations (\ref{eq:DysonR},\ref{eq:DysonA},\ref{eq:DysonRright},\ref{eq:DysonAright})  and keeping in mind that by translational invariance the matrix elements in the rotating reference frame are,
\begin{subequations}
 \begin{eqnarray}
 {[} \hat{G}^X {]}_{ \bd{k} t, \bd{k}^{\prime} t^{\prime} }
& =  & \delta_{ \bd{k} , - \bd {k}^{\prime}} G^X ( \bd{k} , t , t^{\prime} ),
 \label{eq:GXk2}
 \\
{[} \hat{\Sigma}^X {]}_{ \bd{k} t, \bd{k}^{\prime} t^{\prime} }
& =  & \delta_{ \bd{k} , - \bd{k}^{\prime}} \Sigma^X ( \bd{k}^{\prime} , t , t^{\prime} ),
 \label{eq:SigmaXk2}
 \end{eqnarray}
\end{subequations}
with $X = R,A,K$,
we obtain,
\begin{subequations} \label{eq:kinEqGrGa}
  \begin{align}
   & i \partial_t G^{R/A}(\bd{k},t,t')  -  M_{\bd{k}}  G^{R/A}(\bd{k},t,t') = Z \delta(t-t') \nonumber \\
   &\qquad  + \int_{t_0}^{t/t'} dt_1 Z  \Sigma^{R/A}(\bd{k},t,t_1) G^{R/A}(\bd{k},t_1,t') , \\
   & i \partial_{t'} G^{R/A}(\bd{k},t,t')   -  G^{R/A}(\bd{k},t,t') M^T_{\bd{k}} = -  Z \delta(t-t') \nonumber \\
   &\qquad - \int_{t_0}^{t/t'} dt_1  G^{R/A}(\bd{k},t,t_1) \Sigma^{R/A}(\bd{k},t_1,t') Z .
  \end{align}
\end{subequations}
In the same way we obtain from (\ref{eq:GKdyson}, \ref{eq:GKdysonright}) the following kinetic equations for the Keldysh component,
\begin{align}
   &i \partial_{t} G^K(\bd{k},t,t')  - M_{\bd{k}} G^K(\bd{k},t,t')   \nonumber \\
  & \qquad = \int_{t_0}^{t'} dt_1  Z \Sigma^K(\bd{k},t,t_1) G^A(\bd{k},t_1,t') \nonumber \\
  &\qquad  + \int_{t_0}^t dt_1 Z \Sigma^R(\bd{k},t,t_1) G^K(\bd{k},t_1,t') ,
 \label{eq:kinintnonequal}
\end{align}
and
\begin{align}
  &i \partial_{t'} G^K(\bd{k},t,t')  - G^K(\bd{k},t,t') M^T_{\bd{k}}   \nonumber \\
  &\qquad = - \int_{t_0}^{t} dt_1  G^R(\bd{k},t,t_1) \Sigma^K(\bd{k},t_1,t') Z \nonumber \\
  &\qquad \quad  - \int_{t_0}^{t'} dt_1 G^K(\bd{k},t,t_1) \Sigma^A(\bd{k},t_1,t') Z .
 \label{eq:kinintnonequal2}
\end{align}
Finally, in order uniquely define the solution of the set of coupled first-order partial differential equations, proper boundary conditions for the Green functions have to be specified.
From the definitions (\ref{eq:GR},\ref{eq:GA}) and (\ref{eq:DR},\ref{eq:DA}) of the advanced and retarded propagators we find that for infinitesimal $\eta$,
\begin{subequations} \label{eq:grgabound}
\begin{align}
  G^R(\bd{k},t,t-\eta) &= - i Z , \\
  G^A(\bd{k},t,t + \eta) &=  i Z ,
\end{align}
\end{subequations}
and that the Keldysh Green function should reduce to the matrix $G^K(\bd{k},0,0)$ at the reference time $t = t' = 0$.
Note that we have not made any approximations so far and the time evolution is exact provided we insert the exact self-energies.
The evolution equations are causal by construction, since no quantity in the collision integrals on the right-hand side depends
on future states. Interpreting the time derivatives as finite-difference expressions, the solution can be obtained by stepwise propagating
the equations in $t$ and $t'$ direction.
Note that in the non-interacting limit where all self-energies and collision
integrals vanish, our kinetic equations  (\ref{eq:kinintnonequal},\ref{eq:kinintnonequal2}) correctly reduce to
the equation of motion for the non-interacting Keldysh Green function
given in Eq.~(\ref{eq:GKmat3}).

For open systems coupled to an external bath, it is sometimes
convenient to move some of the terms on the right-hand side
of the kinetic equations (\ref{eq:kinintnonequal},\ref{eq:kinintnonequal2}) to the left-hand side such that the remaining
terms on the right-hand side correspond to the
``in-scattering'' and the ``out-scattering-rate'' in the
Boltzmann  equation. To achieve this,
we introduce the average (mean) and the imaginary part
of the retarded and advanced self-energies, \cite{Rammer07}
 \begin{equation}
 \hat{\Sigma}^M  =  \frac{1}{2}[ \hat{\Sigma}^R +  \hat{\Sigma}^A ]
 \quad , \quad
 \hat{\Sigma}^I  =  i [ \hat{\Sigma}^R -  \hat{\Sigma}^A ].
 \label{eq:Asymdef}
 \end{equation}
The inverse relations are
\begin{equation}
 \hat{\Sigma}^R  =   \hat{\Sigma}^M - \frac{i}{2}  \hat{\Sigma}^I
\quad , \quad
\hat{\Sigma}^A  =   \hat{\Sigma}^M + \frac{i}{2}  \hat{\Sigma}^I .
 \label{eq:SigmaAMI}
 \end{equation}
A similar decomposition is also introduced for the
retarded and advanced Green functions,
 \begin{equation}
\hat{G}^M  =  \frac{1}{2} [  \hat{G}^R + \hat{G}^A  ]
\quad , \quad
 \hat{G}^I  =  i [ \hat{G}^R - \hat{G}^A ],
 \label{eq:GIdef}
 \end{equation}
so that
 \begin{equation}
 \hat{G}^R  =  \hat{G}^M - \frac{i}{2} \hat{G}^I
 \quad , \quad
\hat{G}^A  =  \hat{G}^M + \frac{i}{2} \hat{G}^I .
 \end{equation}
Subtracting the Keldysh component of the left and right-hand sides of the Dyson equations  (\ref{eq:GKdyson},\ref{eq:GKdysonright}),  we obtain the (subtracted) kinetic equation
\begin{eqnarray}
& & \hat{Z}   \hat{D}^M \hat{G}^K   -  \hat{G}^K
\hat{D}^M \hat{Z}
-    [  \hat{Z}    \hat{\Sigma}^K \hat{G}^M  -
\hat{G}^M \hat{\Sigma}^K     \hat{Z} ]
 \nonumber
 \\
 &= &
 \hat{C}^{\rm in } - \hat{C}^{\rm out },
 \label{eq:Gkkin}
\end{eqnarray}
with
 \begin{equation}
  \hat{D}^{M} = \hat{D} - \hat{\Sigma}^M =  \hat{D} -  \frac{1}{2}[ \hat{\Sigma}^R +  \hat{\Sigma}^A ].
 \end{equation}
The collision integrals are represented by symmetric matrices,
\begin{subequations}
\begin{eqnarray}
  \hat{C}^{\rm in} & = &
 \frac{i}{2} [
\hat{Z} \hat{\Sigma}^K \hat{G}^I +
\hat{G}^I  \hat{\Sigma}^K \hat{Z} ],
 \label{eq:Ccolin}
 \\
 \hat{C}^{\rm out} & = &  \frac{i}{2}
 [ \hat{Z}\hat{\Sigma}^I \hat{G}^K +
\hat{G}^K \hat{\Sigma}^I \hat{Z} ],
 \label{eq:Ccolout}
 \end{eqnarray}
\end{subequations}
and correspond to the usual ``in-scattering'' and ``out-scattering'' term in the Boltzmann equation. \cite{Kamenev04}
The kinetic equation (\ref{eq:Gkkin}) generalizes the subtracted kinetic equation given in Ref.~[\onlinecite{Rammer07}] to matrix form, which includes also off-diagonal correlations.
In equilibrium both terms
on the right-hand side of Eq.~(\ref{eq:Gkkin}) cancel.

\subsection{Evolution equations for the equal-time Keldysh Green functions}
\label{sec:kindis}

Keeping in mind that in analogy with Eq.~(\ref{eq:Ft})
the diagonal and off-diagonal distribution functions are contained
in the matrix $F (\bd{k}, t )$,
 \begin{equation}
 F (\bd{k},  t ) =
 i Z G^K (\bd{k},   t , t ) Z
=   \left( \begin{array}{cc} - 2 p^{\ast}_{\bd{k}} ( t ) & 2 n_{\bd{k}} ( t ) +1 \\
 2 n_{\bd{k}} ( t ) +1 &  - 2 p_{\bd{k}} ( t ) \end{array} \right),
 \label{eq:Fttoyk}
 \end{equation}
we only need to calculate the time evolution of the equal-time Keldysh Green function in order to obtain the distribution function.
Adding Eqs.~(\ref{eq:kinintnonequal}) and (\ref{eq:kinintnonequal2}) and using
\begin{equation}
   \left.(\partial_{t} + \partial_{t'}) G^K(\bd{k},t,t') \right|_{ t^{\prime} =t }  = \partial_{t} G^K(\bd{k},t,t) ,
\end{equation}
we arrive at the evolution equation for the equal-time Keldysh Green function,
\begin{eqnarray}
      i \partial_t  G^K (\bd{k},  t , t )
& - &
 M_{\bd{k}}  G^K (\bd{k},  t , t ) -
G^K (\bd{k},  t , t ) M^T_{\bd{k}}
 \nonumber
 \\
 & = &    \int_{ t_0}^{t} d t_1 [
 {Z} \Sigma^K (\bd{k},  t, t_1 ) G^A (\bd{k}, t_1 , t)
 \nonumber
 \\
 & & \hspace{9mm}
- G^R (\bd{k}, t , t_1)   \Sigma^K (\bd{k}, t_1 , t) {Z}  ]
 \nonumber
 \\
&+ &
  \int_{ t_0}^{t} d t_1 [
{Z} \Sigma^R (\bd{k}, t , t_1 ) G^K (\bd{k}, t_1, t )
\nonumber
 \\
 & & \hspace{9mm}
-G^K ( \bd{k}, t, t_1 )   \Sigma^A (\bd{k}, t_1 , t ) {Z}
  ] . \hspace{9mm}
\label{eq:kinint}
\end{eqnarray}
Although the left-hand side of Eq.~(\ref{eq:kinint}) involves
the Keldysh Green function only at equal times, the
integrals on the right-hand side depend also on the
Green functions at different times (notice the implicit dependence via the self-energies), so that
Eq.~(\ref{eq:kinint}) is not a closed equation for
$G^K (\bd{k},  t , t )$.
In principle, one has to solve
the more general equations (\ref{eq:kinintnonequal}) and (\ref{eq:kinintnonequal2}) which fully determines
$G^K (\bd{k},  t , t^{\prime} )$ for all time arguments.
An alternative strategy is to close the kinetic equation (\ref{eq:kinint}) for
the equal-time Keldysh Green function
by approximating all Keldysh Green functions
at different times on the right-hand side in terms of
the corresponding equal-time Green function.
This is achieved by means of the so-called generalized Kadanoff-Baym ansatz\cite{Lipavsky86} (GKBA)
which is one  of the standard approximations to derive
kinetic equations for the distribution function from the
Kadanoff-Baym equations of motion for the
two-time non-equilibrium Green functions.\cite{Haug08,Fricke96,Lipavsky86}
 For bosons with diagonal and off-diagonal correlation,
the GKBA ansatz reads
\begin{eqnarray}
 G^K (\bd{k},  t , t^{\prime} ) & \approx &  - i [
 G^R (\bd{k},  t , t^{\prime} ) Z G^K (\bd{k},  t^{\prime} , t^{\prime} )
 \nonumber
 \\
 & &  \hspace{2mm} -
G^K (\bd{k},  t , t ) Z G^A (\bd{k},  t , t^{\prime} )] ,
 \label{eq:KadanoffBaym}
 \end{eqnarray}
which assumes that the exact relation Eq.~(\ref{eq:KadanoffBaym0})
between non-interacting Green functions remains approximately true
also  in the presence of interactions. Note that
Eq.~(\ref{eq:KadanoffBaym}) is a non-trivial $2 \times 2$ matrix
relation in flavor space. In appendix B we discuss the approximations which are
necessary to obtain the GKBA from the exact equations of motion (\ref{eq:kinintnonequal},\ref{eq:kinintnonequal2}) for the two-time
Keldysh Green function.

\section{Non-equilibrium functional renormalization group}
\label{sec:FRG}

The functional renormalization group (FRG) has been
quite successful to study strongly interacting systems
in equilibrium, see Refs.\ [\onlinecite{Bagnuls01,*Berges02,*Pawlowski07,*Delamotte07,*Rosten10,Kopietz10}]
for reviews.
In contrast to conventional renormalization group methods where
only a finite number of coupling constants is considered,
the FRG keeps track of the renormalization group flow of
entire correlation functions which depend on momentum, frequency, or time.
In principle, it should therefore be possible to calculate the non-equilibrium
time evolution
of quantum systems using FRG methods.
Recently, several authors
have generalized the FRG  approach to quantum systems
out of equilibrium. \cite{Gezzi07,Jakobs07,Jakobs10,Gasenzer08,*Gasenzer10,Berges09}
In particular, Gasenzer and Pawlowski \cite{Gasenzer08,*Gasenzer10} have used
FRG methods to obtain the non-equilibrium time evolution of bosons.

Given the Keldysh action defined in
Eqs.~(\ref{eq:Kelaction},\ref{eq:SGauss},\ref{eq:S1kel})
it is straightforward to write down the formally exact hierarchy of
FRG flow equations for the one-particle  irreducible vertices of the
non-equilibrium theory, which we shall do in the following subsection.
The real challenge is to devise sensible cutoff schemes and approximation strategies.
We shall address this problem in Sec.~\ref{sec:toy} using  a simple exactly solvable
toy model to check the accuracy of various approximations.

\subsection{Exact FRG flow equations}

To begin with, we consider an arbitrary bosonic many-body system whose Gaussian action
is determined by some inverse matrix propagator $\mathbf{G}_0^{-1}$, which we
modify
by introducing some cutoff parameter $\Lambda$,
 \begin{equation}
 \mathbf{G}_0 \rightarrow \mathbf{G}_{ 0 , \Lambda}.
 \label{eq:G0cut}
 \end{equation}
Depending on the problem at hand, different choices of
$\Lambda$ may be appropriate. For systems in thermal equilibrium
it is usually convenient to choose $\Lambda$  such that it removes
long-wavelength or low-energy fluctuations.\cite{Kopietz10}
In order to calculate the time evolution of many-body systems, other
choices of $\Lambda$ are more appropriate.
For example,
Gasenzer and Pawlowski  have proposed that $\Lambda$ should be identified
with a time scale $\tau$ which cuts off the time evolution of correlation functions
at long times.  \cite{Gasenzer08,*Gasenzer10} In Sec.~\ref{sec:outscattering} we shall propose an alternative
cutoff scheme which uses an external ``out-scattering rate'' as RG cutoff.

Given the cutoff-dependent  Gaussian propagator (\ref{eq:G0cut}),
the generating functional of all correlation functions
depends on the cutoff.
By taking the derivative of the generating
functional $\Gamma_{\Lambda} [ \Phi ]$ of the irreducible vertices with respect to
the cutoff, we obtain a rather compact closed functional equation for  $\Gamma_{\Lambda} [ \Phi ]$ which is
sometimes called the Wetterich equation.\cite{Wetterich93}
Formally the Wetterich equation is valid also for quantum systems
out of equilibrium provided we use the proper non-equilibrium field theory to describe
the system. \cite{Gasenzer08,*Gasenzer10}
By expanding the generating functional
 $\Gamma_{\Lambda} [ \Phi ]$ in powers of the fields we obtain the
one-particle irreducible vertices of our non-equilibrium theory,
 \begin{equation}
 \Gamma_{\Lambda} [ \Phi ] = \sum_{ n=0}^{\infty}
 \frac{1}{n!} \int_{\alpha_1} \cdots \int_{\alpha_n}
 \Gamma^{(n)}_{ \Lambda, \alpha_1 \ldots \alpha_n}
 \Phi_{\alpha_1} \cdots \Phi_{\alpha_n}.
 \label{eq:funcTaylor}
\end{equation}
Here the collective labels $\alpha_1, \alpha_2 , \ldots$ stand for all
labels  which are necessary to specify the fields. \cite{Schuetz05,Kopietz10} For the boson model
defined in Sec.~\ref{sec:Green} the collective label
$\alpha = (  \lambda, \sigma, \bd{k} , t ) $
represents the Keldysh label
$\lambda \in \{  C, Q \}$, the flavor label $\sigma \in \{ a, \bar{a} \}$, as well as
the momentum and time labels $\bd{k}$ and $ t $.
The corresponding integration symbol is
 \begin{equation}
 \int_{\alpha} = \sum_{ \lambda} \sum_{ \sigma} \sum_{\bd{k}} \int d t.
 \end{equation}
The exact FRG flow equations for the irreducible vertices can be obtained
from the general FRG  flow equations given in
Refs.~[\onlinecite{Kopietz10},\,\onlinecite{Schuetz05},\,\onlinecite{Schuetz06}]
by making the following   substitutions to take into account the
different normalization of the action in the Keldysh formalism,
 \begin{eqnarray}
 \Gamma^{(n)}_{ \Lambda, \alpha_1 \ldots \alpha_n } & \rightarrow & i
\Gamma^{(n)}_{ \Lambda, \alpha_1 \ldots \alpha_n },
 \\
 \mathbf{G}_{\Lambda}  \rightarrow  -i  \mathbf{G}_{\Lambda} &\; \; , \; \; &
 \mathbf{\dot{G}}_{\Lambda}  \rightarrow  -i  \mathbf{\dot{G}}_{\Lambda},
 \end{eqnarray}
where the  single-scale propagator is given by
 \begin{equation}
  \dot{\mathbf{G}}_{\Lambda} = - \mathbf{G}_{\Lambda}
 ( \partial_{\Lambda} \mathbf{G}_{ 0 , \Lambda}^{-1} ) \mathbf{G}_{\Lambda}.
 \label{eq:singlescaleprop}
 \end{equation}
This definition implies that the blocks
of the single-scale propagator for general cutoff are
 \begin{subequations}
 \label{eq:singlescaleproprak}
\begin{eqnarray}
  [ \dot{\mathbf{G}}_{\Lambda} ]^{CQ}  =
 \dot{\hat{G}}^R_{\Lambda} & = & - \hat{G}^R_{\Lambda} [ \partial_{\Lambda}
 ( \hat{G}^R_{ 0, \Lambda} )^{-1} ]  \hat{G}^R_{\Lambda},
 \\
  { [} \dot{\mathbf{G}}_{\Lambda} ]^{QC}  =
 \dot{\hat{G}}^A_{\Lambda} & = & - \hat{G}^A_{\Lambda} [ \partial_{\Lambda}
 ( \hat{G}^A_{ 0, \Lambda} )^{-1} ]  \hat{G}^A_{\Lambda},
 \\
  {[} \dot{\mathbf{G}}_{\Lambda} ]^{CC}  =
 \dot{\hat{G}}^K_{\Lambda} & = & - \hat{G}^R_{\Lambda} [ \partial_{\Lambda}
 ( \hat{G}^R_{ 0, \Lambda} )^{-1} ]  \hat{G}^K_{\Lambda}
 \nonumber
 \\
 &- &  \hat{G}^K_{\Lambda} [ \partial_{\Lambda}
 ( \hat{G}^A_{ 0, \Lambda} )^{-1} ]  \hat{G}^A_{\Lambda}
 \nonumber
 \\
 &- &  \hat{G}^R_{\Lambda} ( \partial_{\Lambda} [
   \mathbf{G}_{ 0, \Lambda}^{-1} ]^{QQ}) \hat{G}^A_{\Lambda} .
 \end{eqnarray}
\end{subequations}
If the expectation values of the field components $\Phi_{\alpha}$ vanish in the absence of sources,
the exact FRG flow equation for the irreducible self-energy (two-point function) is
\begin{eqnarray}
\partial_{\Lambda} \Gamma^{(2)}_{\Lambda, \alpha_{1} \alpha_{2}}
& = & \frac{i}{2} \int_{\beta_1} \int_{\beta_2}
[   \dot{\mathbf{G}}_{\Lambda} ]_{\beta_1 \beta_2}
{\Gamma}_{\Lambda, \beta_2 \beta_1 \alpha_{1} \alpha_{2}}^{(4) }   ,
\label{eq:flowGamma2}
\end{eqnarray}
while the four-point vertex (effective interaction) satisfies
 \begin{eqnarray}
\partial_{\Lambda} \Gamma^{(4)}_{\Lambda, \alpha_{1} \alpha_{2} \alpha_3 \alpha_4}
& = & \frac{i}{2} \int_{\beta_1} \int_{\beta_2}
[   \dot{\mathbf{G}}_{\Lambda} ]_{\beta_1 \beta_2}
{\Gamma}_{\Lambda, \beta_2 \beta_1 \alpha_{1} \alpha_{2} \alpha_3 \alpha_4}^{(6) }
 \nonumber
 \\
 &  & \hspace{-30mm}  + \frac{i}{2} \int_{\beta_1} \int_{\beta_2} \int_{\beta_3} \int_{\beta_4}
[   \dot{\mathbf{G}}_{\Lambda} ]_{\beta_1 \beta_2}
[  {\mathbf{G}}_{\Lambda} ]_{\beta_3 \beta_4}
\nonumber
 \\
 & &  \hspace{-25mm} \times
\Bigl[  \Gamma^{(4)}_{\Lambda, \beta_2 \beta_3 \alpha_3 \alpha_4}
 \Gamma^{(4)}_{\Lambda, \beta_4 \beta_1 \alpha_1 \alpha_2}
 +
\Gamma^{(4)}_{\Lambda, \beta_2 \beta_3 \alpha_1 \alpha_2}
 \Gamma^{(4)}_{\Lambda, \beta_4 \beta_3 \alpha_3 \alpha_4}
 \nonumber
 \\
 & & \hspace{-15mm}
  + ( \alpha_1 \leftrightarrow \alpha_2) +
 ( \alpha_1 \leftrightarrow \alpha_4) \Bigr].
\label{eq:flowGamma4}
\end{eqnarray}
If, in the absence of sources, the field has a finite expectation
value $\Phi^0_{\alpha} \neq 0$, it is convenient to re-define the
vertices in the functional Taylor series (\ref{eq:funcTaylor}) by
expanding in powers of $\delta \Phi = \Phi - \Phi^0$,
\begin{equation}
 \Gamma_{\Lambda} [ \Phi ] = \sum_{ n=0}^{\infty}
 \frac{1}{n!} \int_{\alpha_1} \cdots \int_{\alpha_n}
 \Gamma^{(n)}_{ \Lambda, \alpha_1 \ldots \alpha_n} ( \Phi^0)
 \delta \Phi_{\alpha_1} \cdots \delta \Phi_{\alpha_n}.
 \label{eq:funcTaylor2}
\end{equation}
Then, the odd vertices ($n = 1,3,5,\ldots$) are in general also finite.
The requirement that the one-point vertex vanishes identically leads to the following
 flow equation for the field expectation value,\cite{Schuetz06,Kopietz10}$\!^,$
\footnote{
As pointed out in  Refs.~[\onlinecite{Kopietz10},\,\onlinecite{Schuetz06}],
in equilibrium it is convenient to choose
the cutoff-dependent Gaussian propagator such that
 $
 \mathbf{G}_{0, \Lambda}^{-1} \Phi^0 =0$ and $
[\partial_{\Lambda} \mathbf{G}_{0, \Lambda}^{-1}] \Phi^0 =0$
which may be achieved by introducing suitable counter terms.
Then the flow equation (\ref{eq:floworder}) for the order parameter $\Phi^0$ reduces to
 \begin{equation*}
 - \int_{\beta}  ( \partial_{\Lambda} \Phi^0_{\beta} )
\Gamma^{(2)}_{\Lambda, \beta \alpha}
=   \frac{i}{2} \int_{\beta_1} \int_{\beta_2}
[   \dot{\mathbf{G}}_{\Lambda} ]_{\beta_1 \beta_2}
\Gamma^{(3)}_{\Lambda, \beta_2 \beta_1 \alpha}.
\end{equation*}
Out of equilibrium the conditions
$
 \mathbf{G}_{0, \Lambda}^{-1} \Phi^0 =0$ and $
[\partial_{\Lambda} \mathbf{G}_{0, \Lambda}^{-1}] \Phi^0 =0$
are not satisfied for our cutoff choice, so that we have
to work with the more general order parameter flow equation (\ref{eq:floworder}).
}
 \begin{eqnarray}
 & & \int_{\beta} \left[ ( \partial_{\Lambda} \Phi^0_{\beta} )
[\mathbf{G}_{\Lambda}^{-1}]_{ \beta \alpha}
+      \Phi^0_{\beta}
[  \partial_{\Lambda} \mathbf{G}_{0,\Lambda}^{-1}]_{ \beta \alpha}
 \right]
 \nonumber
 \\
& = &  \frac{i}{2} \int_{\beta_1} \int_{\beta_2}
[   \dot{\mathbf{G}}_{\Lambda} ]_{\beta_1 \beta_2}
\Gamma^{(3)}_{\Lambda, \beta_2 \beta_1 \alpha}.
 \label{eq:floworder}
 \end{eqnarray}
Moreover, the FRG flow equation for the two-point vertex contains
additional terms involving the three-point vertex,
\begin{eqnarray}
\partial_{\Lambda} \Gamma^{(2)}_{\Lambda, \alpha_{1} \alpha_{2}}
& = & \frac{i}{2} \int_{\beta_1} \int_{\beta_2}
[   \dot{\mathbf{G}}_{\Lambda} ]_{\beta_1 \beta_2}
{\Gamma}_{\Lambda, \beta_2 \beta_1 \alpha_{1} \alpha_{2}}^{(4) }
\nonumber
 \\
 & + &  \int_{\beta} ( \partial_{\Lambda} \Phi^0_{\beta} ) \Gamma^{(3)}_{\Lambda, \beta
 \alpha_1 \alpha_2}
\nonumber
 \\
 &  & \hspace{-23mm}
+ i \int_{\beta_1} \int_{\beta_2} \int_{\beta_1^{\prime}} \int_{\beta_2^{\prime}}
[   \dot{\mathbf{G}}_{\Lambda} ]_{\beta_1 \beta_1^{\prime}}
[  {\mathbf{G}}_{\Lambda} ]_{\beta_2 \beta_2^{\prime}}
\Gamma^{(3)}_{\Lambda, \beta_1 \beta_2 \alpha_1}
\Gamma^{(3)}_{\Lambda, \beta_1^{\prime} \beta_2^{\prime} \alpha_2} .
\label{eq:flowGamma2three}
 \nonumber
 \\
 & &
\end{eqnarray}

\subsection{Cutoff schemes}
\label{sec:cutoff}

\subsubsection{General considerations}

The crucial point is now to identify a sensible flow parameter
$\Lambda$. Since we are interested in calculating the time evolution
of the distribution function at long times the flow parameter should
be chosen such that for sufficiently large $\Lambda$ the long-time
asymptotics is simple. This is the case if $\Lambda$ is identified
with a scattering rate which  introduces some kind of damping.
This strategy was already implemented by Jakobs \emph{et al.}\cite{Jakobs10} in their recent FRG study of stationary non-equilibrium states of the Anderson impurity model.
Formally,
such a cutoff can be introduced by replacing the infinitesimal
imaginary part $\eta$  appearing in the retarded and advanced blocks
of the inverse matrix propagator $\mathbf{G}_0^{-1}$ given in
Eqs.~(\ref{eq:GRAM}) by a finite quantity $\Lambda$. This amounts
to the following replacement of the inverse retarded and advanced
propagators by cutoff-dependent quantities,
 \begin{subequations}
 \begin{eqnarray}
 ( \hat{G}_0^R )^{-1} \rightarrow  ( \hat{G}_{0, \Lambda}^R )^{-1} & = &
 ( \hat{G}_0^R )^{-1} - i \Lambda \hat{Z},
 \label{eq:GRLambda}
 \\
  ( \hat{G}_0^A )^{-1} \rightarrow  ( \hat{G}_{0, \Lambda}^A )^{-1} & = &
 ( \hat{G}_0^A )^{-1} + i \Lambda \hat{Z}.
 \label{eq:GALambda}
 \end{eqnarray}
 \end{subequations}
Explicitly, the cutoff-dependent retarded and advanced Green functions are then
 \begin{subequations}
\begin{eqnarray}
 G^R_{ 0, \Lambda} (\bd{k},  t , t^{\prime} ) & = &
G^R_{ 0} ( \bd {k},  t , t^{\prime} ) e^{ - \Lambda ( t - t^{\prime} )},
 \\
 G^A_{ 0, \Lambda} (\bd{k},  t , t^{\prime} ) & = &
G^A_{ 0} (\bd{k},  t , t^{\prime} ) e^{  - \Lambda (  t^{\prime}- t )}.
 \end{eqnarray}
\end{subequations}
As far as the $QQ$-component of the inverse free propagator is concerned
(which for infinitesimal $\eta$  is a pure regularization), we set
 \begin{equation}
 [\mathbf{G}_{0,\Lambda}^{-1}]^{QQ} =
2 i \Lambda \hat{F}_{\ast,\Lambda},
 \label{eq:switchcutgen}
 \end{equation}
where the distribution matrix $\hat{F}_{\ast,\Lambda}$
will be further specified below.
Defining the cutoff-dependent non-interacting
distribution matrix $\hat{F}_{0, \Lambda}$ as
in Eq.~(\ref{eq:kelF}), we have
\begin{equation}
 \hat{G}^K_{0, \Lambda} = \hat{G}^R_{0, \Lambda} \hat{F}_{0,\Lambda} \hat{Z}
-  \hat{Z} \hat{F}_{0, \Lambda} \hat{G}^A_{0,\Lambda},
 \label{eq:kelFLambda}
 \end{equation}
and
 \begin{eqnarray}
 [\mathbf{G}_{0, \Lambda}^{-1}]^{QQ} & = &
 -   (\hat{G}_{0, \Lambda}^R)^{-1}  \hat{G}_{0, \Lambda}^{K} (\hat{G}_{0, \Lambda}^A)^{-1}
 \nonumber
 \\
  & = &    (\hat{G}_{0, \Lambda}^R)^{-1} \hat{Z} \hat{F}_{0,\Lambda} -
\hat{F}_{0,\Lambda}  \hat{Z}
( \hat{G}_{0, \Lambda}^A)^{-1}
 \nonumber
 \\
 & = &  \hat{D} \hat{Z} \hat{F}_{0,\Lambda} -  \hat{F}_{0,\Lambda}  \hat{Z}
 \hat{D} + 2 i \Lambda    \hat{F}_{0, \Lambda}.
 \end{eqnarray}
Comparing this with Eq.~(\ref{eq:switchcutgen}), we see
that our cutoff-dependent distribution matrix satisfies
\begin{equation}
\label{eq:kinfcut}
\hat{D} \hat{Z} \hat{F}_{0,\Lambda} -  \hat{F}_{0, \Lambda}  \hat{Z}
 \hat{D} + 2 i \Lambda    ( \hat{F}_{0,\Lambda} -   \hat{F}_{\ast,\Lambda}) =0.
 \end{equation}
For our cutoff choice given in Eqs.\ (\ref{eq:GRLambda},\ref{eq:GALambda}) we have
 \begin{subequations}
 \begin{eqnarray}
 \partial_{\Lambda}
 ( \hat{G}^R_{ 0, \Lambda} )^{-1} & = & - i \hat{Z} ,
 \\
\partial_{\Lambda}
( \hat{G}^A_{ 0, \Lambda} )^{-1} & = &  i \hat{Z} ,
 \\
\partial_{\Lambda} [\mathbf{G}_{ 0, \Lambda}^{-1} ]^{QQ} & = & 2 i \hat{F}_{\ast,\Lambda} ,
 \end{eqnarray}
 \end{subequations}
so that in this scheme the blocks of the single-scale propagator (\ref{eq:singlescaleproprak}) are
  \begin{subequations}
\begin{eqnarray}
 \dot{\hat{G}}^R_{\Lambda} & = & i  \hat{G}^R_{\Lambda}  \hat{Z}  \hat{G}^R_{\Lambda},
 \label{eq:dotGR}
 \\
 \dot{\hat{G}}^A_{\Lambda} & = & - i \hat{G}^A_{\Lambda}
 \hat{Z}  \hat{G}^A_{\Lambda},
 \label{eq:dotGA}
 \\
 \dot{\hat{G}}^K_{\Lambda} &  = &  i
 \left[
  \hat{G}^R_{\Lambda} \hat{Z}  \hat{G}^K_{\Lambda}
 -  \hat{G}^K_{\Lambda} \hat{Z}   \hat{G}^A_{\Lambda}
- 2 \hat{G}^R_{\Lambda}  \hat{F}_{\ast,\Lambda}  \hat{G}^A_{\Lambda}  \right].
 \hspace{7mm}
 \label{eq:dotGK}
 \end{eqnarray}
\end{subequations}
Let us now discuss two possible choices of $\hat{F}_{\ast,\Lambda}$.

\subsubsection{Out-scattering rate cutoff}
\label{sec:outscattering}

The simplest possibility is to choose
\begin{equation}
 \hat{F}_{\ast,\Lambda} = \frac{\eta }{\Lambda} \hat{F}_{0,\Lambda} \rightarrow 0,
\end{equation}
so that the cutoff-dependent distribution function defined via Eq.~(\ref{eq:kinfcut}) satisfies
\begin{equation}
\hat{D} \hat{Z} \hat{F}_{0,\Lambda} -  \hat{F}_{0, \Lambda}  \hat{Z}
 \hat{D} + 2 i \Lambda    \hat{F}_{0,\Lambda} =0.
 \label{eq:difdistribution}
 \end{equation}
In this case the $QQ$-component of the inverse free propagator is chosen to be the following
cutoff-dependent infinitesimal regularization,
\begin{equation}
 [\mathbf{G}_{0,\Lambda}^{-1}]^{QQ}  =
2 i \eta \hat{F}_{0, \Lambda}.
 \end{equation}
The term $2 i \Lambda    \hat{F}_{0,\Lambda}$ in Eq.~(\ref{eq:difdistribution})
amounts to the following substitution for the time-derivative
in the  equations of motion for the distribution function,
 \begin{equation}
 \partial_t \rightarrow \partial_t + 2 \Lambda.
 \end{equation}
The time-diagonal element of the non-interacting distribution function is then modified as follows,
 \begin{equation}
 F_{0, \Lambda}  (\bd{k},  t ) = e^{ - 2 \Lambda t }  F_{0}  ( \bd{k}, t ),
 \end{equation}
whereas the cutoff-dependent non-interacting Keldysh Green function is now given by
 \begin{equation}
G^K_{ 0, \Lambda} (\bd{k},  t , t^{\prime} ) = e^{ - \Lambda ( t + t^{\prime} ) }
 G^K_{ 0} (\bd{k},  t , t^{\prime} ).
 \end{equation}
The occupation numbers therefore decrease exponentially
with rate $\Lambda$ for large time, which
justifies the name ``out-scattering cutoff scheme.''
Because for $\Lambda \rightarrow \infty$ all propagators vanish in this scheme,
the FRG flow equations should be integrated with the initial condition
 \begin{equation}
 \label{eq:gamma0}
 \lim_{\Lambda \rightarrow \infty} \Gamma^{(n)}_{\Lambda, \alpha_1 \ldots \alpha_n}
=0 \quad \text{if } n \neq 4,
 \end{equation}
and the limit of $\Gamma^{(4)}_{\Lambda, \alpha_1 \alpha_2 \alpha_3  \alpha_4}$ is given by the bare interaction Eq. (2.51).

\subsubsection{Hybridization cutoff}

Alternatively, we may choose $\hat{F}_{\ast,\Lambda} = \hat{F}_{0, \Lambda}$,
so that the distribution function satisfies
\begin{equation}
\hat{D} \hat{Z} \hat{F}_{0,\Lambda} -  \hat{F}_{0, \Lambda}  \hat{Z}
 \hat{D} =0.
\label{eq:inoutkin}
 \end{equation}
This equation agrees exactly with the cutoff-independent non-interacting kinetic equation
(\ref{eq:nullrelation}), so that we may identify $\hat{F}_{0, \Lambda} = \hat{F}_0$
with the cutoff-independent non-interacting distribution function.
Obviously, this cutoff choice amounts to replacing the
infinitesimal $\eta$ appearing in Eq.~(\ref{eq:pureregu}) by the
running cutoff $\Lambda$, so that
 \begin{equation}
 [\mathbf{G}_{0,\Lambda}^{-1}]^{QQ} =
2 i \Lambda \hat{F}_{0},
 \label{eq:switchcut}
 \end{equation}
where $\hat{F}_0$ is the distribution function for infinitesimal $\eta$, which is
determined by the same equation as for $\Lambda =0$.
With this cutoff choice
all propagators at non-equal times vanish for $\Lambda \rightarrow \infty$.
Explicitly, we obtain for the non-interacting $2 \times 2$ Green functions
in flavor space
in this cutoff scheme,
 \begin{eqnarray}
G^K_{ 0, \Lambda} (\bd{k},  t , t^{\prime} ) & = &
 G^R_{ 0, \Lambda} (\bd{k},  t , t^{\prime} ) F_0 ( \bd{k}, t^{\prime} ) Z
 \nonumber
 \\
 &- & Z F_0 (\bd{k},  t )  G^A_{ 0, \Lambda} (\bd{k}, t , t^{\prime} ).
 \end{eqnarray}
Because for large $\Lambda \rightarrow \infty$
all propagators at non-equal times
are suppressed, each time integration in loops
yields a factor of $1/\Lambda$. For $\Lambda \rightarrow \infty$
only the Hartree-Fock contribution to the self-energy survives because it depends
only on the equal-time component of the Green function.
The FRG flow equations in this cutoff scheme should therefore be integrated with the
boundary condition that for $\Lambda \rightarrow \infty$ the irreducible self-energy
is given by the self-consistent Hartree-Fock approximation.

The finite value of the $QQ$-block of the inverse propagator
can be considered to be a part of the Keldysh self-energy, which
in turn is related to an in-scattering rate.\cite{Kamenev04}
Compared to the out-scattering rate cutoff introduced before, this cutoff scheme contains both in and out scattering contributions, such that the
bare distribution function is cutoff independent.
Essentially the same cutoff scheme has recently been proposed and tested by
Jakobs, Pletyukhov and Schoeller. \cite{Jakobs10a,Jakobs10,Karrasch10a}
In particular in Ref.~[\onlinecite{Jakobs10}]  they used this scheme to study
stationary non-equilibrium states of the Anderson impurity model. In this case they identified the cutoff parameter $\Lambda$ with the hybridization energy arising from the coupling to the bath of free electrons.
Following their suggestion we shall therefore refer to this scheme as the ``hybridization cutoff scheme.''

\subsubsection{Alternative cutoff schemes}

At this point it is not clear which cutoff choice is superior.
By construction, both schemes do no violate causality for any value of the running
cutoff $ \Lambda$. Moreover, to describe
systems close to thermal equilibrium it might be important
to require that in thermal equilibrium
the fluctuation-dissipation theorem relating the Keldysh Green function
to its retarded and advanced counter-parts is satisfied for any value of the
running cutoff.
Using  the spectral representation of the Green functions, it is
easy to show that for our model the fluctuation-dissipation theorem
can be written as the following relation between the Fourier transforms
of the $2 \times 2$ matrix Green functions in
flavor space,
 \begin{equation}
 G^K ( \bd{k} , \omega ) = [
 G^R ( \bd{k} , \omega ) -   G^A ( \bd{k} , \omega ) ]
 \left[ \frac{2}{e^{\beta \omega} -1} +1 \right].
 \end{equation}
While this relation is manifestly violated in the
out-scattering cutoff scheme,
in the hybridization cutoff scheme it remains valid
in the non-interacting limit.

Finally, let us point out that in certain situations other choices of the distribution
matrix $\hat{F}_{\ast,\Lambda}$ in the $QQ$-block of the regularized
inverse propagator given in Eq.~(\ref{eq:switchcutgen}) may be advantageous.
For example, for describing the approach to thermal equilibrium,
it might be useful to identify $\hat{F}_{\ast,\Lambda}$ with the true
equilibrium distribution.
To see this, let us approximate the Keldysh Green function $\hat{G}^K$
appearing in the Keldysh-block of the single-scale propagator
(\ref{eq:dotGK}) by the generalized Kadanoff-Baym ansatz
(\ref{eq:KadanoffBaym}),
which
assumes that the non-interacting relation (\ref{eq:KadanoffBaym0})
remains approximately valid for the interacting system.
Introducing the flowing distribution function
in analogy with Eq.~(\ref{eq:Ft}),
\begin{equation}
 F_\Lambda (\bd{k},  t ) =
 i Z G^K_\Lambda (\bd{k},  t , t ) Z,
 \label{eq:FGtilde}
 \end{equation}
the generalized Kadanoff-Baym ansatz (\ref{eq:KadanoffBaym})
can also be written as
\begin{eqnarray}
G^K_{  \Lambda} (\bd{k},  t , t^{\prime} ) & = &
 G^R_{  \Lambda} (\bd{k},  t , t^{\prime} ) F_{\Lambda} ( \bd{k}, t^{\prime} ) Z
 \nonumber
 \\
 &- & Z F_{\Lambda} (\bd{k},  t )  G^A_{  \Lambda} (\bd{k}, t , t^{\prime} ).
 \end{eqnarray}
Substituting this approximation into Eq.~(\ref{eq:dotGK}), we obtain for
the Keldysh block of the single-scale propagator
at equal times
 \begin{eqnarray}
  \dot{G}^K_{\Lambda} ( \bd{k} , t ,t )
 & = & 2 i \int_{t_0}^t d t_1 G^R_{\Lambda} ( \bd{k}, t , t_1 ) [
 F_{\Lambda} ( \bd{k}, t_1 ) - F_{\ast,\Lambda} ( \bd{k} ) ]
 \nonumber
 \\
 & & \hspace{10mm} \times
 G^A_{\Lambda} ( \bd{k}, t_1 , t ).
\label{eq:genCutoff}
 \end{eqnarray}
Obviously, this expression vanishes if the flowing  distribution matrix
$F_{\Lambda} ( \bd{k}, t_1 )$ approaches the equilibrium distribution
$F_{\ast,\Lambda} ( \bd{k} )$.

\subsection{Combining FRG flow equations with quantum kinetic equations}

The FRG flow equation (\ref{eq:flowGamma2}) relates the derivative  of the
self-energy $[\mathbf{\Sigma}_{ \Lambda} ]_{\alpha_1 \alpha_2} \equiv
\Gamma^{(2)}_{\Lambda, \alpha_{1} \alpha_{2}}$
with respect to the flow parameter $\Lambda$
to the flowing Green function $\mathbf{G}_{\Lambda}$
and to the flowing effective interaction
${\Gamma}_{\Lambda, \beta_2 \beta_1 \alpha_{1} \alpha_{2}}^{(4) } $.
Our final goal is to obtain a  closed equation for the Keldysh block
$\hat{G}_{\Lambda}^K$ of the Green function matrix at equal times
(or alternatively the distribution function $\hat{F}_{\Lambda}$),
from which we can extract the time evolution of the
diagonal and off-diagonal distribution functions given in Eq.~(\ref{eq:nkdef},\ref{eq:pkdef}).
Therefore we have to solve the FRG flow equation
(\ref{eq:flowGamma2}) simultaneously with the cutoff-dependent quantum kinetic equation,
which can be derived analogously to Sec.~\ref{sec:quantumkinetic} from the cutoff-dependent Dyson equation,
\begin{equation}
 \mathbf{G}^{-1}_{\Lambda} = \mathbf{G}^{-1}_{0, \Lambda} -
 \mathbf{\Sigma}_{\Lambda}.
 \end{equation}
The cutoff-dependent kinetic  equation for the Keldysh
block can be derived in the same way as in
Sec.~\ref{sec:quantumkinetic}, and we thus obtain for the equal-time
Keldysh Green function,
\begin{eqnarray}
      i \partial_t  G^K_{\Lambda} (\bd{k},  t , t )
& - &
 M_{\Lambda, \bd{k}}  G^K_{\Lambda} (\bd{k},  t , t ) -
G^K_{\Lambda} (\bd{k},  t , t ) M_{\Lambda, \bd{k}}^T
 \nonumber
 \\
 & = &    \int_{ t_0}^{t} d t_1 [
 {Z} \Sigma^K_{\Lambda} (\bd{k},  t, t_1 ) G^A_{\Lambda} (\bd{k}, t_1 , t)
 \nonumber
 \\
 & & \hspace{9mm}
- G^R_{\Lambda} ( \bd{k}, t , t_1)   \Sigma^K_{\Lambda} ( \bd{k}, t_1 , t) {Z}  ]
 \nonumber
 \\
&+ &
  \int_{ t_0}^{t} d t_1 [
{Z} \Sigma^R_{\Lambda} ( \bd{k}, t , t_1 ) G^K_{\Lambda} ( \bd{k}, t_1, t )
\nonumber
 \\
 & & \hspace{9mm}
-G^K_{\Lambda} (\bd{k},  t, t_1 )   \Sigma_{\Lambda}^A (\bd{k},  t_1 , t ) {Z}
  ] , \hspace{9mm}
 \label{eq:kinintLambdaequal}
\end{eqnarray}
where $M_{\Lambda, \bd{k}}$ is a cutoff dependent deformation of the
matrix $M_{\bd{k}}$ defined in Eq.~(\ref{eq:tildeMkdef}). The
explicit form of $M_{\Lambda, \bd{k}}$ depends on the cutoff scheme.
For the out-scattering cutoff scheme, it follows from
Eq.~(\ref{eq:difdistribution}) and (\ref{eq:FGtilde}) that
$M_{\Lambda, \bd{k}} = M_{\bd{k}} - i \Lambda I$. For the hybridization cutoff scheme,  Eqs.\ (\ref{eq:inoutkin}) and (\ref{eq:FGtilde}) imply that $M_{\Lambda, \bd{k}} = M_{\bd{k}}$,
which is identical to Eq.~(\ref{eq:tildeMkdef}). The general form of
the kinetic equation (\ref{eq:kinintLambdaequal}) is of course
similar to Eq.~(\ref{eq:kinint}), except that now all Green
functions and self-energies depend on the  cutoff parameter
$\Lambda$. Together with the FRG flow equation
(\ref{eq:flowGamma2}), this equation forms a system of coupled
first-order partial integro-differential equations with two
independent variables  $t$ and $\Lambda$, which have to be solved
simultaneously. Because the flow equation (\ref{eq:flowGamma2})
depends on the effective interaction which satisfies the flow
equation (\ref{eq:flowGamma4}), the simplest truncation is to neglect the flow of the interaction.
However, the resulting system of kinetic
and flow equations given by Eqs.\ (\ref{eq:kinintLambdaequal}) and
(\ref{eq:flowGamma2}) is not closed because the flow and the kinetic
equation contains integrals involving the two-time Keldysh Green
function. To reduce the complexity and to close the system of
equations, the usual approximation strategies of quantum kinetics
can  now be made. For example, on the right-hand side of the quantum
kinetic equation (\ref{eq:kinintLambdaequal}), one could express
the Keldysh Green function for non-equal times in terms of the
corresponding equal-time Keldysh Green function using the
generalized Kadanoff-Baym ansatz (\ref{eq:KadanoffBaym}). Further
simplifying approximations such as the Markov-approximation where
the time-arguments of all Keldysh Green functions on the right-hand
side of  Eq.~(\ref{eq:kinintLambdaequal}) are replaced by the
external time $t$ might also be useful.
For the solution to be unique, we have to further specify the boundary conditions.
For our system of kinetic and flow equation it is sufficient to define the distribution $F_\Lambda(t_0)$ at the initial time $t_0$  for arbitrary cutoff $\Lambda$, and the self-energy $\Sigma_{\Lambda_0}(t)$ at the initial cutoff scale $\Lambda_0$ for arbitrary time $t$.
We shall illustrate this choice of boundary conditions and the
approximations mentioned above in the following section within the framework of a
simple exactly solvable toy model.

\section{Exactly solvable toy model}
 \label{sec:toy}

Although the functional renormalization group approach for bosons out
of equilibrium developed in Sec.~\ref{sec:FRG} is rather general, at
this point it is perhaps not so clear whether this approach is
useful in practice to calculate the non-equilibrium time evolution
of interacting bosons. One obvious problem which we have not
addressed so far is that truncation strategies of the
formally exact hierarchy of FRG flow equations have to be
constructed which correctly describe the long-time asymptotics.

As a first step in this direction, we shall consider in this section
a simplified version of our boson Hamiltonian (\ref{eq:Hpr}) which
is obtained by retaining only the operators associated with the
$\bd{k} =0$ mode. Setting $a_{\bd{k} =0} =a$, $\epsilon_{\bd{k} =0}
= \epsilon$, $\gamma_{\bd{k}=0} = \gamma$ and $U ( 0,0; 0,0)/V = u$,
our boson Hamiltonian (\ref{eq:Hpr}) thus reduces to the following
bosonic ``toy model'' Hamiltonian,
 \begin{eqnarray}
 {\cal{H}} ( t ) & = & \epsilon a^{\dagger} a + \frac{1}{2}\left[ \gamma e^{ - i \omega_0 t}
 a^{\dagger} a^{\dagger} + \gamma^{\ast} e^{ i \omega_0  t } a  a \right]
 \nonumber
 \\
 &+ & \frac{u}{2} a^{\dagger} a^{\dagger} a a.
 \label{eq:Htoy0}
 \end{eqnarray}
In the rotating reference frame Eq.~(\ref{eq:Htoy0}) becomes
\begin{equation}
 \tilde{\cal{H}}  =  \left(\epsilon - \frac{\omega_0}{2}\right) a^{\dagger} a + \frac{|\gamma|}{2}\left[
 a^{\dagger} a^{\dagger} +  a  a \right]
+ \frac{u}{2} a^{\dagger} a^{\dagger} a a.
 \label{eq:Htoy}
 \end{equation}
For notational simplicity, we redefine again$^{40}$  $\epsilon - \frac{\omega_0}{2} \rightarrow
\epsilon$. This simplified model describes a single anharmonic
quantum mechanical oscillator subject to a time-dependent external
field which creates and annihilates pairs of excitations. Although
this toy model  does not describe relaxation and dissipation
processes, it does captures some aspects
of the physics of parametric resonance in dipolar
ferromagnets.~\cite{Kloss10}

The non-equilibrium dynamics of the Hamiltonian (\ref{eq:Htoy}) can be easily
determined numerically by directly solving the time-dependent
Schr\"{o}dinger equation.
Expanding the time-dependent states $ | \psi ( t ) \rangle$
of the  Hilbert space associated with Eq.~(\ref{eq:Htoy})
in the basis of eigenstates $ | n \rangle$ of the particle number
operator $ a^{\dagger} a$,
 \begin{equation}
| \psi ( t ) \rangle = \sum_{n=0}^{\infty} \psi_n ( t ) | n \rangle,
 \end{equation}
the time-dependent Schr\"{o}dinger equation assumes the form
 \begin{eqnarray}
 i  {\partial}_t \psi_n ( t ) & = & \left[ \epsilon n + \frac{u}{2} n ( n-1 ) \right]
\psi_n ( t )
 \nonumber
 \\
& + &
 \frac{| \gamma| }{2}
 \Bigl[
\sqrt{ n ( n-1 ) } \psi_{ n-2} ( t )
 \nonumber
 \\
&  & \hspace{7mm} +
  \sqrt{ (n+2) ( n+1 ) } \psi_{ n+2} ( t ) \Bigr].
 \label{eq:DGLrot}
 \end{eqnarray}
This system of equations is easily solved numerically. From the solution
we may construct the normal and anomalous distribution functions,
 \begin{eqnarray}
 n ( t ) & =  & \langle \psi ( t ) | a^{\dagger} a | \psi ( t ) \rangle
 = \sum_{n=0}^{\infty} n | \psi_n ( t ) |^2,
\label{eq:nexact}
 \\
 p ( t ) & = &  \langle \psi ( t ) | a a | \psi ( t ) \rangle
 \nonumber
 \\
& = &
 \sum_{n=0}^{\infty} \sqrt{ (n+2)(n+1)}  \psi_n^{\ast} ( t ) \psi_{ n+2} ( t ).
 \label{eq:pexact}
\end{eqnarray}
We have prepared the coefficients at the initial time as
\begin{equation}
  \psi_n(t_0) = \delta_{n,1},
\end{equation}
so that the normal and anomalous pair-correlators have the initial values $n (0) =1$ and $p (0 ) =0$.
With this choice all other correlators vanish at the initial time.
We have solved the Schr\"odinger equation (\ref{eq:DGLrot}) numerically by both integrating it directly and by calculating the matrix exponential $\exp[-i \tilde{\mathcal{H}} (t-t_0)]$ and using
\begin{equation}
 \psi_n ( t ) = \sum_{j=0}^{n_\text{max}-1}  [e^{-i \tilde{\mathcal{H}} (t-t_0)}]_{nj} \psi_j ( t_0
),
\label{eq:matexp}
\end{equation}
where the Hamiltonian $\mathcal{H}$ has the matrix elements
 \begin{align}
 [\mathcal{H}]_{nm}  &=  \left[ \epsilon n + \frac{u}{2} n ( n-1 ) \right] \delta_{n,m}
+ \frac{| \gamma| }{2}
 \Bigl[
\sqrt{ n ( n-1 ) } \delta_{n-2,m}
 \nonumber
 \\
 &+
  \sqrt{ (n+2) ( n+1 ) } \delta_{n+2,m} \Bigr].
 \end{align}
We found identical results with both methods. A total number of
$n_\text{max} = $ 20 basis coefficients was sufficient for convergence.

\subsection{Time-dependent Hartree-Fock approximation}
\label{sec:hf}

\begin{figure}[tb]
  \centering
\includegraphics[width=80mm]{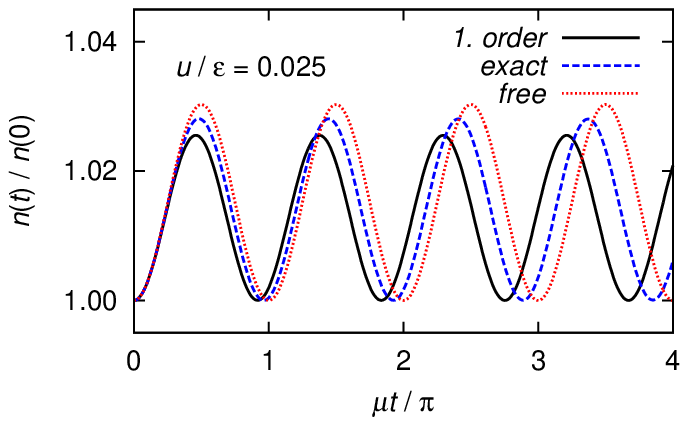}
\\
\includegraphics[width=80mm]{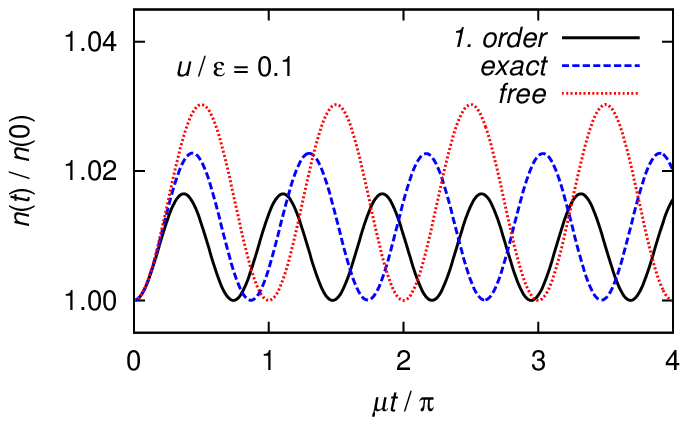}
\\
\includegraphics[width=80mm]{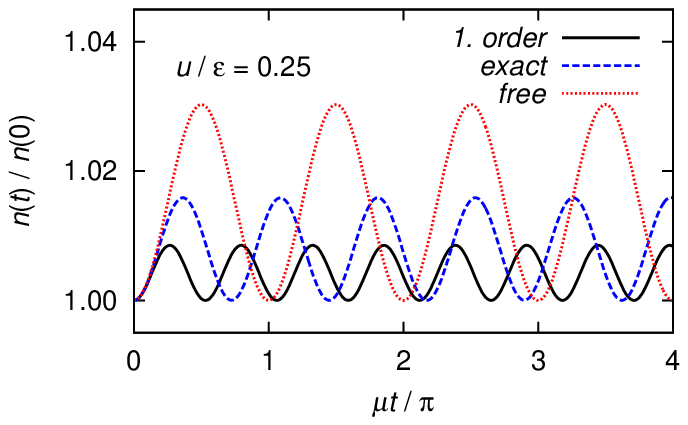}
  \vspace{-4mm}
  \caption{%
(Color online) Time evolution of the diagonal
distribution function for  initial conditions  $n (0) =1$ and $p (0 ) =0$.
We have chosen $|\gamma| / \epsilon =  0.1$ and three different values
of the interaction strength: $u / \epsilon = 0.025$ (top), 
$u / \epsilon = 0.1$ (middle), and  $u / \epsilon = 0.25$ (bottom). The frequency
$\mu = \sqrt{ \epsilon^2 - | \gamma |^2 }$ determines the oscillation period in the non-interacting limit.
We compare the  result of the
self-consistent Hartree-Fock approximation (solid line) with the exact solution (dashed line)
and the time evolution in the non-interacting limit (dotted line).
}
  \label{fig:mean-field}
\end{figure}

\begin{figure}[tb]
  \centering
\includegraphics[width=80mm]{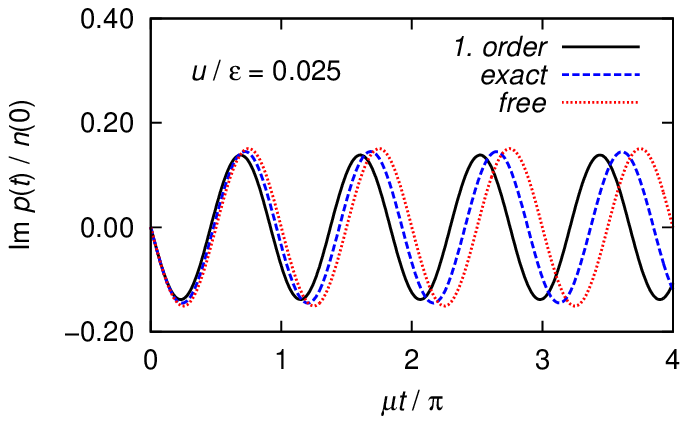}
\\
\includegraphics[width=80mm]{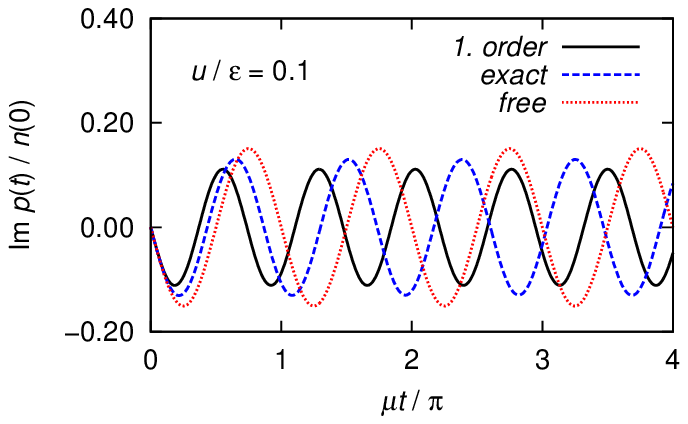}
\\
\includegraphics[width=80mm]{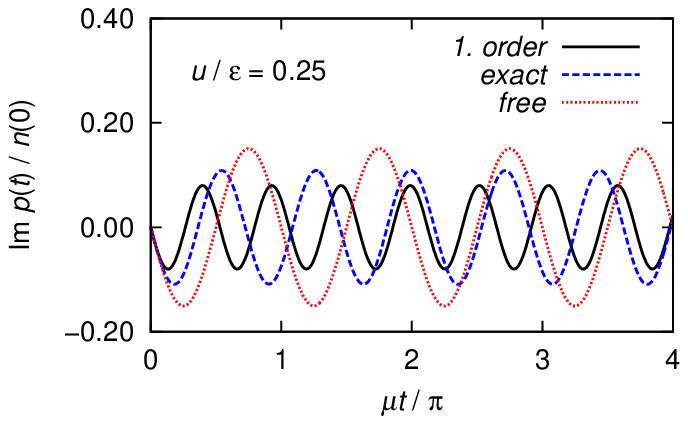}
  \vspace{-4mm}
  \caption{%
(Color online) Time evolution of the off-diagonal
distribution function for increasing interaction strength (from top to bottom).
The parameters and initial conditions are the same as in Fig.~\ref{fig:mean-field}.
We compare the  result of the
self-consistent Hartree-Fock approximation (solid line) with the exact solution (dashed line)
and the time evolution in the non-interacting limit (dotted line).
}
  \label{fig:mean-field-anom}
\end{figure}

As a reference, let is briefly discuss the self-consistent Hartree-Fock approximation
for our toy model. In the context of parametric resonance of magnons in yttrium-iron garnet, this approximation
is also referred to as ``S-theory.'' \cite{Zakharov70,Lvov94} Within this approximation
the self-energy matrix is  time-diagonal,
 \begin{equation}
 [ \mathbf{\Sigma} ]_{ \sigma t , \sigma^{\prime} t^{\prime} }^{ \lambda \lambda^{\prime} } = \delta ( t - t^{\prime} )   \Sigma_{ \sigma \sigma^{\prime}}^{ \lambda \lambda^{\prime}} ( t ) .
 \end{equation}
With the help of the symmetrized interaction vertex defined
in Eq.~(\ref{eq:ubare}) we may write
 \begin{eqnarray}
 \Sigma_{ \sigma \sigma^{\prime} }^{ \lambda \lambda^{\prime}} ( t )  &= &
\frac{1}{2}
\sum_{ \sigma_1 \sigma_2} \sum_{ \lambda_1  \lambda_2}
 U^{  \lambda_1  \lambda_2  \lambda  \lambda^{\prime}}_{ \sigma_1 \sigma_2 \sigma
\sigma^{\prime} }
\langle
\Phi_{ \sigma_1}^{\lambda_1} ( t ) \Phi_{ \sigma_2}^{\lambda_2} ( t )
\rangle
 \nonumber
 \\
 &   = & \frac{i}{2}
\sum_{ \sigma_1 \sigma_2}
 \sum_{ \lambda_1  \lambda_2}
 U^{  \lambda_1  \lambda_2  \lambda  \lambda^{\prime}}_{ \sigma_1 \sigma_2 \sigma
\sigma^{\prime}  }
 G^{\lambda_1 \lambda_2}_{ \sigma_1 \sigma_2} ( t , t ).
 \label{eq:selfsym1}
 \end{eqnarray}
Recall that by definition ${\Sigma}^{QC} \equiv {\Sigma}^{R}$ and  ${\Sigma}^{CQ} \equiv {\Sigma}^{A}$,
so that we obtain for the time-diagonal
elements of the retarded and advanced self-energy,
 \begin{align}
   {\Sigma}_1 ( t ) \equiv \Sigma^R ( t ) &= \Sigma^A ( t ) 
= i u \left( \begin{array}{cc}
\frac{1}{2} G^K_{ \bar{a} \bar{a}} ( t,t ) &   G^K_{ a \bar{a}} ( t,t ) \\
 G^K_{ \bar{a} a} ( t,t ) & \frac{1}{2} G^K_{  {a} {a}} ( t,t ) \end{array} \right) \nonumber \\
&  = u \left( \begin{array}{cc}
 p^{\ast} ( t ) & 2 n (t) + 1 \\
2 n ( t ) +1 & p ( t ) \end{array} \right)    .
 \label{eq:SigmaStheory}
\end{align}
The Keldysh component of the self-energy vanishes in this
approximation,
 \begin{equation}
  {\Sigma}^{QQ} ( t ) \equiv   {\Sigma}^K ( t )  =0.
 \end{equation}
Actually, there is an additional time-independent interaction correction
$-u $ to the normal component of the advanced and retarded self-energy which arises from the
symmetrization of the Hamiltonian, as discussed
in Sec.~\ref{sec:func}.
According to Eq.\ (\ref{eq:epsilonshift}) this contribution simply leads to a constant shift $-u$ in the energy in Eq.~(\ref{eq:tildeMkdef}).
Taking this shift into account, we find that
our kinetic equation (\ref{eq:kinint})
reduces to the following $2 \times 2$ matrix equation,
\begin{equation}
  i \partial_t  F ( t ) =  - M^T(t)  F ( t ) - F ( t ) M ( t )    ,
 \label{eq:kinS}
\end{equation}
where
 \begin{equation}
  M ( t ) =
 M
 + Z \Sigma_1 ( t )  =
\left( \begin{array}{cc} \epsilon ( t ) & \gamma ( t ) \\
 - \gamma^{\ast} ( t ) & - \epsilon ( t ) \end{array} \right),
 \end{equation}
with
 \begin{equation}
 M =
\left( \begin{array}{cc} \epsilon -u & | \gamma | \\
 - | \gamma | & - (\epsilon -u )  \end{array} \right),
 \label{eq:tildeMdef}
 \end{equation}
and
\begin{subequations}
 \begin{eqnarray}
 \epsilon ( t ) & = & \epsilon + 2 u n ( t ) ,
 \label{eq:epst}
 \\
  \gamma ( t ) & = & | \gamma | + u p ( t ).
 \label{eq:gammat}
 \end{eqnarray}
 \end{subequations}
Recall that according to Eq.~(\ref{eq:Ft}) the $2 \times 2$ distribution matrix
is given by
 \begin{equation}
 F (  t ) =
 i Z G^K (  t , t ) Z
=   \left( \begin{array}{cc} - 2 p^{\ast} ( t ) & 2 n ( t ) +1 \\
 2 n ( t ) +1 &  - 2 p ( t ) \end{array} \right).
 \label{eq:Fttoy}
 \end{equation}
At this level of approximation
the kinetic equation (\ref{eq:kinS}) has the same structure as the
corresponding equation (\ref{eq:kinfree}) in the absence of interactions.
From Eqs.~(\ref{eq:Fttoy}) and (\ref{eq:kinS})
we obtain the following
kinetic equations for the diagonal and off-diagonal distribution functions, \cite{Kloss10}
 \begin{subequations}
 \begin{eqnarray}
 i \partial_t n ( t ) & = & -  \gamma^{\ast} (t)  p (t ) +  \gamma ( t )
p^{\ast} (t) ,
 \label{eq:kinn1}
 \\
 i  \partial_t p ( t ) & = & 2 \epsilon ( t ) p (t )  +
 \gamma ( t )  [ 2 n (t ) +1 ] .
 \label{eq:kinp1}
\end{eqnarray}
 \end{subequations}
In Figs.~\ref{fig:mean-field} and \ref{fig:mean-field-anom} we compare
the numerical solution of these equations with the exact result
obtained from Eqs.~(\ref{eq:DGLrot}--\ref{eq:pexact}), and with the
time evolution in the non-interacting limit.
Because our simple toy model does not
account for damping and dissipative effects, the
time evolution is purely oscillatory.
However, the true oscillation period lies  between the
non-interacting oscillation period $ T_0 = \pi / \mu = \pi / \sqrt{ \epsilon^2 - | \gamma |^2 }$ and the smaller
oscillation period predicted by the self-consistent Hartree-Fock
approximation. A similar phenomenon is also observed for the oscillation amplitudes. As expected the deviation between the three curves  increases with increasing interaction strength.
We thus conclude that the Hartree-Fock approximation is only capable for moderate interaction strength and short times as illustrated in the middle panel of
Figs.\ \ref{fig:mean-field} and \ref{fig:mean-field-anom}. 
In this case the Hartree-Fock result up to times of the order
$T_0/4$ is reliable. However,  in this regime the free time evolution is also  fairly accurate.

\subsection{Kinetic equation with self-energy up to second order}
\label{sec:pert}
Let us consider again the quantum kinetic equation for the
Keldysh Green function $G^K ( t , t^{\prime})$, which for our toy
model can be obtained by simply omitting the momentum label in
Eqs.~(\ref{eq:kinintnonequal}) and (\ref{eq:kinintnonequal2}). Substituting on the right-hand side of this equation the
self-energies up to second order in the interaction given in
Eq.~(\ref{eq:selfsym1}) (first order self-energy) and in
appendix C (second order self-energy), we obtain an equation of
motion for the two-time Keldysh Green function $G^K ( t , t^{\prime}
)$. Together with the corresponding retarded and advanced components
of the Dyson equation, this equation forms a closed system of
partial differential equations, which can in principle be solved
numerically. To simplify the numerics, we will focus here only on
the evolution equation for the equal-time Keldysh Green function
$G^K ( t , t )$ which can be obtained from the kinetic equation
(\ref{eq:kinint}) by omitting the momentum labels,
\begin{eqnarray}
      i \partial_t  G^K (  t , t )
& - &
 M  G^K (  t , t ) -
G^K (  t , t ) M^T
 \nonumber
 \\
 & =  &    \int_{ t_0}^{t} d t_1 [
 {Z} \Sigma^K (  t, t_1 ) G^A ( t_1 , t)
 \nonumber
 \\
 & & \hspace{9mm}
- G^R ( t , t_1)   \Sigma^K ( t_1 , t) {Z}  ]
 \nonumber
 \\
&+ &
  \int_{ t_0}^{t} d t_1 [
{Z} \Sigma^R ( t , t_1 ) G^K ( t_1, t )
\nonumber
 \\
 & & \hspace{9mm}
-G^K (  t, t_1 )   \Sigma^A ( t_1 , t ) {Z}
  ] . \hspace{9mm}
\label{eq:kinnok}
\end{eqnarray}
Since the two-time function $G^K ( t , t^{\prime} )$ appears again on the right hand side of this equation
let us make three additional standard approximations to close the system of equations:

\begin{enumerate}

\item {\it{Generalized Kadanoff-Baym ansatz:}}
As discussed in Sec.~\ref{sec:kindis}, with the
help of the generalized Kadanoff-Baym ansatz (\ref{eq:KadanoffBaym})
we may derive a closed integral equation
for the equal-time Keldysh Green function $G^K ( t , t )$.
For our toy model, the generalized Kadanoff-Baym ansatz reads
\begin{eqnarray}
 G^K ( t , t^{\prime} )  & \approx & - i [
 G^R ( t , t^{\prime} ) Z G^K ( t^{\prime} , t^{\prime} )
 \nonumber
 \\
& & \hspace{2mm}
-
G^K ( t , t ) Z G^A ( t , t^{\prime} )] .
 \label{eq:KadanoffBaymtoy}
 \end{eqnarray}
This ansatz amounts to approximating
the distribution matrix in the collision integrals by its diagonal elements,
 \begin{eqnarray}
 [\hat{F}]_{tt'} \equiv F ( t, t^{\prime} ) & \approx & \delta ( t - t^{\prime} ) F ( t )
 \nonumber
 \\
& = &
\delta ( t - t^{\prime} ) i Z
 G^K ( t,t) Z.
 \label{eq:genKB2}
 \end{eqnarray}

\item {\it{Markov approximation}}:
To reduce the integro-differential equation for the
equal-time Keldysh Green function to an ordinary differential equation,
we replace  under the integral
 \begin{eqnarray}
 \int_{t_0}^{t} d t_1  G^K ( t_1 , t_1 )  \ldots & \rightarrow& G^K ( t , t )
 \int_{t_0}^{t} d t_1 \ldots .
 \end{eqnarray}

\item {\it{Free advanced and retarded propagators:}}
Finally, we neglect the self-energy corrections of the advanced and
retarded propagators in Eq.~(\ref{eq:kinnok}) and thus replace
$G^R(t,t')$ and $G^A(t,t')$ by the free propagators, which can be obtained by omitting the momentum labels in
Eqs.~(\ref{eq:GRres},\ref{eq:GAres}).

\end{enumerate}

After these approximations, the collision integrals in Eq.~(\ref{eq:kinnok}) can be calculated analytically and the non-equilibrium distribution functions
are easily obtained by numerically solving a system of two coupled ordinary
differential equations.
For the numerical solution, the time grid was
chosen equally spaced with $ \Delta t \,\epsilon = 1.3 \times
10^{-2} $ and the differential equations were solved using a fourth-order Runge-Kutta algorithm. \cite{Hankebook}

\begin{figure}[tb]
  \centering
\includegraphics[width=80mm]{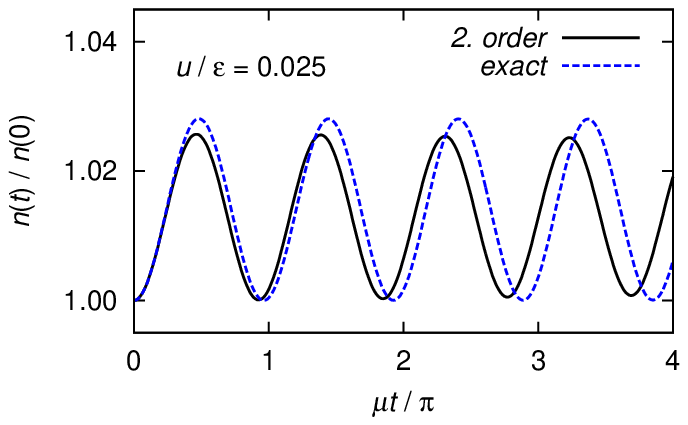}
\\
\includegraphics[width=80mm]{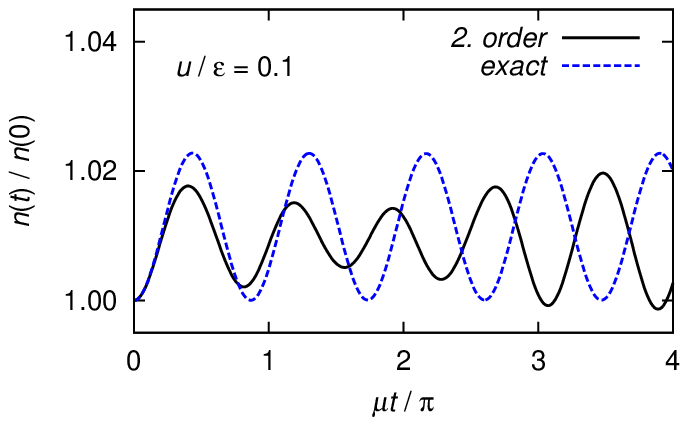}
\\
\includegraphics[width=80mm]{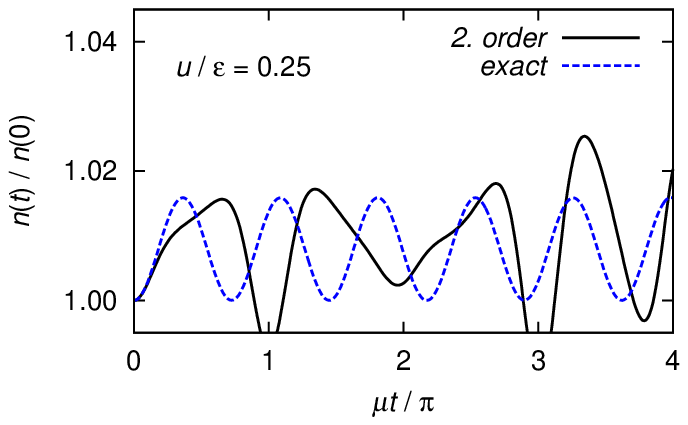}
  \vspace{-4mm}
  \caption{%
(Color online)
Time evolution of the diagonal distribution function of the toy model.
The parameters and initial conditions are the same as in Fig.~\ref{fig:mean-field}.
The solid line is the solution of the kinetic equation with second order self-energies,
simplified using the generalized Kadanoff-Baym ansatz and the Markov approximation.
For comparison we also show the exact solution (dashed line).
}
  \label{fig:pert}
\end{figure}
\begin{figure}[tb]
  \centering
\includegraphics[width=80mm]{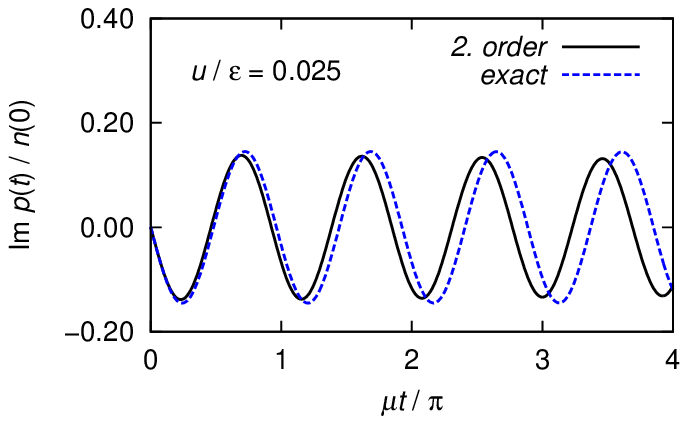}
\\
\includegraphics[width=80mm]{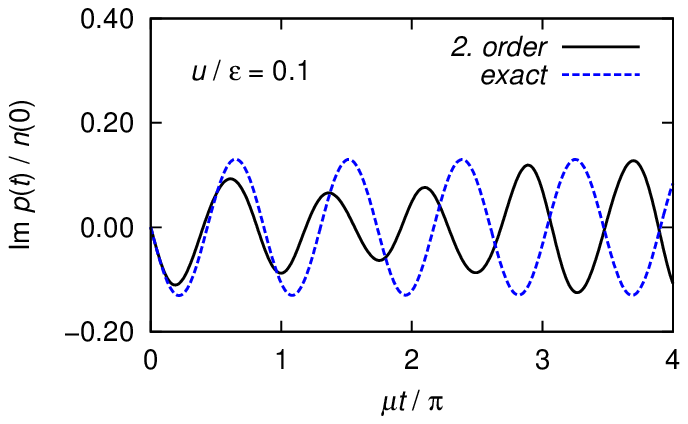}
\\
\includegraphics[width=80mm]{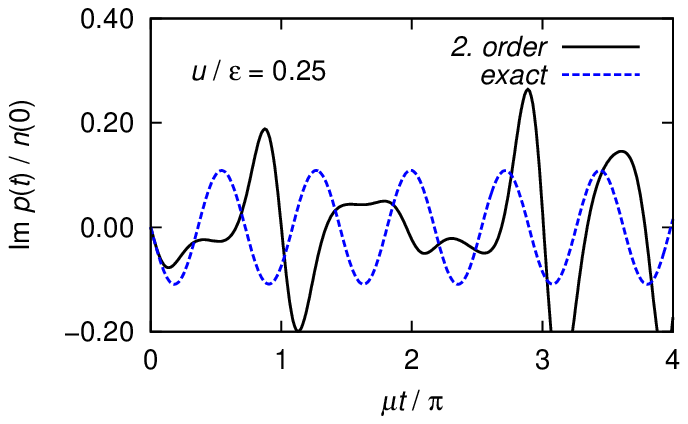}
  \vspace{-4mm}
  \caption{%
(Color online)
Time evolution of the off-diagonal
distribution function for increasing interaction strength (from top to bottom)  of the toy model.
The parameters and initial conditions are the same as in Fig.~\ref{fig:mean-field}.
The solid line is the solution of the kinetic equation with second order self-energies,
simplified using the generalized Kadanoff-Baym ansatz and the Markov approximation.
The dashed line is again the exact solution.
}  \label{fig:pert-anom}
\end{figure}
The result for the same parameters and initial conditions
as in Figs.~\ref{fig:mean-field} and \ref{fig:mean-field-anom}  is shown in 
Figs.~\ref{fig:pert} and \ref{fig:pert-anom}.
Obviously, for moderate interaction strength the inclusion of second-order
corrections indeed improves the agreement with the exact solution up to times of order $T_0 = \pi / \mu = \pi / \sqrt{
\epsilon^2 - | \gamma |^2 }$.
However, for stronger interactions and for times exceeding $T_0$, the solution of the kinetic equation with
second-order corrections to the self-energy disagrees even more
drastically from the exact solution than the time-dependent
Hartree-Fock approximation shown in Figs.~\ref{fig:mean-field} and \ref{fig:mean-field-anom}.
In addition we found secular behavior and unphysical divergences of the pair correlators for long times (not shown in  Figs.~\ref{fig:pert} and \ref{fig:pert-anom}).
By numerically solving the kinetic equation without making the above
approximations (see appendix D), we have checked that the strong disagreement of the
time evolution beyond one oscillation period $T_0$ with the exact
result is not an artifact of the Kadanoff-Baym ansatz, the Markov
approximation or the neglected renormalization of the retarded and
advanced propagators.

\subsection{First order truncation of the FRG hierarchy}


We now show that a very simple truncation of the non-equilibrium FRG
flow equation for the self-energy where the flow of the effective
interaction is neglected leads to much better results  for the time
evolution than the previous two approximations. A similar truncation
has also been made by Gezzi {\it{et al.}}\cite{Gezzi07} in their FRG
study of stationary non-equilibrium states of the Anderson
impurity model. In the exact FRG flow equation (\ref{eq:flowGamma2})
for the self-energy we replace the flowing effective
interaction by the bare interaction (recall that for our toy model the collective labels
$\alpha_i$ represents $( t_i, \lambda_i, \sigma_i )$),
 \begin{equation}
 \Gamma^{(4)   }_{ \Lambda , \alpha_1 \alpha_2 \alpha_3 \alpha_4}
 \approx
 \delta ( t_1 - t_2 ) \delta ( t_2 - t_3 ) \delta ( t_3 - t_4 )
 U^{ \lambda_1   \lambda_2  \lambda_3
\lambda_4 }_{\sigma_1 \sigma_2 \sigma_3 \sigma_4},
 \end{equation}
where up to permutation of the indices the symmetrized
bare interaction is given by (see Eq.~(\ref{eq:ubare}))
 \begin{eqnarray}
& & U_{\;  \bar{a} \, \bar{a} \,  a \, a}^{CQCC}
=
 U_{\;  \bar{a} \, \bar{a} \,  a \, a}^{CQQQ} =
 U_{\;  \bar{a} \, \bar{a} \,  a \, a}^{CCCQ}  =
U_{\;  \bar{a} \, \bar{a} \,  a \, a}^{QQCQ} =u   .
 \hspace{7mm}
 \label{eq:ubaretoy}
\end{eqnarray}
In this approximation the two-point function is  time-diagonal,
 \begin{equation}
 \Gamma^{(2) \lambda \lambda^{\prime}}_{\Lambda, \sigma t \, \sigma^{\prime} t^{\prime}}
= \delta ( t - t^{\prime} ) \Sigma^{\lambda \lambda^{\prime}}_{\Lambda, \sigma \sigma^{\prime}} ( t ),
 \end{equation}
where the self-energies satisfy the FRG flow equation
 \begin{equation}
 \partial_{\Lambda} \Sigma^{\lambda \lambda^{\prime}}_{\Lambda, \sigma \sigma^{\prime}} ( t )
 = \frac{i}{2} \sum_{ \lambda_1 \sigma_1} \sum_{ \lambda_2 \sigma_2}
U^{ \lambda  \lambda^{\prime}
\lambda_1  \lambda_2 }_{\sigma \sigma^{\prime} \sigma_1 \sigma_2}
 \dot{{G}}_{\Lambda,\sigma_1 \sigma_2}^{ \lambda_1 \lambda_2} ( t , t ).
 \end{equation}
With the bare interaction given by
 Eq.~(\ref{eq:ubaretoy}), this leads to the FRG flow equations
for the retarded
($\lambda \lambda^{\prime} = QC$)
and advanced
($\lambda \lambda^{\prime} = CQ$)
self-energies,
 \begin{align}
 \partial_{\Lambda}  \Sigma^R_{ \Lambda} ( t ) &= \partial_{\Lambda}  \Sigma^A_{ \Lambda} ( t ) \equiv
 \partial_{\Lambda}  \Sigma_{ \Lambda} ( t ) \nonumber \\
 &=
 i u \left( \begin{array}{cc}
\frac{1}{2} \dot{G}^K_{\Lambda, \bar{a} \bar{a}} ( t,t ) &   \dot{G}^K_{\Lambda, a \bar{a}} ( t,t ) \\
 \dot{G}^K_{\Lambda, \bar{a} a} ( t,t ) & \frac{1}{2} \dot{G}^K_{ \Lambda, {a} {a}} ( t,t ) \end{array} \right)  .
\end{align}
For the Keldysh component of the self-energy
corresponding to $\lambda \lambda^{\prime} = QQ$ we obtain
 \begin{subequations}
 \begin{align}
 \partial_{\Lambda} \Sigma^K_{\Lambda, a \bar{a}} ( t ) & = i u
 \left[ \dot{G}^R_{\Lambda, a \bar{a}} ( t,t ) + \dot{G}^A_{\Lambda, a \bar{a}} ( t,t ) \right],
 \label{eq:sigmakel1}
 \\
\partial_{\Lambda} \Sigma^K_{\Lambda, a {a}} ( t ) & = i \frac{u}{2}
 \left[ \dot{G}^R_{\Lambda, \bar{a} \bar{a}} ( t,t )
+ \dot{G}^A_{\Lambda, \bar{a} \bar{a}} ( t,t ) \right].
 \label{eq:sigmakel2}
 \end{align}
\end{subequations}
From the definitions~(\ref{eq:dotGR},\ref{eq:dotGA})
of the retarded and advanced components of the single-scale propagators
it is easy to see that
at equal times $\dot{G}^R_\Lambda ( t,t )=0 = \dot{G}^A_\Lambda ( t,t )$, so that within our truncation
the right-hand sides of the flow-equations (\ref{eq:sigmakel1},\ref{eq:sigmakel2})
for the Keldysh self-energy vanish. Because the initial Keldysh self-energy is zero,
it remains zero during the entire RG flow within our truncation.

At this point we specify our cutoff procedure. It turns out that from
the two cutoff schemes discussed in Sec.~\ref{sec:cutoff}, the
out-scattering rate scheme described in Sec.~\ref{sec:outscattering} is superior. Recall that this scheme amounts to setting
$\hat{F}_{\ast,\Lambda}=0$ in Eq.~(\ref{eq:dotGK}). For our toy model the
Keldysh component of the single-scale propagator at equal times is
then given by the following $2 \times 2$ matrix equation,
 \begin{eqnarray}
 \dot{G}_{\Lambda}^K ( t , t ) & = & i \int_{ t_0}^{t} d t_1
 \Bigl[ G_\Lambda^R ( t, t_1 ) Z G_{\Lambda}^K ( t_1 , t )
 \nonumber
 \\
 & & \hspace{10mm}
- G_\Lambda^K ( t, t_1 ) Z G_{\Lambda}^A ( t_1 , t )
 \Bigr].
 \label{eq:Gdottoy}
 \end{eqnarray}
Note that this equation still contains memory effects.
We simplify Eq.~(\ref{eq:Gdottoy}) using the same approximations
as in the previous subsection:
First of all, we use the generalized Kadanoff-Baym ansatz
(\ref{eq:KadanoffBaymtoy}) to express the Keldysh Green functions
on the right-hand side of Eq.~(\ref{eq:Gdottoy}) in terms of
the corresponding equal-time Green function.
Introducing the cutoff-dependent distribution function
 \begin{equation}
 F_{\Lambda} ( t ) = i Z G_{\Lambda}^K ( t , t ) Z,
 \end{equation}
we obtain
\begin{equation}
 \dot{G}_{\Lambda}^K ( t , t ) \approx 2 i \int_{ t_0}^{t} d t_1
 G_\Lambda^R ( t, t_1 ) F_{\Lambda} ( t_1 ) G_{\Lambda}^A ( t_1 , t ).
 \label{eq:Gdottoy2}
 \end{equation}
Next, we replace $F_{\Lambda} ( t_1 ) \rightarrow F_{\Lambda} ( t )$
under the integral sign (Markov approximation).
Substituting the advanced and retarded propagators by their free counterparts (which can be obtained from Eq.~(\ref{eq:GRres},\ref{eq:GAres}) by omitting the momentum labels) we finally
arrive at the following simple expression for the Keldysh component of the
single-scale propagator in the out-scattering rate cutoff scheme,
\begin{equation}
 \dot{G}_{\Lambda}^K ( t , t )  \approx  2 i
\int_{ t_0}^{t} d t_1
 G_{0,\Lambda}^R ( t, t_1 ) F_{\Lambda} ( t ) G_{0,\Lambda}^A ( t_1 , t ).
 \label{eq:Gdottoy3}
 \end{equation}
At this point we have arrived at a system of two coupled partial differential equations (PDE)
for the cutoff-dependent distribution matrix $F_{\Lambda} ( t )$ and the self-energy matrix $\Sigma_\Lambda(t)$.
The former contains the normal
($n_{\Lambda} ( t )$) and anomalous ($p_{\Lambda} ( t )$) distributions
as in Eq.~(\ref{eq:Fttoy}),
\begin{equation}
 F_{\Lambda}  (  t )
=   \left( \begin{array}{cc} - 2 p_{\Lambda}^{\ast} ( t ) & 2 n_{\Lambda} ( t ) +1 \\
 2 n_{\Lambda} ( t ) +1 &  - 2 p_{\Lambda} ( t ) \end{array} \right).
 \label{eq:FLambdattoy}
 \end{equation}
The time evolution of this distribution matrix
is determined by a kinetic equation which is formally identical
to the corresponding kinetic equation (\ref{eq:kinS})
within time-dependent Hartree-Fock approximation,
\begin{equation}
  i \partial_t  F_{\Lambda} ( t ) =  - M_{\Lambda}^T(t) F_{\Lambda} ( t ) - F_{\Lambda} ( t ) M_{\Lambda} ( t ) .
 \label{eq:kinSLambda}
 \end{equation}
The cutoff-dependent matrix $M_{\Lambda} ( t )$ is
 \begin{align}
  &M_{\Lambda} ( t )  =
 M - i \Lambda I
 + Z \Sigma_{\Lambda} ( t )
 \nonumber
 \\
& =
\left( \begin{array}{cc}
\epsilon - u - i \Lambda + \Sigma_{ \Lambda, \bar{a} a} ( t ) &
| \gamma  | + \Sigma_{ \Lambda, \bar{a} \bar{a} }   (t)  \\
 - |\gamma | -   \Sigma_{ \Lambda, {a} {a} }   (t)    &
- [\epsilon - u + i \Lambda +     \Sigma_{ \Lambda, a \bar{a} } ( t ) ]  \end{array} \right),
 \end{align}
and depends on the bare matrix $M$ defined in Eq.~(\ref{eq:tildeMdef}), on the cutoff parameter $\Lambda$, and on the
cutoff-dependent self-energy $\Sigma_{\Lambda}(t)$.
The flowing self-energy matrix satisfies
 \begin{equation}\label{eq:flowToy}
 \partial_{\Lambda} \Sigma_{\Lambda} ( t )
 =  i u \left( \begin{array}{cc}
\frac{1}{2}\dot{G}_{\Lambda, \bar{a} \bar{a}}^K ( t ,t) &
\dot{G}_{\Lambda, a \bar{a}}^K ( t ,t) \\
\dot{G}_{\Lambda,  \bar{a}a }^K ( t,t )& \frac{1}{2} \dot{G}_{\Lambda,  {a} {a} }^K ( t ,t)
  \end{array} \right),
 \end{equation}
where the matrix $\dot{G}_{\Lambda}^K ( t,t )$ is given by
 Eq.~(\ref{eq:Gdottoy3}).

\begin{figure}[tb]
  \centering
\includegraphics[width=60mm]{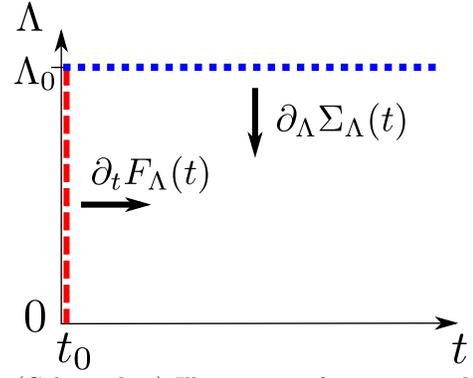}
  \vspace{-4mm}
  \caption{%
(Color online) Illustration of our approach to solve the system (\ref{eq:genDiff1}) and (\ref{eq:genDiff2}) of partial differential equations. The kinetic equation
(\ref{eq:genDiff1})
 describes the propagation of $F_\Lambda(t)$ in $t$, where the flow equation (\ref{eq:genDiff2}) describes the propagation of
$\Sigma_\Lambda(t)$ in $\Lambda$-direction. The boundary conditions define
the distribution function $F_\Lambda(t_0)$ at the initial time $t_0$ (dashed line)
and the self-energy $\Sigma_{\Lambda_0}(t)$ at the initial
$\Lambda_0$ (dotted line).
}
  \label{fig:bound}
\end{figure}

Mathematically, the problem is now reduced to the solution of a system of first order partial differential equations in two independent variables $t$ and $\Lambda$.
To illustrate the structure more clearly, we rewrite the system (\ref{eq:kinSLambda}) and
(\ref{eq:flowToy}) as
 \begin{subequations}
 \begin{align}
 \partial_{t} F_{\Lambda} ( t ) & =  A ( F_{\Lambda} ( t ),\Sigma_{\Lambda} ( t ),\Lambda),
\label{eq:genDiff1}
 \\
 \partial_{\Lambda} \Sigma_{\Lambda} ( t ) & =  B ( F_{\Lambda} ( t
),\Lambda,t),
\label{eq:genDiff2}
 \end{align}
\end{subequations}
where the explicit form of the matrix functions $A$ and $B$ follows from the right-hand sides of Eq.~(\ref{eq:kinSLambda}) and
(\ref{eq:flowToy}).
Note, that the system
is not fully symmetric in the variables $\Lambda$ and $t$, because
the flow equation contains causal memory integrals over the time
$t$. Without Markov approximation, the right-hand side of the flow
equation (and in higher-order truncations also the kinetic equation)
is a functional of the distribution matrix $F_\Lambda(t)$ and depends
on the distribution matrix at earlier times. Using the
Markov approximation the distribution matrix can be pulled out
of the integral and the functional $B$ reduces to an ordinary function of the
distribution $F_\Lambda(t)$.
To define the solution of Eqs.~(\ref{eq:genDiff1},\ref{eq:genDiff2}) uniquely we note that the boundary conditions fix the distribution matrix $F_\Lambda(t_0)$ at the initial time $t_0$ and arbitrary cutoff $\Lambda$, and the self-energy matrix $\Sigma_{\Lambda_0}(t)$ at the initial cutoff $\Lambda_0$ and arbitrary time $t$.
In fact, within our truncation, the boundary condition for the distribution matrix is $F_\Lambda(t_0) = F(t_0)$.
Since for a large cutoff all one-particle irreducible vertices (with the exception of $\Gamma^{(4)}$) vanishes due to Eq.~(\ref{eq:gamma0}), the boundary condition for the self-energy matrix at sufficiently large initial cutoff $\Lambda_0$ is $\Sigma_{\Lambda_0}(t) = 0$.
The standard method of dealing with this kind of first order PDEs is
the method of characteristics. \cite{*[{See, for example, }][{}] Zachmanoglou}
However, in our case the characteristic curves coincide with the curves where the
boundary conditions are specified, so that the standard procedure is not applicable.
 Nevertheless, it is easy to see that the solution with the proper boundary conditions can be obtained by means of the following algorithm:
We first note that the kinetic equation (\ref{eq:genDiff1})
 describes the propagation of $F_{\Lambda}(t)$ in $t$, and that the flow equation (\ref{eq:genDiff2}) gives the propagation of
$\Sigma_{\Lambda}(t)$ in $\Lambda$ direction, as illustrated in Fig.~\ref{fig:bound}.
Solving the kinetic equation (\ref{eq:genDiff1}) for an infinitesimally small time step $dt$, the resulting
distribution function at $t+dt$ can be used to integrate the flow equation (\ref{eq:genDiff2}) at fixed $t+dt$ over $\Lambda$.
Repeating these two steps allows to obtain the solution of $F_{\Lambda}(t)$ and $\Sigma_{\Lambda}(t)$ in the entire
$(t,\Lambda)$-plane.

In the following we explain our approach to numerically solve the coupled set of first order partial differential equations.
We focus on the out-scattering cutoff scheme but generalizations to other cases are straightforward. We consider a discretization of the two variables $t$ and $\Lambda$
in the form
\begin{subequations}
\begin{align}
  t \rightarrow t_m &\in \{t_0, \ldots ,t_{M-1} \}  ,   \\
  \Lambda \rightarrow \Lambda_n &\in \{\Lambda_0, \ldots ,\Lambda_{N-1} \},
\end{align}
\end{subequations}
with $m \in \{0,\ldots,M-1\}$ and $n \in \{0,\ldots,N-1\}$. The discretized grid points are ordered as $t_n < t_{n+1}$ and $\Lambda_n > \Lambda_{n+1}$.
Both grids do not need to be equally spaced. Moreover, the number of points $M$ and $N$  can be chosen arbitrarily.
The discretized functions are written as
\begin{equation}
  F_{\Lambda_n}(t_m) =  F_{n \, m} \quad , \quad \Sigma_{\Lambda_n}(t_m) = \Sigma_{n \, m} ,
\end{equation}
and
\begin{equation}
  \dot{G}^K_{\Lambda_n}(t_m,t_m) =  \dot{G}^K_{n\,m} .
\end{equation}
The derivatives are  approximated by first-order finite-difference expressions,
\begin{subequations}
\begin{align}
   \partial_t F_{\Lambda_n}(t_m)   &\approx    \frac{F_{n\,m+1} - F_{n\,m}}{t_{m+1} - t_m} ,  \\
   \partial_\Lambda \Sigma_{\Lambda_n}(t_m)  &\approx    \frac{\Sigma_{n+1\,m} - \Sigma_{n\,m}}{\Lambda_{n+1} - \Lambda_n}.
\end{align}
\end{subequations}
The discretized version of the kinetic equation (\ref{eq:kinS}) then follows as
\begin{equation}
  F_{n\,m+1} = F_{n\,m} + i (t_{m+1} - t_m) ( M_{n\,m}^T F_{n\,m}  +  F_{n\,m} M_{n\,m} ) ,
\label{eq:kinNum}
\end{equation}
with the time-dependent coefficient matrix
\begin{equation}
  M_{n\,m} = M - i \Lambda_n I  + Z \Sigma_{n\,m} .
\end{equation}
In the same way the discretized flow equation (\ref{eq:flowToy}) for the self-energy is
\begin{align}
\label{eq:flowNum}
&  \Sigma_{n+1\,m} = \Sigma_{n\,m} \nonumber \\
& \quad +
  i u (\Lambda_{n+1} - \Lambda_n)
\left( \begin{array}{cc}
\frac{1}{2}\dot{G}_{n\, m, \bar{a} \bar{a}}^K  &
\dot{G}_{n\, m, a \bar{a}}^K  \\
\dot{G}_{n\, m,  \bar{a}a }^K & \frac{1}{2} \dot{G}_{n\, m,  {a} {a} }^K
  \end{array} \right) .
\end{align}
According to Eq.\ (\ref{eq:Gdottoy3}) the single-scale propagator $\dot{G}_{n\,m}^K$ on the right-hand side is defined as
\begin{equation}
\label{eq:singlescalenum}
 \dot{G}^K_{n\,m}
 =   2 i \int_{t_0}^{t_m} dt' G^R_{0,\Lambda_n}( t_m, t') F_{n\,m} G^A_{0,\Lambda_n}(t',t_m) .
 \end{equation}
From the structure of the discretized equations it is obvious that causality is preserved since each time step can be calculated from  the previous ones and does not depend on quantities at later times.
Starting from the initial values which specify the distribution matrix $F_{n\,0}$ with $n \in \{0,\ldots,N-1\}$ and the self-energy matrix $ \Sigma_{0\,m}$ with $m \in \{0,\ldots,M-1\}$ on the boundaries, one can obtain the entire solution by stepwise propagating in
$t$- and in $\Lambda$-direction in terms of basic Euler steps.
One Euler step from $t_{m}$ to $t_{m+1}$ contains two parts:
First, with the solution $F_{n\,m}$ where $n \in \{0,\ldots,N-1\}$ from the previous step and the initial self-energy $ \Sigma_{0\,m}$ on the boundary, the flow equation (\ref{eq:flowNum}) at fixed time $t_{m}$ can be integrated
in $N$ sub-steps from $\Lambda_0$ to $\Lambda_{N-1}$ to obtain $\Sigma_{n\,m}$ on all points $n \in \{0,\ldots,N-1\}$.
Next, by using the kinetic equation (\ref{eq:kinNum}), $F_{n\,m+1}$ can be derived from $F_{n\,m}$ and $\Sigma_{n\,m}$.
This completes one basic Euler step since the distribution function $F_{n\,m+1}$ is now known at time $t_{m+1}$. Repeating this procedure $M$-times yields the full solution up to time $t_{M-1}$.
Numerically, the first-order finite difference derivatives are not
accurate enough unless the grid spacing becomes very small which is
not feasible in practice. Therefore a fourth-order Runge-Kutta
method for the propagation in $t$- and the second-order Heun method
for the propagation in $\Lambda$-direction \cite{Hankebook} is used.
One Runge-Kutta step from $F_{n\,m}$ to $F_{n\,m+1}$ consists
of four Euler steps of the form described above. The integral
(\ref{eq:singlescalenum}) was solved analytically using the free
retarded and advanced Green functions (given by (\ref{eq:GRres},
\ref{eq:GAres}) without $\bd{k}$ dependence). The time grid was
chosen similar to the Hartree-Fock and the second-order case described in Sec.~\ref{sec:pert}.
The $\Lambda$-grid ranges between $\Lambda_0 / \epsilon =
8.1$ and $\Lambda_{299} / \epsilon = 2.1 \times 10^{-7}$ and was
adjusted in such a way that for $\Lambda / \epsilon < 1$,  the resolution of the grid spacing was increased to take
into account the higher curvature of the self-energy in this region.

\begin{figure}[tb]
  \centering
\includegraphics[width=80mm]{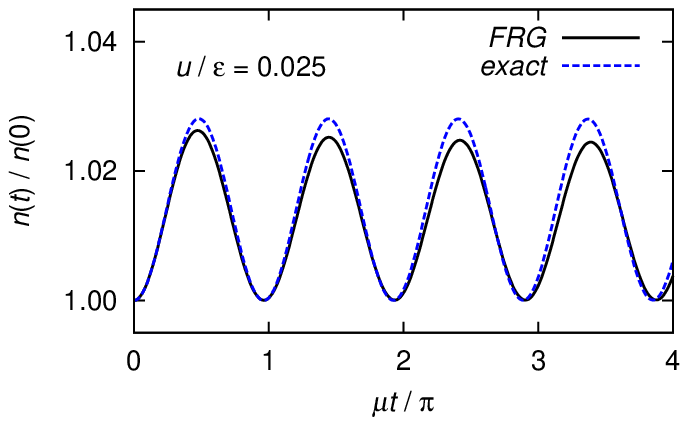}
\\
\includegraphics[width=80mm]{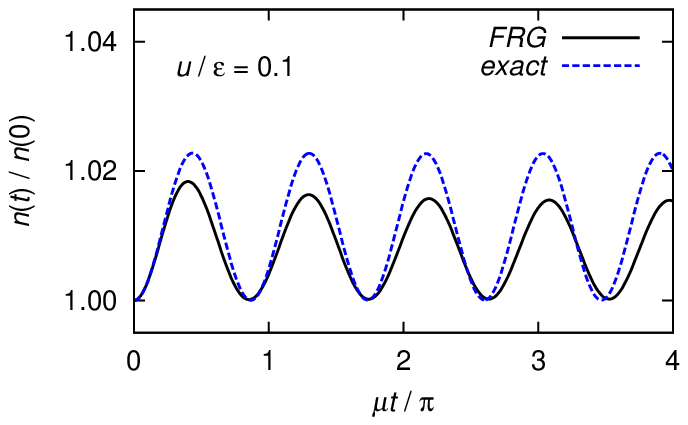}
\\
\includegraphics[width=80mm]{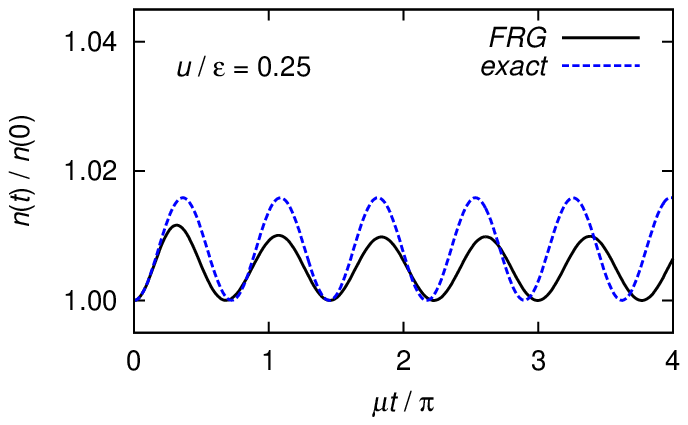}
  \vspace{-4mm}
  \caption{%
(Color online)
The solid lines are our
FRG results
for the diagonal
distribution function for increasing interaction strength (from top to bottom).
The parameters and initial conditions are the same as in Figs.~\ref{fig:mean-field} and \ref{fig:pert}.
For comparison we also show the exact solution (dashed line).
}
  \label{fig:FRG}
\end{figure}

\begin{figure}[tb]
  \centering
\includegraphics[width=80mm]{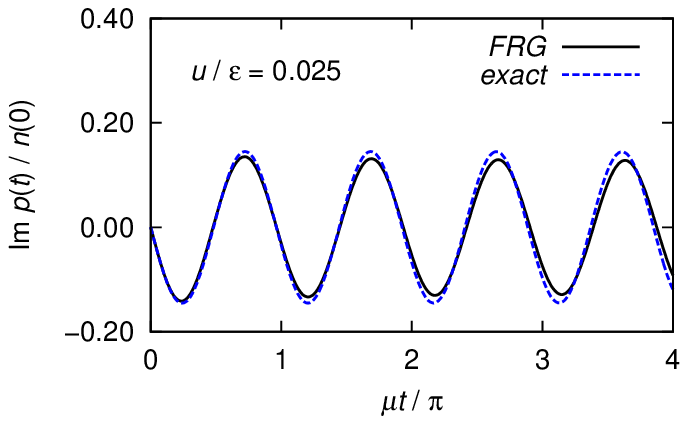}
\\
\includegraphics[width=80mm]{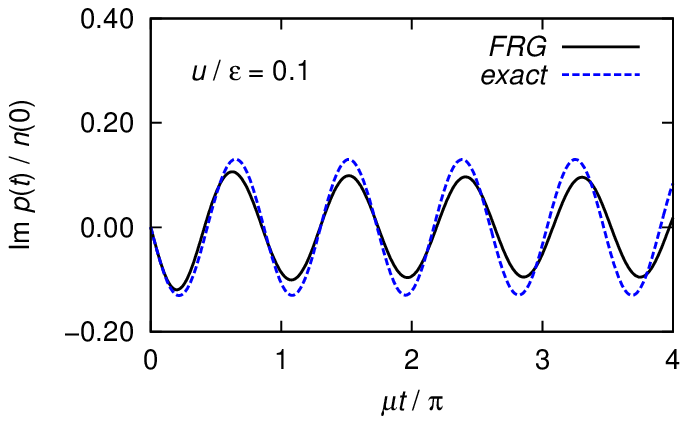}
\\
\includegraphics[width=80mm]{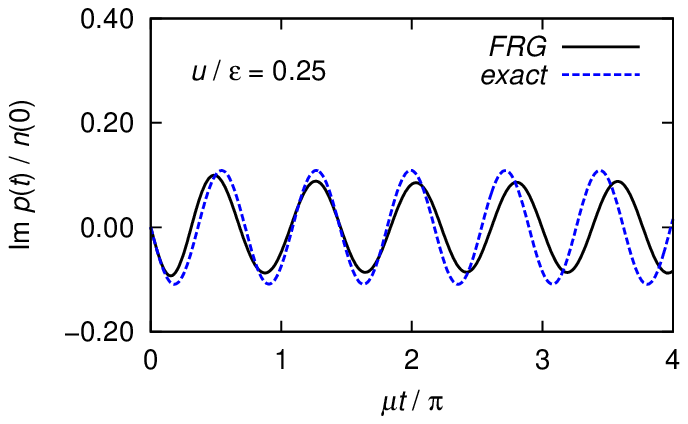}
  \vspace{-4mm}
  \caption{%
(Color online)
The solid lines are our
FRG results for the off-diagonal
distribution function for increasing interaction strength (from top to bottom).
The parameters and initial conditions are the same as before.
The dashed line is again the exact solution.
}
  \label{fig:FRG-anom}
\end{figure}

The result for the FRG approach with the out-scattering cutoff scheme for the same parameters and initial conditions as in
perturbation theory (compare Figs.\ \ref{fig:mean-field}, \ref{fig:mean-field-anom} and \ref{fig:pert}, \ref{fig:pert-anom})
is shown in Figs.\ \ref{fig:FRG} and \ref{fig:FRG-anom}.
For the oscillation period of  the pair correlators, the FRG treatment clearly improves the results compared to perturbative approaches.
Up to time $T_0 = \pi / \mu = \pi / \sqrt{ \epsilon^2 - | \gamma |^2}$ the period of the oscillation is nearly identical with the exact result. Even after longer times of the order $4 T_0$ (middle panel) the deviation from the exact solution is small.
The oscillation is regular and we found no secular behavior even at long times.
However, the amplitude of the pair correlators is underestimated and is comparable to the perturbative mean-field result shown in Figs.\ \ref{fig:mean-field} and \ref{fig:mean-field-anom}.
In contrast, with the hybridization cutoff scheme, we were not able to obtain any reasonable results for the pair-correlator dynamics.
This suggests that in practice the out-scattering cutoff scheme works better than the hybridization cutoff scheme.


\section{Summary and conclusions}
\label{sec:summary}
We have developed a real-time
functional renormalization group (FRG) approach
to calculate the
time evolution of interacting bosons out of equilibrium.
To be specific, we have developed our formalism
in the context  of the interacting time-dependent
boson Hamiltonian (\ref{eq:Hpr})
describing the non-equilibrium dynamics  of
magnons in dipolar magnets such as yttrium-iron garnet\cite{Cherepanov93}
subject to an oscillating microwave field.\cite{Zakharov70,Lvov94}
To take into account the off-diagonal correlations inherent in this model
we have introduced an efficient matrix notation  which facilitates
the derivation of quantum kinetic equations for both the normal and anomalous
components of the Green functions in the Keldysh formalism.
We have also extended the generalized Kadanoff-Baym ansatz \cite{Lipavsky86}
to include both diagonal and off-diagonal correlations on the same footing.

In our FRG approach the time evolution
of the diagonal and off-diagonal distribution functions is obtained
by solving a quantum kinetic equation
with cutoff-dependent collision integrals simultaneously with
a renormalization group flow equation for cutoff-dependent
non-equilibrium self-energies appearing in the
collision integrals.
To implement this procedure, we proposed a new cutoff scheme where the infinitesimal imaginary part
defining the boundary conditions
of the inverse advanced and retarded
propagators is replaced
by a finite scale acting as a running cutoff.
We have called this cutoff procedure
out-scattering rate cutoff scheme because the cutoff-dependent
imaginary parts in the retarded and advanced propagators
lead to an exponential decay of the occupation numbers.
In principle one can also replace the
infinitesimal imaginary part appearing
in the Keldysh component of the inverse free propagator
by a cutoff-dependent finite quantity which leads to the hybridization cutoff scheme proposed by Jakobs \emph{et al.}. \cite{Jakobs10a} For the toy model we  presented evidence that
it is better to work with the out-scattering cutoff scheme, keeping the
Keldysh component of the inverse free propagator
infinitesimal.

We have explicitly tested our FRG approach for a
simplified toy model which is obtained from the Hamiltonian (\ref{eq:Hpr})
by retaining only a single momentum mode.
Although this simplified model
does not contain damping and dissipative effects,
it does describe some aspects of the magnon dynamics
in yttrium-iron garnet.\cite{Kloss10}
Since the non-equilibrium time evolution
of our toy model can be obtained
exactly by direct numerical integration of the time-dependent
Schr\"{o}dinger equation, our toy model allows us to
test the quality of various approximations.
Specifically, we have studied the following approximations:

\begin{enumerate}
\item Self-consistent Hartree-Fock approximation, which is also called S-theory
in the context of non-equilibrium dynamics of magnons.\cite{Zakharov70,Lvov94}

\item A perturbative approach based on the calculation of the
non-equilibrium self-energies to second order in the interaction,
in combination with
the generalized Kadanoff-Baym ansatz and the Markov approximation.

\item A FRG approach based on the simultaneous solution
of a coupled system of kinetic equations
and renormalization group flow equations for the
scale-dependent self-energies, using a simple truncation
of the FRG flow equations for the non-equilibrium self-energies
where the flow of the interaction is neglected.

\end{enumerate}

For each approach we have calculated the time dependence of the
normal and anomalous distribution function for some representative
values of the interaction and compared the result with the exact solution.
It turns out that the first two approaches do not give
reliable predictions for the time evolution beyond one
oscillation period. Although inclusion of second order
self-energy corrections somewhat improves the agreement
for short times, the time dependence beyond a single
oscillation period disagrees even more strongly with the exact
solution than the prediction of
the self-consistent Hartree-Fock approximation.
In fact, in Appendix~D we shall show that
the numerical solution of the two-time quantum kinetic equations 
with second order self-energies does not lead to a better agreement
with the exact solution of the toy model than calculations which in addition rely
on the Kadanoff-Baym ansatz and the Markov-approximation.
The perturbative approaches are therefore not able to reproduce
the real-time dynamics of our toy model and do not allow for systematic improvements.
The failure of perturbation theory to predict the long-time behavior of correlation
functions is not unexpected.\cite{Berges04}
In contrast, our simple truncation of the
FRG flow equations
in combination with the out-scattering cutoff scheme leads to
quite good agreement with the exact solution over many oscillation
periods.
Note, however, that
our  FRG approach is numerically more costly than the
other two methods, because one has to solve a coupled system of
partial differential equations in two independent variables,
the time $t$ and the cutoff-parameter $\Lambda$.
Moreover, due to our simple truncation of the FRG flow equations
the oscillation amplitudes are still underestimated.
Nevertheless, from all approximation strategies we have tested
our first order FRG approach with out-scattering cutoff scheme
clearly gives the most satisfactory results for the time-evolution of our toy model.
One should keep in mind, however, that the
toy model does not have any intrinsic dissipation and therefore
does not relax towards a stationary state at long times.
In fact, we expect that for models with intrinsic dissipation
 other cutoff schemes such as the hybridization cutoff\cite{Jakobs10a,Jakobs10}
discussed in Sec.~\ref{sec:cutoff} is superior, because
the hybridization cutoff retains the balance between in-scattering and
out-scattering terms in the collision integral which is crucial
to describe the relaxation towards a stationary state. 
The hybridization cutoff scheme also preserves the fluctuation-dissipation theorem 
during the entire flow\cite{Jakobs10} which is advantageous close to thermal equilibrium.

Our work can be extended in several directions:
First of all, it should be interesting
to use our FRG approach to calculate the time evolution
of  infinite or open quantum systems which
exhibit relaxation and dissipative processes.
We expect that 
for such systems
standard approximations such as the Kadanoff-Baym ansatz or the
Markov approximation may have different regimes of validity
than for our toy model. Recall 
that for weakly
correlated systems such as semiconductors the Kadanoff Baym ansatz
has been shown to be quite useful and accurate.\cite{Haug08}
It should also be interesting to extend our study of the toy model
to the regime of strong pumping where the original vacuum state is unstable.\cite{Kloss10}
Moreover, it would be even more interesting to apply our
non-perturbative FRG method
to study the non-equilibrium dynamics of
the time-dependent boson Hamiltonian
(\ref{eq:Hpr}) in the regime of strong pumping.
It is well known \cite{Zakharov70,Lvov94} that for sufficiently large
values of the pumping parameter $\gamma_{\bd{k}}$ the system exhibits
the phenomenon of parametric resonance. The magnon operators
acquire than finite expectation values and the system approaches
a non-trivial time-independent non-equilibrium state which is
dominated by interactions.\cite{Zakharov70,Lvov94}$^{,}$\cite{Hick10}
Although this state has been studied at the level of time-dependent Hartree-Fock approximation\cite{Zakharov70,Lvov94} (S-theory)
it would be interesting to describe the time evolution into stationary non-equilibrium states
non-perturbatively,
and check if the states exhibit non-thermal scaling properties as predicted in Ref.~[\onlinecite{Berges09}].

\section*{ACKNOWLEDGMENTS}
The authors thank J.~Berges and J.~Hick for useful discussions. Financial support by
SFB/TRR49, FOR 723, the CNRS, and the Humboldt foundation is gratefully
acknowledged.

\begin{appendix}

\renewcommand{\theequation}{A\arabic{equation}}
\section*{APPENDIX A: TRANSFORMATION TO THE ROTATING REFERENCE FRAME}
\setcounter{equation}{0}

In order to simplify the calculations, it is useful to remove the explicit time-dependence from the
Hamiltonian $\mathcal{H} ( t )$ in Eq.~(\ref{eq:Hpr}). This can be achieved by means
of a unitary transformation to the rotating reference frame, as discussed
in Sec.~\ref{sec:intro}.
In the rotating reference frame the Hamiltonian
does not explicitly depend on time, see Eqs.~(\ref{eq:tildeHdef}).
To distinguish quantities in the original- and the corresponding rotating frame, we put in this appendix an extra tilde over  Green functions in the rotating frame.
Introducing the  unitary $2 \times 2$  matrix
 \begin{equation}
 U_{\bd{k}}( t ) = \left( \begin{array}{cc}  e^{  -\frac{i}{2} ( \omega_0 t - \varphi_{\bd{k}} ) } & 0 \\
 0 & e^{  \frac{i}{2} ( \omega_0 t - \varphi_{\bd{k}} ) } \end{array} \right),
 \end{equation}
the relations between the elements
of the matrix Green-functions $\hat{G}^R$,
  $\hat{G}^A$,   $\hat{G}^K$ defined in
Eqs.~(\ref{eq:GRmat}--\ref{eq:GKmat}) and the corresponding quantities
in the rotating reference frame (with the tilde)
is
 \begin{equation}
 G^X (\bd{k},  t , t^{\prime}  )   = U_{\bd{k}} ( t )   \tilde{{G}}^X (\bd{k}, t, t^{\prime}  )
 U_{\bd{k}} ( {t^{\prime}} ) ,
 \label{eq:Grotate}
 \end{equation}
where $X = R,A,K$, and we have defined the $2\times2$-matrices in
flavor space,
 \begin{equation}
G^X (\bd{k},  t , t^{\prime}  ) = [ \hat{G}^X ]_{\bd{k} t , -\bd{k} t^{\prime} }.
 \end{equation}
Introducing the diagonal matrix
 \begin{equation}
\hat{U} = U ( t ) \otimes \hat{1},
 \end{equation}
with matrix elements
 $
[ \hat{U} ]_{\bd{k} t ,\bd{k'} t^{\prime}} =  \delta_{\bd{k},\bd{k'}} \delta ( t - t^{\prime} )  U_{\bd{k}} ( t )
$,
we can rewrite Eq.~(\ref{eq:Grotate}) in the more compact form
 \begin{equation}
\hat{G}^X = \hat{U} \hat{\tilde{G}}^X \hat{U}.
 \end{equation}
The
distribution function matrix $\hat{F}$ defined via Eq.\ (\ref{eq:kelF}) is
related to its counterpart    $\hat{\tilde{F}}$
in the rotating reference frame via
  \begin{equation}
\hat{F} = \hat{U}^{\dagger} \hat{\tilde{F}} \hat{U}^{\dagger}.
 \label{eq:Frotate}
 \end{equation}
Taking matrix elements in  the time-labels and using the fact that
in the non-interacting limit the distribution function matrix is time-diagonal (see Eqs.~(\ref{eq:Fdiag})),
the relation (\ref{eq:Frotate}) implies the
$2\times 2$ matrix equation,
 \begin{equation}
 F_0 (\bd{k}, t ) =  U^{\dagger}_{\bd{k}} ( t ) \tilde{F}_0 (\bd{k}, t ) U^{\dagger}_{\bd{k}} ( t ) ,
\label{eq:F0rot}
\end{equation}
where we have used the fact that
 \begin{equation}
U^{\dagger}_{\bd{k}} ( t )   = Z^{T} U_{\bd{k}} ( t )  Z = Z  U_{\bd{k}} ( t )  Z^{T}.
 \end{equation}

\renewcommand{\theequation}{B\arabic{equation}}
\section*{APPENDIX B: GENERALIZED KADANOFF-BAYM ANSATZ}
\setcounter{equation}{0}
\label{sec:gkba}

The generalized Kadanoff-Baym ansatz (GKBA) is an approximate relation between matrix elements of the
Keldysh Green function at different times and its equal-time counterparts.
To derive the matrix form of the GKBA given in (\ref{eq:KadanoffBaym}) formally and to
identify the terms which are neglected if one uses this ansatz,
we follow the derivation by  Lipavsk\'y {\it{et al.}}. \cite{Lipavsky86}
Without any loss of generality we will neglect the momentum labels for a moment and concentrate
on the time dependence only.
In addition we use the short-hand notation $G_{tt'} \equiv [\hat G]_{tt'}$  for the matrix elements in the time labels.
We introduce
 \begin{subequations}
 \begin{eqnarray}
 G^{KR}_{tt'} = G^{KR} ( t , t^{\prime} ) & = & \Theta ( t - t^{\prime} ) G^K ( t , t^{\prime} ),
 \label{eq:GKR}
 \\
 G^{KA}_{tt'} = G^{KA} ( t , t^{\prime} ) & = & \Theta (  t^{\prime} - t ) G^K ( t , t^{\prime} ),
 \end{eqnarray}
 \end{subequations}
so that by definition
 \begin{equation}
 G^K_{tt'} =  G^{KR}_{tt'} + G^{KA}_{tt'}.
 \end{equation}
Note that the above functions are $2 \times 2$ matrices in flavor space.
Acting with $ (\hat{G}^R )^{-1}$ from the left on Eq.~(\ref{eq:GKR})
and using the left Dyson equation (\ref{eq:DysonR}) for the retarded Green function
we obtain
 \begin{align}
 & [ (\hat{G}^R)^{-1} \hat{G}^{KR} ]_{ t  t^{\prime} }  =  - i \delta ( t - t^{\prime} )
Z G^{K}_{ t  t^{\prime} }
 \nonumber
 \\
 & \qquad
+ \Theta ( t - t^{\prime} )  \left(
[ (\hat{G}_0^R)^{-1} \hat{G}^K]_{ t t^{\prime}}
 -
\int_{ t^{\prime}}^{t} d t_1
 \Sigma^R_{ t  t_1} G^K_{ t_1 t^{\prime} } \right).
 \label{eq:GKRmotion}
 \end{align}
A similar relation can be derived for the advanced component of the Keldysh
Green function,
 \begin{align}
 & [ \hat{G}^{KA}    (\hat{G}^A)^{-1}   ]_{ t  t^{\prime} }  =   i \delta ( t - t^{\prime} )
G^{K}_{ t  t^{\prime} } Z
 \nonumber
 \\
 & \qquad
+ \Theta ( t^{\prime} - t )  \left(
[ \hat{G}^K  (\hat{G}_0^A)^{-1}   ]_{ t t^{\prime}}
 -
\int_{ t}^{t^{\prime}} d t_1 G^K_{ t t_1  }
 \Sigma^A_{   t_1 t^{\prime}}  \right).
 \label{eq:GKAmotion}
 \end{align}
Using the Keldysh components of the left and right
Dyson equations given in Eqs.~(\ref{eq:GKdyson}) and (\ref{eq:GKdysonright}),
the action of the free inverse propagators on $\hat{G}^K$ can be written  as
\begin{equation}\label{eq:gr0gk}
   [ (\hat{G}_0^R)^{-1} \hat{G}^{K} ]_{ t  t^{\prime} }   =
   \int_{-\infty}^{t} \! \! dt_1  \Sigma^R_{t t_1}G^K_{t_1 t^{\prime}} + \int_{-\infty}^{t^{\prime}} \!\!
dt_1  \Sigma^K_{t t_1}G^A_{t_1 t^{\prime}},
\end{equation}
\begin{equation} \label{eq:gkga0}
    [ \hat{G}^K  (\hat{G}_0^A)^{-1}   ]_{ t t^{\prime}}
 =
   \int_{-\infty}^{t^{\prime}} \!\! d t_1  G^K_{t t_1}\Sigma^A_{ t_1 t^{\prime}}
+ \int_{-\infty}^{t} \!\! dt_1  G^R_{t t_1}\Sigma^K_{t_1 t^{\prime}}.
\end{equation}
Substituting these expressions into Eqs.~(\ref{eq:GKRmotion}) and (\ref{eq:GKAmotion})
and solving for $\hat{G}^{KR}$ and $\hat{G}^{KA}$ we obtain
  \begin{align}
 &  [ \hat{G}^{KR} ]_{ t  t^{\prime} }  =  - i G^{R}_{ t t^{\prime}}  Z
 G^{K}_{ t^{\prime}  t^{\prime} }
+ \Theta ( t - t^{\prime} )
 \nonumber
 \\
 & \quad \,
 \times \int_{ t^{\prime}}^{t} d t_1 \int_{ - \infty}^{t^{\prime}}
d t_2 \, G^R_{ t t_1} \left[
 \Sigma^R_{ t_1  t_2} G^K_{ t_2 t^{\prime} }
 + \Sigma^K_{ t_1  t_2} G^A_{  t_2 t^{\prime} }
\right],
 \label{eq:GKRresult}
 \end{align}
  \begin{align}
&  [ \hat{G}^{KA} ]_{ t  t^{\prime} }  =   i    G^{K}_{ t  t }    Z G^{A }_{ t t^{\prime}}  + \Theta (  t^{\prime} - t )
 \nonumber
 \\
 & \quad \,
\times \int^{ t^{\prime}}_{t} d t_1 \int_{ - \infty}^{t}
d t_2
  \left[
 G^K_{ t  t_2} \Sigma^A_{  t_2 t_1 }
 + G^R_{ t  t_2} \Sigma^K_{ t_2 t_1 }
\right] G^A_{  t_1 t^{\prime}}  .
 \label{eq:GKAresult}
 \end{align}
Adding Eqs.~(\ref{eq:GKRresult}) and (\ref{eq:GKAresult}) we obtain  the following exact integral equation for the Keldysh component of the Green function,
  \begin{align}
& [ \hat{G}^{K} ]_{ t  t^{\prime} }  =  - i \left[ G^{R}_{ t t^{\prime}}  Z
 G^{K}_{ t^{\prime}  t^{\prime} } -   G^{K}_{ t  t }    Z G^{A }_{ t t^{\prime}}   \right]
 \nonumber
 \\
 & \quad
+ \Theta ( t - t^{\prime} )  \int_{ t^{\prime}}^{t}\!  d t_1 \! \! \int_{ - \infty}^{t^{\prime}} \!\!\!
d t_2 \,
 G^R_{ t t_1} \! \left[
 \Sigma^R_{ t_1  t_2} G^K_{ t_2 t^{\prime} }
 + \Sigma^K_{ t_1  t_2} G^A_{  t_2 t^{\prime} }
\right]
 \nonumber
 \\
 & \quad
+ \Theta (  t^{\prime} - t )  \int^{ t^{\prime}}_{t}\!\! d t_1 \! \! \int_{ - \infty}^{t} \!\!\!
d t_2
  \left[
 G^K_{ t  t_2} \Sigma^A_{  t_2 t_1 }
 + G^R_{ t  t_2} \Sigma^K_{ t_2 t_1 }
\right] G^A_{  t_1 t^{\prime}}  .
 \label{eq:GKresult}
 \end{align}
To rewrite this equation in a more compact form we introduce the functions
\begin{subequations}
\begin{eqnarray}
  && \hspace{-9ex}  W^R_{tt'} = \Theta(t - t') \int_{-\infty}^{\infty} dt_2
 (\Sigma^R_{t t_2} G^{KA}_{t_2 t'} + \Sigma^K_{t t_2}G^A_{t_2 t'}),  \\
  && \hspace{-9ex}  W^A_{tt'} = \Theta(t' - t) \int_{-\infty}^{\infty} dt_2 (G^{KR}_{t t_2}\Sigma^A_{t_2 t'}
+ G^R_{t t_2}\Sigma^K_{t_2 t'}).
\end{eqnarray}
\end{subequations}
Then we may write
  \begin{eqnarray}
 [ \hat{G}^{KR} ]_{ t  t^{\prime} } & = & - i G^{R}_{ t t^{\prime}}  Z
 G^{K}_{ t^{\prime}  t^{\prime} } + \Theta ( t - t^{\prime} )
 \int_{ - \infty}^{\infty} d t_1 G^R_{ t t_1} W^R_{ t_1 t^{\prime}}
 \nonumber
  \\
 & = & - i G^{R}_{ t t^{\prime}}  Z
 G^{K}_{ t^{\prime}  t^{\prime} } +
 \int_{ - \infty}^{\infty} d t_1 G^R_{ t t_1} W^R_{ t_1 t^{\prime}}
 \nonumber
 \\
 & = & - \int_{ - \infty}^{\infty} d t_1 G^R_{ t t_1} Z F^R_{ t_1 t^{\prime}}
 \nonumber
 \\
 & = &
- [ \hat{G}^R \hat{Z} \hat{F}^R ]_{ t t^{\prime}},
 \end{eqnarray}
  \begin{eqnarray}
 [ \hat{G}^{KA} ]_{ t  t^{\prime} } & = &
 i    G^{K}_{ t  t }    Z G^{A }_{ t t^{\prime}}
 + \Theta (  t^{\prime} - t )
 \int_{ - \infty}^{\infty} d t_1 W^A_{ t t_1} G^A_{ t_1 t^{\prime}}
 \nonumber
  \\
 & = &
 i    G^{K}_{ t  t }    Z G^{A }_{ t t^{\prime}}
 +
 \int_{ - \infty}^{\infty} d t_1 W^A_{ t t_1} G^A_{ t_1 t^{\prime}}
 \nonumber
 \\
 & = &  \int_{ - \infty}^{\infty} d t_1 F^A_{ t t_1} Z {G}^A_{ t_1 t^{\prime}}
 \nonumber
 \\
 & = &
 [  \hat{F}^A \hat{Z} \hat{G}^A   ]_{ t t^{\prime}},
 \end{eqnarray}
and hence
 \begin{equation}
\hat{G}^{K} = - \hat{G}^R \hat{Z} \hat{F}^R + \hat{F}^A \hat{Z} \hat{G}^A.
 \label{eq:GKFgen}
 \end{equation}
 Here the retarded and advanced component of the distribution function matrix
is defined by
\begin{subequations}
\begin{align}
  \hat{F}^R &= \hat{F}^D + \hat{Z} \hat{W}^R , \\
  \hat{F}^A &= \hat{F}^D - \hat{W}^A \hat{Z},
\end{align}
\end{subequations}
with the time-diagonal part given by
\begin{equation}
  [\hat{F}^D]_{tt'} = i \delta ( t - t^{\prime} ) G^K(t,t) .
\end{equation}
One easily verifies that the blocks have the following symmetries,
\begin{subequations}
\begin{align}
 (\hat{G}^{KR})^T &= \hat{G}^{KA} ,\\
 (\hat{W}^R)^T   &= \hat{W}^A , \\
 (\hat{F}^R)^T   &= \hat{F}^A ,\\
 (\hat{F}^D)^T   &= \hat{F}^D .
\end{align}
\end{subequations}
The above relations are all exact.
Comparing Eq.~(\ref{eq:kelF})
with Eq.~(\ref{eq:GKFgen}), we conclude that the parametrization in Eq.~(\ref{eq:kelF}) is indeed correct, and that
 \begin{equation}
 \hat{F} = \hat{Z} \hat{F}^R \hat{Z}.
 \end{equation}
The GKBA amounts to retaining only the diagonal part
$\hat{F}^D$ of the distribution function. Then the matrix elements
of the general relation (\ref{eq:GKFgen}) reduce to
\begin{equation} \label{eq:GKBA}
 G^K (t ,t^{\prime} )  = - i [
G^R ( t , t^{\prime} )   Z  G^K ( t^{\prime} , t^{\prime} )
- G^K ( t , t )   Z   G^A ( t  , t^{\prime} ) ] .
\end{equation}
This is identical to the GKBA ansatz (\ref{eq:KadanoffBaymtoy})
which was used to study the toy model. Repeating the above calculation
in the same fashion including the full momentum dependence leads to
the relation (\ref{eq:KadanoffBaym}).

 \onecolumngrid

\renewcommand{\theequation}{C\arabic{equation}}
\section*{APPENDIX C: SECOND ORDER SELF-ENERGY OF THE TOY MODEL}
\setcounter{equation}{0}

In this appendix we explicitly give the matrix elements of the
non-equilibrium self-energies $\Sigma_{2} ( t , t^{\prime} )$
of our toy model introduced in Sec.~\ref{sec:toy} to second order in the interaction.
Ignoring Hartree type of diagrams which are implicitly taken into account
by imposing self-consistency in the first order calculation,
the non-equilibrium self-energy to second order in the interaction
is in the contour basis ($p,p' \in \{ +,- \}$) given by
 \begin{align}
 [ \hat{\Sigma}_2 ]^{ p p^{\prime}}_{ t t^{\prime}}  =  \left(
 \begin{array}{cc} \Sigma^{ p p^{\prime}}_{2, aa} ( t , t^{\prime} ) &
\Sigma^{ p p^{\prime}}_{2, a \bar{a}} ( t , t^{\prime} ) \\
 \Sigma^{ p p^{\prime}}_{2, \bar{ a} a} ( t , t^{\prime} ) &
\Sigma^{ p p^{\prime}}_{2, \bar{a} \bar{a}} ( t , t^{\prime} )
 \end{array}
 \right)
  =  - 2 u^2 p p^{\prime}
 \left(
 \begin{array}{c|cc}
 G^{ p p^\prime}_{ a a}  ( G^{ p p^\prime}_{ \bar{a} \bar{a}}  )^2
 + 2 G^{ p p^\prime}_{ \bar{a} \bar{a}}    G^{ p p^\prime}_{a \bar{a}}
         G^{ p p^\prime}_{ \bar{a} a}
           & &
 G^{ p p^\prime}_{ a \bar{a}}  ( G^{ p p^\prime}_{ \bar{a} {a}}  )^2
 + 2    G^{ p p^\prime}_{ \bar{a} {a}}   G^{ p p^\prime}_{ a {a}}
G^{ p p^\prime}_{  \bar{a} \bar{a}}
\\
\hline
 G^{ p p^\prime}_{ \bar{a} {a}}  ( G^{ p p^\prime}_{ a \bar{a} }  )^2
 + 2  G^{ p p^\prime}_{  {a} \bar{a}}
G^{ p p^\prime}_{ \bar{a} \bar{a}}    G^{ p p^\prime}_{  {a} {a}}
     & &
 G^{ p p^\prime}_{ \bar{a} \bar{a}}  ( G^{ p p^\prime}_{ {a} {a}}  )^2
 + 2   G^{ p p^\prime}_{ {a} {a}}   G^{ p p^\prime}_{ \bar{a} a}
         G^{ p p^\prime}_{ a \bar{a} }
 \end{array}
 \right), \nonumber \\
 \label{eq:selfmatcontour}
 \end{align}
 where the time labels of all Green functions are $(t, t^{\prime} )$.
 Using the relations (\ref{eq:Rpm}--\ref{eq:Kpm}) we obtain
for the normal part of the matrix elements of the self-energy in the Keldysh (RAK)-basis,
 \begin{subequations}
 \begin{align}
 \Sigma^{ R}_{2, a \bar{a}} ( t , t^{\prime} )  = &
 - \frac{u^2}{2} \Bigl\{  G^R_{ a \bar{a} } [ (G^R_{ \bar{a} a })^2 + (G^K_{ \bar{a} a })^2 ]
 +  2 G^K_{ a \bar{a} } G^R_{ \bar{a} a } G^K_{ \bar{a} a }
 + 2 G^R_{ \bar{a} a }  [ G^R_{aa} G^R_{ \bar{a} \bar{a} } + G^K_{ aa} G^K_{ \bar{a} \bar{a} } ]
+ 2 G^K_{ \bar{a} a }  [ G^R_{aa} G^K_{ \bar{a} \bar{a} } +   G^K_{ aa}  G^R_{ \bar{a} \bar{a} }   ]
\Bigl\},
 \\
 \Sigma^{ A}_{2,  a \bar{a}} ( t , t^{\prime} )  = & \left\{ \mbox{replace $R \rightarrow A$
 in the above expression for $\Sigma^{R}_{2, a \bar{a}} (t,t')$ } \right\},
\\
\Sigma^{K}_{2, a \bar{a}} ( t , t^{\prime} )  = &
  - \frac{u^2}{2} \Bigl\{ G^K_{ a \bar{a} } [ (G^R_{ \bar{a} a })^2 +    (G^A_{ \bar{a} a })^2 +
 (G^K_{ \bar{a} a })^2 ]
 +  2 [ G^R_{ a \bar{a} } G^R_{ \bar{a} a } +   G^A_{ a \bar{a} } G^A_{ \bar{a} a }   ]  G^K_{ \bar{a} a }
 \nonumber
 \\
 & +
   2 G^K_{ \bar{a} a} [ G^R_{ aa} G^R_{ \bar{a} \bar{a}} +
G^A_{ aa} G^A_{ \bar{a} \bar{a}} + G^K_{ aa} G^K_{ \bar{a} \bar{a}} ]
+ 2 G^R_{ \bar{a} a} [ G^K_{ aa} G^R_{ \bar{a} \bar{a}} + G^R_{ aa} G^K_{ \bar{a} \bar{a}} ]
+ 2 G^A_{ \bar{a} a} [ G^K_{ aa} G^A_{ \bar{a} \bar{a}} + G^A_{ aa} G^K_{ \bar{a} \bar{a}} ]
\Bigr\}
 \nonumber
 \\
  = &   \frac{u^2}{2} \Bigl\{ G^K_{ a \bar{a} } [ (G^I_{ \bar{a} a })^2 -
 (G^K_{ \bar{a} a })^2 ] + 2  G^I_{ a \bar{a} }
  G^I_{ \bar{a} a }     G^K_{ \bar{a} a }
 + 2 G^K_{ \bar{a} a} [ G^I_{ aa}  G^I_{ \bar{a} \bar{a}} - G^K_{ aa} G^K_{ \bar{a} \bar{a}} ]
 + 2   G^I_{ \bar{a} a}  [ G^K_{ aa} G^I_{ \bar{a} \bar{a}}
+   G^I_{ {a} a } G^K_{ \bar{a} \bar{a}}] \Bigr\},
 \end{align}
 \end{subequations}
where $G^I_{\sigma \sigma^{\prime}} ( t , t^{\prime})$ is the matrix element
of the matrix $\hat{G}^I$ defined in Eq.~(\ref{eq:GIdef}), i.e.,
 \begin{equation}
 G^I_{ \sigma \sigma^{\prime} } ( t , t^{\prime} ) = i [
G^R_{ \sigma \sigma^{\prime} } ( t , t^{\prime} ) -
G^A_{ \sigma \sigma^{\prime} } ( t , t^{\prime} )],
 \end{equation}
and we have used the fact that $G^R  ( t , t^{\prime} ) G^A  ( t , t^{\prime} )=0$.
The corresponding self energies  $\Sigma^{ R}_{2, \bar{a} {a}} (t,t')$,
 $\Sigma^{ A}_{ 2,\bar{a} {a}} (t,t')$, and  $\Sigma^{ K}_{2, \bar{a} {a}} (t,t')$ can be obtained by simply
exchanging $a \leftrightarrow \bar{a}$ in the above expressions.
The anomalous components of the self-energy are
 \begin{subequations}
 \begin{align}
 \Sigma^{ R}_{2, a {a}} ( t , t^{\prime} )  = &
 - \frac{u^2}{2} \Bigl\{  G^R_{ a {a} } [ (G^R_{ \bar{a} \bar{a} })^2 + (G^K_{ \bar{a} \bar{a} })^2 ]
 +  2 G^K_{ a {a} } G^R_{ \bar{a} \bar{a} } G^K_{ \bar{a} \bar{a} }
 + 2 G^R_{ \bar{a} \bar{a} }  [ G^R_{a \bar{a}} G^R_{ \bar{a} {a} } + G^K_{ a \bar{a}}
G^K_{ \bar{a} {a} } ]
+ 2 G^K_{ \bar{a} \bar{a} }  [ G^R_{a \bar{a}} G^K_{ \bar{a} {a} } +   G^K_{ a \bar{a}}
G^R_{ \bar{a} {a} }   ]
\Bigl\},
 \label{eq:Sigma2Raa}
 \\
 \Sigma^{ A}_{2, a {a}} ( t , t^{\prime} )  = & \left\{ \mbox{replace $R \rightarrow A$
 in the above expression for $\Sigma^{R}_{2, a {a}} (t,t')$} \right\},
 \label{eq:Sigma2Aaa}
\\
\Sigma^{ K}_{2, a {a}} ( t , t^{\prime} )  = &
 - \frac{u^2}{2} \Bigl\{ G^K_{ a {a} } [ (G^R_{ \bar{a} \bar{a} })^2 +    (G^A_{ \bar{a} \bar{a} })^2 +
 (G^K_{ \bar{a} \bar{a} })^2 ]
 +  2 [ G^R_{ a {a} } G^R_{ \bar{a} \bar{a} } +   G^A_{ a {a} } G^A_{ \bar{a} \bar{a} }   ]  G^K_{ \bar{a} \bar{a} }
 \nonumber
 \\
 &
 +  2 G^K_{ \bar{a} \bar{a}} [ G^R_{ a \bar{a}} G^R_{ \bar{a} {a}} +
G^A_{ a \bar{a}} G^A_{ \bar{a} {a}} + G^K_{ a \bar{a}} G^K_{ \bar{a} {a}} ]
+ 2 G^R_{ \bar{a} \bar{a}} [ G^K_{ a \bar{a}} G^R_{ \bar{a} {a}} + G^R_{ a \bar{a}} G^K_{ \bar{a} a} ]
+ 2 G^A_{ \bar{a} \bar{a}} [ G^K_{ a \bar{a}} G^A_{ \bar{a} {a}}
 + G^A_{ a\bar{a}} G^K_{ \bar{a} {a}} ]
\Bigr\}
 \nonumber
 \\
  = &
  \frac{u^2}{2} \Bigl\{ G^K_{ a {a} }
[ (G^I_{ \bar{a} \bar{a} })^2 -   (G^K_{ \bar{a} \bar{a} })^2 ]
 +  2 G^I_{ a {a} } G^I_{ \bar{a} \bar{a} }   G^K_{ \bar{a} \bar{a} }
 + 2 G^K_{ \bar{a} \bar{a}} [ G^I_{ a \bar{a}}  G^I_{ \bar{a} {a}} - G^K_{ a\bar{a}} G^K_{ \bar{a} {a}} ]
 + 2   G^I_{ \bar{a} \bar{a}}  [ G^K_{ a \bar{a}} G^I_{ \bar{a} {a}}
+   G^I_{ {a} \bar{a} } G^K_{ \bar{a} {a}}] \Bigr\}.
 \label{eq:Sigma2Kaa}
\end{align}
 \end{subequations}
Finally, the conjugate anomalous self-energies
 $\Sigma^{ R}_{ 2,\bar{a} \bar{a}} (t,t') $,
 $\Sigma^{ A}_{ 2, \bar{a} \bar{a}} (t,t')$, and  $\Sigma^{ K}_{2, \bar{a} \bar{a}} (t,t')$ can be obtained by
exchanging $a \leftrightarrow \bar{a}$ on both sides of
Eqs.~(\ref{eq:Sigma2Raa}-\ref{eq:Sigma2Kaa}).
In order to calculate the ``out-scattering term'' (\ref{eq:Ccolout})
in the kinetic equation, we need only the difference between retarded and advanced self-energies,
which to second order in  the interaction  can be written as
 \begin{eqnarray}
\Sigma^{ I}_{2, a \bar{a}} ( t , t^{\prime}  )  \equiv  i [
  \Sigma^{R}_{2, a \bar{a}} ( t , t^{\prime}  )
-  \Sigma^{A}_{2,  a \bar{a}} ( t , t^{\prime}  )  ] & = &
 \frac{u^2}{2} \Bigl\{  G^I_{ a \bar{a} } [ (G^I_{ \bar{a} a })^2 - (G^K_{ \bar{a} a })^2 ]
 -  2 G^K_{ a \bar{a} } G^I_{ \bar{a} a } G^K_{ \bar{a} a }
 \nonumber
 \\
& &\hspace{5mm}
 + 2 G^I_{ \bar{a} a }  [ G^I_{aa} G^I_{ \bar{a} \bar{a} } - G^K_{ aa} G^K_{ \bar{a} \bar{a} } ]
- 2 G^K_{ \bar{a} a }  [ G^I_{aa} G^K_{ \bar{a} \bar{a} } +   G^K_{ aa}  G^I_{ \bar{a} \bar{a} }   ]
\Bigl\},
\label{eq:Aaabar2out}
\\
 \Sigma^{ I}_{2, a {a}} ( t , t^{\prime}  )  \equiv  i [
  \Sigma^{R}_{2, a {a}} ( t , t^{\prime}  )
-  \Sigma^{A}_{2, a {a}} ( t , t^{\prime}  )  ] & = &
  \frac{u^2}{2} \Bigl\{  G^I_{ a {a} } [ (G^I_{ \bar{a} \bar{a} })^2 - (G^K_{ \bar{a} \bar{a} })^2 ]
 -  2 G^K_{ a {a} } G^I_{ \bar{a} \bar{a} } G^K_{ \bar{a} \bar{a} }
 \nonumber
 \\
& &\hspace{5mm}
 + 2 G^I_{ \bar{a} \bar{a} }  [ G^I_{a \bar{a}} G^I_{ \bar{a} {a} } - G^K_{ a \bar{a}}
G^K_{ \bar{a} {a} } ]
- 2 G^K_{ \bar{a} \bar{a} }  [ G^I_{a \bar{a}} G^K_{ \bar{a} {a} } +   G^K_{ a \bar{a}}
G^I_{ \bar{a} {a} }   ]
\Bigl\}.
 \label{eq:Aaa2out}
\end{eqnarray}
The functions
$\Sigma^{I}_{  2,\bar{a}a} (t,t')$ and
$\Sigma^{ I}_{2, \bar{a} \bar{a}} (t,t')$ can again be obtained by exchanging $a \leftrightarrow \bar{a}$
in the above expressions.

\twocolumngrid

\renewcommand{\theequation}{D\arabic{equation}}
\section*{APPENDIX D: SOLUTION OF THE TWO-TIME KINETIC EQUATIONS FOR THE TOY MODEL}
\label{sec:appendixD}
\setcounter{equation}{0}

\begin{figure}[tb]
  \centering
\includegraphics[width=80mm]{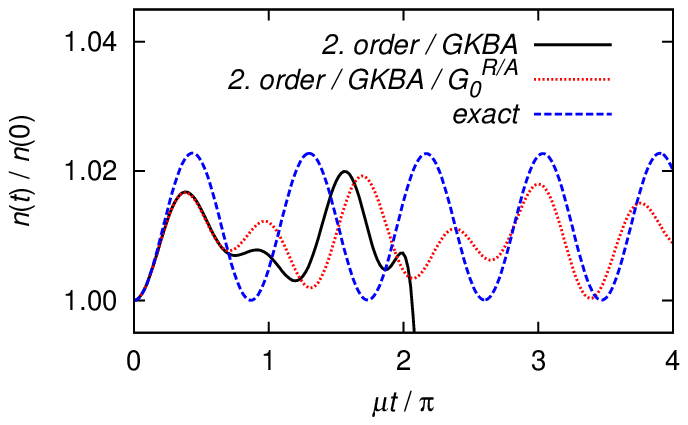}
\\
\includegraphics[width=80mm]{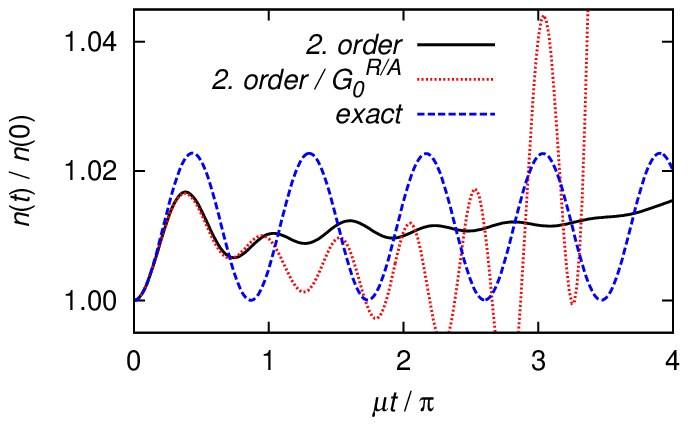}
  \vspace{-4mm}
  \caption{%
(Color online) Time evolution of the diagonal
distribution function for the second-order self-energy without Markov approximation.
We compare the  result with GKBA (top) and the full two-time result without 
GKBA (bottom).
For the retarded and advanced propagators we used the renormalized (solid line) or the free propagators (dashed line).
The parameters and initial conditions are the same as in the middle panel in Fig.~\ref{fig:mean-field}.
The dashed line is again the exact solution.
}
  \label{fig:two-time}
\end{figure}

In this appendix we  examine the validity of the  three
approximations  made in Sec.~\ref{sec:pert}
to obtain a closed system of equations for the equal-time 
Keldysh-Green function of our toy model:
the Kadanoff-Baym ansatz, the Markov
approximation, and the neglected renormalization of the retarded and
advanced propagators.
For simplicity, we focus on the kinetic equations for our toy model 
with the perturbative second-order self-energy. We compare 
different combinations of these approximations and their influence on the results. 
For our toy model we can obtain the quantum dynamics
without relying on any of these  approximations by solving
the two-time quantum dynamic partial differential equations  numerically. 
Technically, this is almost as simple as solving 
a system of 
 ordinary differential equations in the equal-time formalism,
 except that now we have to propagate in two different time directions $t$ and $t'$. 
The collision integrals were calculated numerically using the trapezoidal rule.
To shorten the presentation, we will concentrate on the dynamics of the normal 
pair-correlator with the interaction strength $u/\epsilon = 0.1$ 
corresponding to the middle panel in Fig.\ \ref{fig:pert}. 
The pumping strength $|\gamma|$ and the initial conditions are the same as before.

To begin with, we have solved the kinetic equation (\ref{eq:kinnok}) for $G^K$  
using the GKBA but 
without Markov approximation. The results are shown in the upper panel 
of Fig.\ \ref{fig:two-time} where we compare two different variants, depending how the retarded and advanced Green functions entering the collision integral in Eq.\ (\ref{eq:kinnok}) are handled. In the first case we solved the kinetic equation (\ref{eq:kinnok}) for $G^K$ together with the equations for the renormalized 
retarded and the advanced Green functions, which 
can be obtained from Eq.\ (\ref{eq:kinEqGrGa}) 
by simply omitting the momentum labels. The set of kinetic equations was solved self-consistently using only the GKBA  in the collision integrals. This combination turned out to be quite unstable and the solutions of the equations diverge slightly above the time $\mu t/ \pi = 2$. We have checked that the divergence is not an artifact of our grid discretization.
In the second case we used the free retarded and advanced Green functions in the collision integral of the kinetic equation (\ref{eq:kinnok}). The solution is again stable but shows an irregular dynamics in comparison with the exact solution. 
Compared to the analogous results 
relying in addition on the  Markov approximation 
shown in  Fig.\ \ref{fig:pert} we did not
 find any improvement.

Next, we additionally avoided the GKBA. In the lower panel of 
Fig.\ \ref{fig:two-time} we compare the two different variants using the 
full renormalized or the free retarded and advanced propagators in the collision integral. 
In the first case with the full renormalized quantities, the set of kinetic equations (\ref{eq:kinEqGrGa},\ref{eq:kinintnonequal},\ref{eq:kinintnonequal2}) without momentum labels was solved simultaneously. The dynamics is stable, but the oscillation amplitude disappears nearly completely. In the variant with the free advanced and retarded Green functions the dynamics changes completely and the solution shows large oscillations of the pair-correlator amplitude.
Again we do not observe any improvements towards the correct solution.

In summary,
despite the known limitations of the GKBA and the Markov approximations, 
we have not found any improvements in the full two-time approach. 
Note that
a similar comparison of the GKBA with full and free propagators,
respectively, was performed for semiconductors in Ref.\ [\onlinecite{Kwong98}], with the
conclusion that the full GKBA performs rather well. 
At this point it is not clear to us if the different behavior in the present paper is due to a 
breakdown of standard perturbation theory 
or simply a special feature of the toy model, which contains no intrinsic dissipation.

\end{appendix}

 \bibliography{lit} 

\end{document}